\documentclass[onecolumn,floatfix,showpacs,rmp,aps]{revtex4}
\pdfoutput=1
\usepackage{amssymb}
\usepackage{enumerate}
\usepackage{braket}
 \usepackage{amsmath} 
\usepackage{subfigure}
\usepackage{lipsum}
\usepackage{amsfonts} 
\usepackage{amssymb, mathrsfs}
\usepackage{amsmath}
\usepackage{bm}
\usepackage[pdftex]{graphicx}
\usepackage{footnote}
\def\beq{\begin{equation}}
\def\eeq{\end{equation}}
\def\bsp{\begin{split}}
\def\esp{\end{split}}
\def\bea{\begin{eqnarray}}
\def\eea{\end{eqnarray}}
\def\ba{\begin{array}}
\def\ea{\end{array}}
\def\lb{\left(}
\def\rb{\right)}

\DeclareMathOperator\sn{sn}
\DeclareMathOperator\cn{cn}
\DeclareMathOperator\dn{dn}
\DeclareMathOperator\cd{cd}
\DeclareMathOperator\arccosh{arccosh}
\DeclareMathOperator\arctanh{arctanh}
\DeclareMathOperator\arccot{arccot}
\begin{document}
\title{Macroscopic quantum tunneling and quantum-classical phase transitions of the escape rate in large spin systems}

\author{S. A. Owerre}
\email{solomon.akaraka.owerre@umontreal.ca}
\author{M. B. Paranjape} 
\email{paranj@lps.umontreal.ca}
\affiliation{Groupe de physique des particules, D\'epartement de physique,
Universit\'e de Montr\'eal,
C.P. 6128, succ. centre-ville, Montr\'eal, 
Qu\'ebec, Canada, H3C 3J7 }

\begin{abstract}
This article presents a review on the theoretical and the experimental developments on macroscopic quantum tunneling and quantum-classical phase transitions of the escape rate in large spin systems.  A substantial amount of research work has been done in this area of research over the years, so this article does not cover all the  research areas that have been studied, for instance the effect of dissipation is not discussed and can be found in other review articles.  We present the basic ideas with simplified calculations so that it is readable to both specialists and nonspecialists in this area of research. A brief derivation of the  path integral formulation of quantum mechanics in its original form using the orthonormal position and momentum basis is reviewed. For tunneling of a particle into the classically forbidden region, the imaginary time (Euclidean) formulation of path integral  is useful, we review this formulation and apply it to the problem of tunneling in a double well potential. For spin systems such as single molecule magnets, the formulation of path integral requires the use of non-orthonormal spin coherent states in $(2s+1)$ dimensional Hilbert space, the coordinate independent and the coordinate dependent form of the spin coherent state path integral are derived. These two (equivalent) forms of spin coherent state path integral are applied to the tunneling of single molecule magnets through a magnetic anisotropy barrier. Most experimental and numerical results are presented. The suppression of tunneling for half-odd integer spin (spin-parity effect) at zero magnetic field is derived using both forms of spin coherent state path integral, which shows that this result (spin-parity effect) is independent of the choice of coordinate.  At nonzero magnetic field we present both the experimental and the theoretical results of the oscillation of tunneling splitting as a function of the applied magnetic field applied along the spin hard anisotropy axis direction. The  experimental and the theoretical results of the tunneling in antiferromagnetic exchange coupled dimer model are also reviewed.  As the spin coherent state path integral formalism is a semi-classical method, an alternative exact mapping of a spin Hamiltonian to a particle Hamiltonian with  a  potential field (effective potential method) is derived. This effective potential method allows for the investigation of quantum-classical phase transitions of the escape rate in large spin systems. We present different methods for investigating quantum-classical phase transitions of the escape rate in large spin systems. These methods are applied to different spin models.
\end{abstract}
\maketitle
\tableofcontents

\section{Introduction}
\label{sec:intro}
One of the remarkable manifestations of quantum mechanics is the concept of quantum tunneling. This involves the presence of a potential barrier, that is  the region where the potential energy is greater than the energy of the particle. In classical mechanics, the tunneling  of a particle through this barrier is prohibited as it requires the particle to have a negative kinetic energy, however, in quantum mechanics we find a nonzero probability for finding the particle in the classically forbidden region. Thus, a quantum particle can tunnel through the barrier. In one dimensional systems, the tunneling amplitude (whose modulus squared gives the probability) is usually computed using  two fundamental methods, namely,  the Wentzel-Kramers-Brillouin (WKB) method \cite{ll1} and the ``instanton'' method\cite{cole1,cole3,poly,langer,dhn,ges,gjs,jr} via the Feynman path integral formulation\cite{fey} of quantum mechanics. The term ``instanton'' refers to classical solutions of the equations of motion when the time coordinate has been continued to Euclidean time, $t\rightarrow -i\tau$.  For particle in a double well potential with two degenerate minima, the basic understanding is that in the absence of tunneling the classical ground states of the system, which correspond to the minima of the potential, remain degenerate. Tunneling lifts this degeneracy and the true ground state and the first excited state become the symmetric and antisymmetric linear superposition of the classical ground states with an energy splitting between them\cite{ll1,cole3}. In some cases the two minima of the potential are not degenerate. The state with lower energy is the true vacuum, while the state with higher energy is the false vacuum, which is then rendered unstable due to quantum tunneling. In this case one looks for the decay rate of the false vacuum\cite{cole1,cole2}.  Such a scenario plays a vital role in cosmology, especially in the theory of early universe and inflation. Additionally, in some quantum systems, tunneling does not involve the splitting of the classical ground states or the decay of the false vacuum, but rather a dynamic oscillation of the (phase) difference between two macroscopic order parameters \cite{coo},  which are separated by a thin normal layer, through tunneling of the microscopic effective excitations, such as Cooper pairs as in Josephson effect\cite{jo,manu}. 

In the last few decades, the tunneling phenomenon has been extended to other branches of physics. Tunneling has been predicted in single, molecular,  large magnetic spin systems such as MnAc$_{12}$,  Mn$_{12}$ and Fe$_{8}$\cite{chud1,van,em, wern}.  These single molecule magnets (SMMs) are composed of several molecular magnetic ions, whose spins are coupled by intermolecular interactions giving rise to  an effective single giant spin, which can tunnel through its magnetic anisotropy barrier, hence the name ``macroscopic quantum spin tunneling\footnote{In most literature, macroscopic quantum tunneling refers to tunneling in a bias (metastable) potential while macroscopic quantum coherence refers to tunneling in a potential with degenerate minima \cite{aj}. We will use the former to refer to both systems.}". \textcite{van} first studied the tunneling  in a uniaxial ferromagnetic spin model with an applied magnetic field using the WKB method.  \textcite{em} considered a biaxial model with a magnetic field using instanton technique, subsequently, \textcite{chud1} studied a more general biaxial spin model by solving the instanton trajectory of the Landau Lifshitz equation. These studies were based on a semi-classical description, that is by representing the spin operator as a unit vector parameterized by  spherical coordinates. In this description,  the spin  is represented by a particle on a two-dimensional sphere $\mathcal{S}^2$, however, in the presence of a topological term, called the Berry's phase term or Wess-Zumino action \cite{berry84,wesszu,witt}, which effectively corresponds to the magnetic field of a magnetic monopole at the centre of the two sphere.  Based on this semi-classical description, it was predicted that for integer spins tunneling is allowed, while for half-odd integer spin tunneling is completely suppressed   at zero (external)  magnetic field \cite{loss1, hd}.  The vanishing of tunneling for half-odd integer spins is understood as a consequence of destructive interference between tunneling paths, which is directly related to Kramers degeneracy\cite{kram,mess} due to the time reversal invariance of the Hamiltonian. In the presence of a magnetic field applied along the spin hard axis, \textcite{anu2} showed that the tunneling splitting does not vanish for half-odd integer spins, but rather oscillates with the field and only vanishes at a certain critical value of the field, which was later observed experimentally in Fe$_8$ molecular cluster \cite{wern,wern0,sess}. In this case tunneling suppression is not related to  Kramers degeneracy due to the presence of a magnetic field. 

An exact mapping of spin system was considered by 
\textcite{swh} and \textcite{zas1,zas2}. They studied the exact mapping of a spin system unto a particle in a potential field in contrast to   the semi-classical approach. This method, which is called the effective potential method, deals with an exact correspondence between a spin Hamiltonian and a particle in a potential field. It gives the possibility for investigating spin tunneling just like a particle in a one-dimensional double well potential. In recent years spin tunneling effect has been observed in many small ferromagnetic spin particles such as Fe$_8$ \cite{san}, Mn$_{12}$Ac \cite{foss1,zha1,thm},  in ferrimagnetic nanoparticles \cite{wern1} and also in antiferromagnetic particles \cite{asg,gider,tej},antiferromagnetic exchange coupled dimer [Mn$_4$]$_2$ \cite{hill,tiron} and antiferromagnetic ring clusters with even number of spins \cite{me,ml2001, taft}. These molecular magnets also play a decisive role in quantum computing \cite{ll2001, tej1}. An extensive review on the experimental analysis of  SMMs can be found in \cite{gs}.
\begin{figure}[ht]
\centering
\subfigure[ ]{%
\includegraphics[scale=0.35]{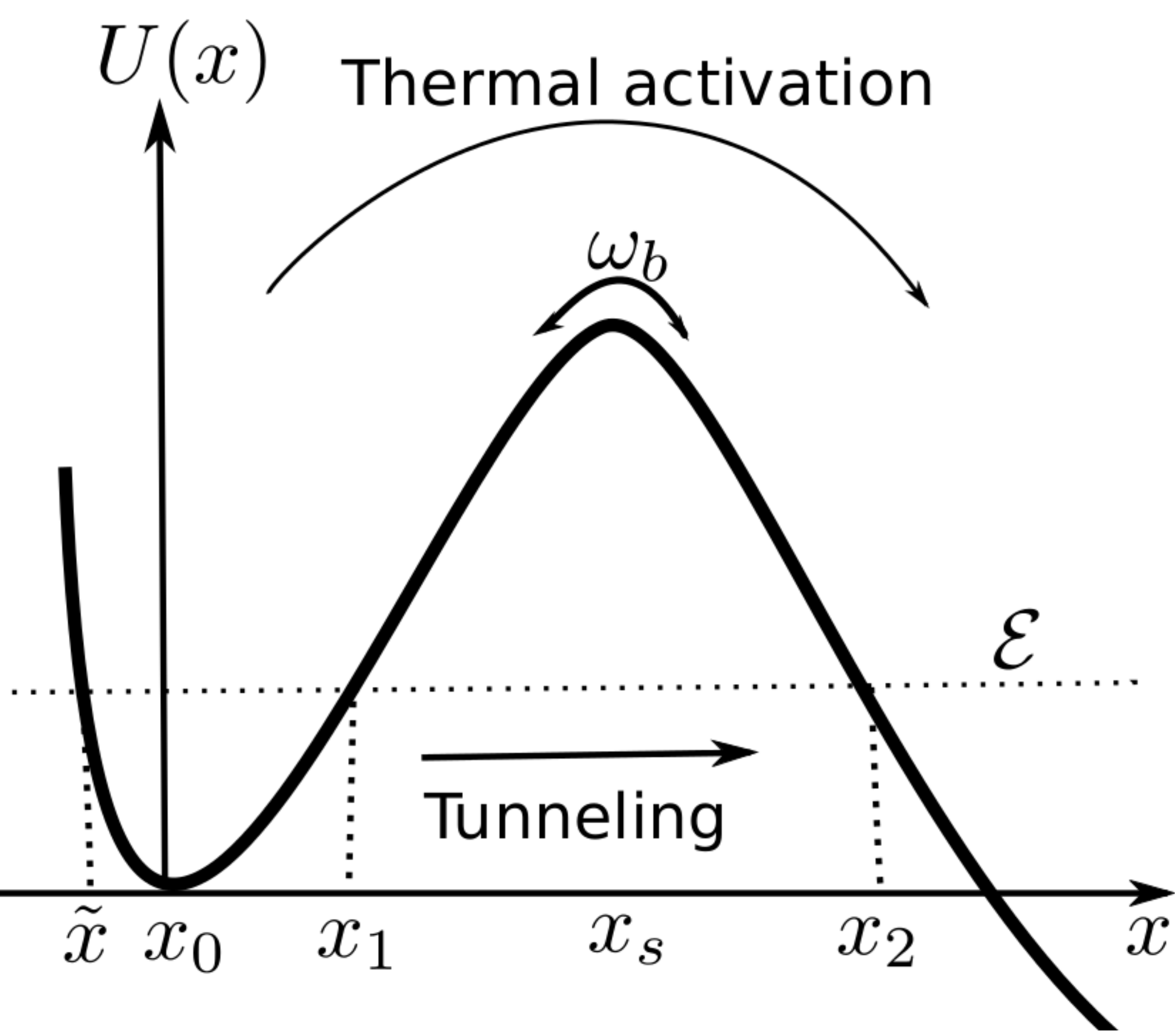}}
\subfigure[]{%
\includegraphics[scale=0.35]{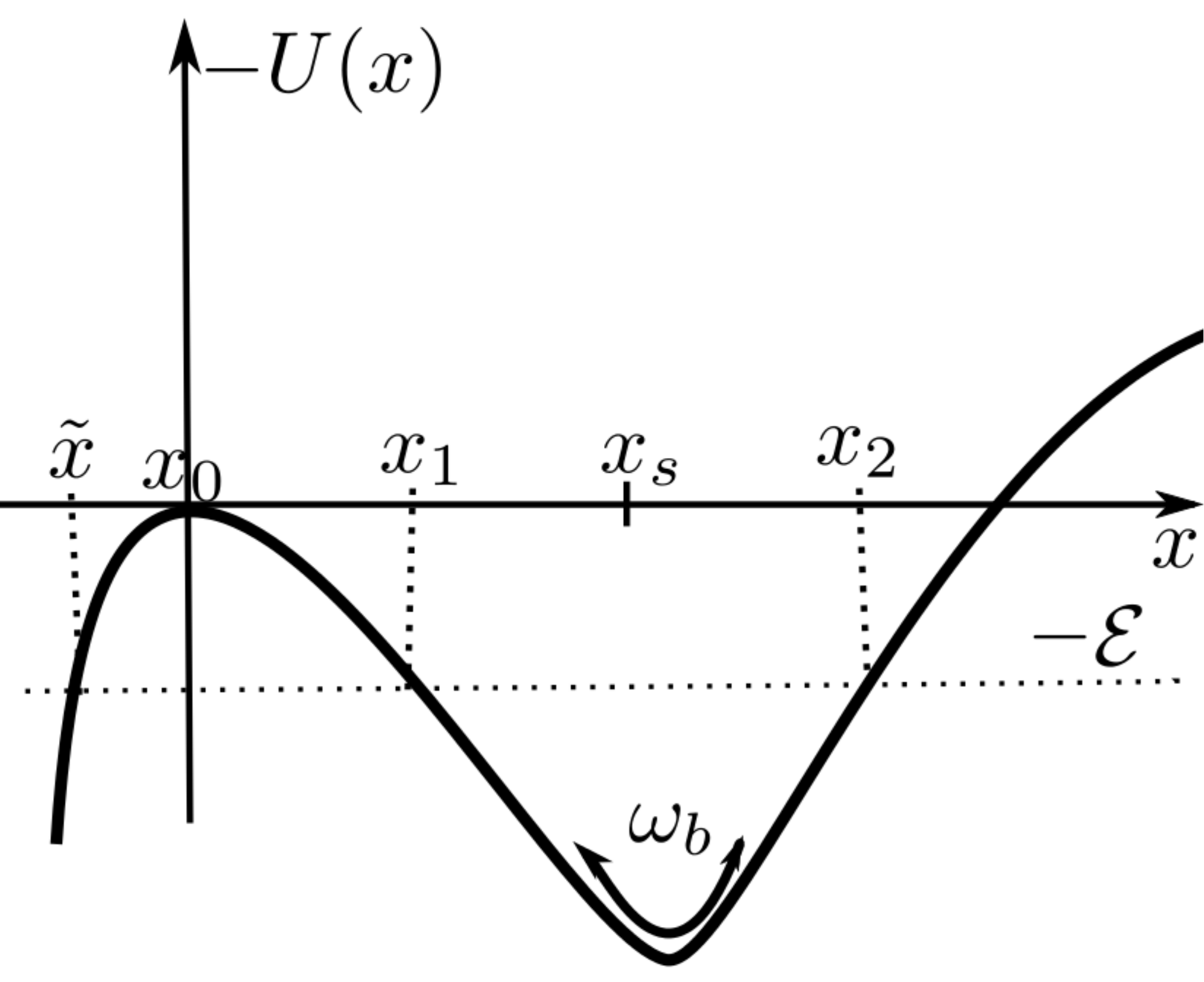}
\label{pot29}}
\caption{(a): A sketch of a metastable potential showing the regions of quantum tunneling at low temperature and classical thermal activation at high temperature. (b): The inverted potential. The coordinates $x_1$ and $x_2$ are the classical turning points.}
\label{pot23}
\end{figure}

The possibility of quantum tunneling, which is mediated by a vacuum instanton trajectory, requires a very low temperature $T\rightarrow 0$. For pure quantum tunneling, the transition amplitude in the stationary phase approximation is $\Gamma= \mathscr{A}e^{-B}$, where $B$ is the vacuum instanton action and $\mathscr{A}$ is a pre-factor. At nonzero temperature, quantum tunneling becomes inconsequential, then the particle has the possibility of crossing over the barrier, a process called classical thermal activation (see Fig.\eqref{pot23}). The study of thermal activation dates back to the work of \textcite{kram1} for the diffusion of a particle over the barrier. A review of this subject for both particle and spin system can be found in the existing literature\cite{han,scb,ckw}. In this case the transition is governed by the Van't Hoff-Arrhenius Law \cite{han} $\Gamma= \mathscr{B}e^{-\beta\Delta U}$, where $\Delta U$ is the height of the potential barrier , $\beta$ is the inverse temperature and $\mathscr{B}$ is a pre-factor. 

The basic understanding of quantum-classical phase transitions of the escape rate is as follows: for a particle in a metastable cubic potential or double well quartic parabolic potential $U(x)$, with no environmental influence (dissipation), transition at finite temperature is dominated by thermon (periodic) instanton trajectory\footnote{This is simply the solution of the imaginary time classical equation of motion with an energy $\mathcal E$.}, whose action is given by $\mathcal{S}_p(\mathcal E)$ \cite{chud6}, where $\mathcal{E}$ is the energy of the particle in the inverted potential $-U(x)$. The escape rate is defined by taking the Boltzmann average over tunneling probabilities at finite energy\cite{aff}.   At the bottom of the barrier we have $\mathcal{S}_p(\mathcal E) \to S(U_\text{min})$, where $S(U_\text{min})$ is the action at the bottom of the barrier, while at the top of the barrier $\mathcal{S}_p(\mathcal E) \to \mathcal{S}_0=\beta\Delta U$, which is the action of a constant trajectory at the top of the barrier. 

Now, if we compare the plot of the thermon action $\mathcal{S}_p$ and that of the thermodynamic action $\mathcal{S}_0$ against temperature\cite{chud6}, there exist a critical temperature $T^{c}$ at which the thermodynamic action  crosses the thermon action. If this intersection is \textit{sharp}, the critical temperature $T^{c}$ can be thought of as a first-order  ``phase transition" (crossover) temperature  from classical (thermal) to quantum regimes. At this temperature  $T^{c}=T_0^{(1)}$, there is a discontinuity in the first-derivative of the action $\mathcal{S}_p$\cite{blatter}. The approximate form of this crossover temperature can be estimated by comparing the quantum action $S(U_\text{min})$ at the bottom of the barrier and that of the classical action at the top of the barrier $\mathcal{S}_0$ \cite{scb}\footnote{Actually, the thermon action is defined over the whole period of oscillation of a particle in the inverted potential. In other words, the particle crosses the barrier twice. Thus, $B=S(U_\text{min})/2$ as the vacuum instanton is defined by half of the whole period. }:
\bea
T_0^{(1)}=\frac{1}{\beta_0^{(1)}} = \frac{\Delta U}{S(U_\text{min})}= \frac{\Delta U}{2B}.
\label{temp0}
\eea
For a particle with a constant mass, the physical understanding for a sharp first-order phase transition to occur is that the top of the barrier should be flat\cite{chud2}. This condition is not widely accepted. It has been argued that the necessary condition for a sharp first-order phase transition to occur is that the top of the barrier should be wider so that tunneling through the barrier from  the ground state is more auspicious than that from the excited states\cite{zha3}.  For a particle with a position dependent mass, the necessary condition for a sharp first-order phase transition to occur requires the mass of the particle at the top of the barrier to be heavier than that at the bottom of the barrier. In this case tunneling from higher excited states is inauspicious. Thus, thermal activation competes with ground state tunneling leading to first-order phase transition. Thermally assisted tunneling (TAT), that is tunneling from excited states which reduces to ground state tunneling  at $T=0$ occurs for temperatures below $T_0^{(1)}$. In this case the particle tunnels through the barrier at the most favourable energy $\mathcal{E}(T)$, which goes from the top of the barrier to the bottom of the barrier as the temperature decreases \cite{chud2}. 
 
However, if the  intersection of  the thermon action $\mathcal{S}_p$ and that of the thermodynamic action $\mathcal{S}_0$ is \textit{smooth}, the critical temperature is said to be of second-order $T^{c}=T_0^{(2)}$. The second derivative of the thermon action in this case has a jump at $T_0^{(2)}$. This crossover temperature is defined as \cite{gold,gold1}
\bea
T_0^{(2)} =\frac{1}{\beta_0^{(2)}}=\frac{\omega_b}{2\pi},
\label{temp1}
\eea
where $\omega_b$ is the frequency of oscillation at the bottom of the inverted potential $-U(x)$, that is $\omega_b^2= -\frac{U^{\prime\prime}(x_{s})}{m}$.  This formula follows from equating the Van't Hoff-Arrhenius exponential factor $\beta \Delta U$ at finite nonzero temperature and the approximate form of the WKB exponential factor $2\pi \Delta U/\omega_b$ at zero temperature. 

Using functional integral approach, \textcite{aff} and \textcite{ll2,ll3} demonstrated that, in the regime $T<T_0^{(2)}$, there is a competing effect between thermal activation and quantum tunneling leading to TAT.  For $T \gg T_0^{(2)}$, quantum tunneling is suppressed and assisted thermal activation becomes the dominant factor in the escape rate. For $T\approx T_0^{(2)}$, the two regimes smoothly join with a jump of the second derivative of the escape rate. Thus, $T_0^{(2)}$ corresponds to the crossover temperature from thermal regime to TAT.  In term of the potential, for a constant mass particle a smooth second-order crossover is favourable with a potential with a parabolic barrier top. An alternative  criterion for the first- and the second-order quantum-classical phase transitions was demonstrated by \textcite{chud6} based on the shape of the potential.  He showed that for a first-order phase transition, the period of oscillation $\beta(\mathcal{E})$ is nonmonotonic function of $\mathcal{E}$, in other words, $\beta(\mathcal{E})$ has a minimum at some point $\mathcal{E}_0<\Delta U$ and then rises again, while for second-order phase transition $\beta(\mathcal{E})$ is monotonically increasing with decreasing $\mathcal{E}$. \textcite{mull4} derived a general criterion formula for investigating  first- and second-order phase transitions, which is similar to the criterion  formula derived by \textcite{kim1}.
 
In this report, we will review the theoretical and  the experimental developments on macroscopic quantum tunneling and quantum-classical phase transitions of the escape rate in large spin systems. The article is organized as follows.  In section\eqref{sec:spinpa}, we will introduce the basic idea of path integral for a one-dimensional particle from Feynman point of view and review its application to the tunneling of a particle in a double well potential. In section\eqref{sec:spinco} we will apply this idea to spin systems using spin coherent states. The path integral for spin systems will be derived in the the coordinate independent form. We will show the steps on how to move from  coordinate independent to coordinate dependent form. In section \eqref{mactun} we will then apply this coordinate dependent formalism to  tunneling problem of SMMs. The quantum phase interference (quenching of tunneling splitting) will be derived and some experimental results will be presented.  Due to lack of solution of these models in coordinate independent form in most of the literature, we will show that both the instanton trajectory and the quantum phase interference can be recovered using the coordinate independent formalism. We will further extend our consideration to tunneling in an exchange coupled dimer model and to an antiferromagnetic spin model in general.  Section\eqref{epmth} deals with the effective potential method, we will review the mapping of a large spin model onto a particle Hamiltonian that consists of a potential energy and a mass. In section\eqref{pt} we will present different methods for studying the quantum-classical phase transitions of the escape rate. We will also apply these methods to both SMMs and exchange coupled dimer model. Theoretical, numerical and experimental results will be presented.  In section\eqref{con} we will summarize our analysis and comment on their significance. 

\section{Path integral formulation}

\subsection{Position state path integral}
\label{sec:spinpa}

In this section we start with a brief review of path integral formulation of quantum mechanics. This formulation is an elegant alternative method of quantum mechanics. It reproduces the Schr\"odinger  formulation of quantum mechanics and the principle of least action in classical mechanics.  In this method the classical action enters into the calculation of a quantum object, the transition amplitude, thereby allowing for a quantum interpretation of a solution of the classical equations of motion. The basic idea of the path integral is that unlike a classical particle with a unique trajectory or path, a quantum particle follows an infinite set of possible trajectories to go from an initial state say $\ket{x }$ at $t=0$ to a final state say $\ket{x^{\prime}}$ at time $t= t^\prime$. The sum over all the possible paths (histories of the particle) appropriately weighted, determines the quantum amplitude of the transition. The weight for each path is exactly the phase  corresponding  to the exponential of the classical action of the path,  multiplied by the imaginary number $i$. 
Consider a particle moving in one dimension, the Hamiltonian of this system is of  usual form:
\bea
\hat H =\frac{\hat{p}^2}{2m} + U(\hat{x}).
\label{onedin}
\eea
Let us introduce the complete, orthonormal eigenstates of the position  $\hat{x}$ and the momentum $\hat{p}$ operators:
\begin{align}
\hat{x}\ket{x}&=x\ket{x}, \quad \hat{p}\ket{p}=p\ket{p}\label{prop0},\\\braket{x^{\prime}|x}&=\delta(x^{\prime}-x), \quad \braket{p^{\prime}|p}=\delta(p^{\prime}-p),
\label{prop1}
\end{align}
with
\bea
\braket{x|p}= e^{ipx/\hbar}.
\label{prop2}
\eea
The resolution of identities are 
\bea
\int dx\ket{x}\bra{x}=\bold{\hat I}=\int \frac{dp}{2\pi\hbar}\ket{p}\bra{p}.
\label{posre}
\eea
Expressing the unitary operator $e^{-i\hat H t}$ as $[e^{-i\hat H t/N}]^N$  and using  Eqs.\eqref{onedin}--\eqref{posre}, the transition  amplitude in the limit $N\rightarrow \infty$ is given by \cite{fey,fey1}
\begin{equation}
\mathcal{A}(x^{\prime},t^{\prime};x,0)=\braket{x^{\prime}|e^{-i\hat H t^{\prime}/\hbar }|x}=\int \mathcal{D}x(t) \thinspace e^{iS[x(t) ]/\hbar},
\label{qua}
\end{equation}
where $\mathcal{D}x(t)$ is the measure for integration over all possible classical paths $x(t)$ that satisfy the boundary conditions $x(0)=x$ and $x(t^{\prime})=x^{\prime}$, where 
\begin{equation}
S[x(t) ]=\int_{0}^{t^{\prime}} dtL, \quad L=\frac{1}{2} m\lb\frac{dx}{dt}\rb^2 - U(x),
 \label{class}
\end{equation}
is the classical action and the Lagrangian of the system. We have written down the path integral for a one-dimensional particle, generalization to higher dimensions is straightforward.

The well-known classical equation of motion can be derived in a very simple way. In the  semiclassical limit, {\it i.e.,} $\hbar\rightarrow 0$, the phase $e^{iS[x(t) ]/\hbar}$ oscillates very rapidly in such a way that nearly all paths cancel each other. The main contribution to the path integral comes from the paths for which the action is stationary, {\it i.e.,} $\delta S[x(t)]=0$, which yields the classical equation of motion. 
\subsubsection{Imaginary time path integral formalism}
The main motivation of imaginary time propagator comes from the partition function in statistical mechanics, which is given by
\bea
Z= Tr(e^{-\beta\hat{H}}),
\label{partition}
\eea
where $\beta=1/T$ is the inverse temperature of the system. Inserting the position resolution of identity in Eqn.\eqref{posre} into the RHS of Eqn.\eqref{partition} gives
\bea
Z=\int dx \mathcal{A}(x,\beta;x,0),\eea
where
\bea
\mathcal{A}(x,\beta;x,0)=\braket{x|e^{-\beta\hat{H}}|x}.
\label{partt}
\eea
Suppose we consider the time in Eqn.\eqref{qua} to be purely imaginary, which can be written as $t^{\prime}=-i\beta$, where $\beta$ is real. Then, substituting into Eqn.\eqref{qua} we obtain the propagator evaluated at imaginary time \cite{poly, ww, mackin}:

\begin{equation}
\mathcal{A}_E=\braket{x^{\prime}|e^{-\beta\hat H /\hbar }|x}=\int \mathcal{D}x(\tau) \thinspace e^{-S_E[x]/\hbar},
\label{quapa}
\end{equation}
where the action is now given by the appropriate analytical continuation of the action, nominally defined as
\begin{equation}
S_E[x]=\int_{0}^{-i\beta} dt\bigg[\frac{1}{2} m\lb\frac{dx}{dt}\rb^2 - U(x)\bigg].
 \label{class1}
 \end{equation}
Then setting $x^{\prime}=x$ in Eqn.\eqref{quapa} yields the partition function Eqn.\eqref{partt}. Thus, the propagator continued to imaginary time gives the partition function. This method is very useful in finding the ground state of a physical system in statistical physics and condensed matter physics. The analytical continuation is obtained by defining a real variable $\tau=it$, which is called the ``imaginary or Euclidean time'', we see that $\tau$ and $t$ are related as follows: $t:0 \to -i\beta$, $\tau:0 \to \beta$. Thus, $S_E[x(\tau) ]=-iS[x(t\rightarrow -i\tau) ]$. Typically,  if $S[x(t)]=\int dt (T-V)$,  the Euclidean action is given by $S_E[x(\tau)]=\int d\tau (T+V)$, as the kinetic energy changes sign with the continuation to imaginary time.
The Euclidean action and the Lagrangian  are
\bea
 S_E[x(\tau) ]=\int_{-\beta/2}^{\beta/2} d\tau L_E; \quad L_E=\frac{1}{2} m\lb\frac{dx}{d\tau}\rb^2 + U(x),
 \label{eucl}
\eea
using time translation invariance.  The  boundary conditions for the imaginary time propagator are $x(-\beta/2)=x$ and $x(\beta/2)=x^{\prime}$. 
This analysis of imaginary time propagator plays a decisive role in tunneling problems, such as that of a particle in a one dimensional double well potential, since the period of oscillation or the momentum of the particle is imaginary in the tunneling region $\mathcal{E}<\Delta U$\cite{ll1,ww}, which is neatly compensated by the imaginary time. Thus, it is almost always convenient to use imaginary time corresponding to  the replacement $t\rightarrow -i\tau$ \cite{ww,poly} when considering tunnelling problems.

\subsubsection{Instantons in the double well potential}
\label{dwp}
\begin{figure}[ht]
\includegraphics[width=2.5in]{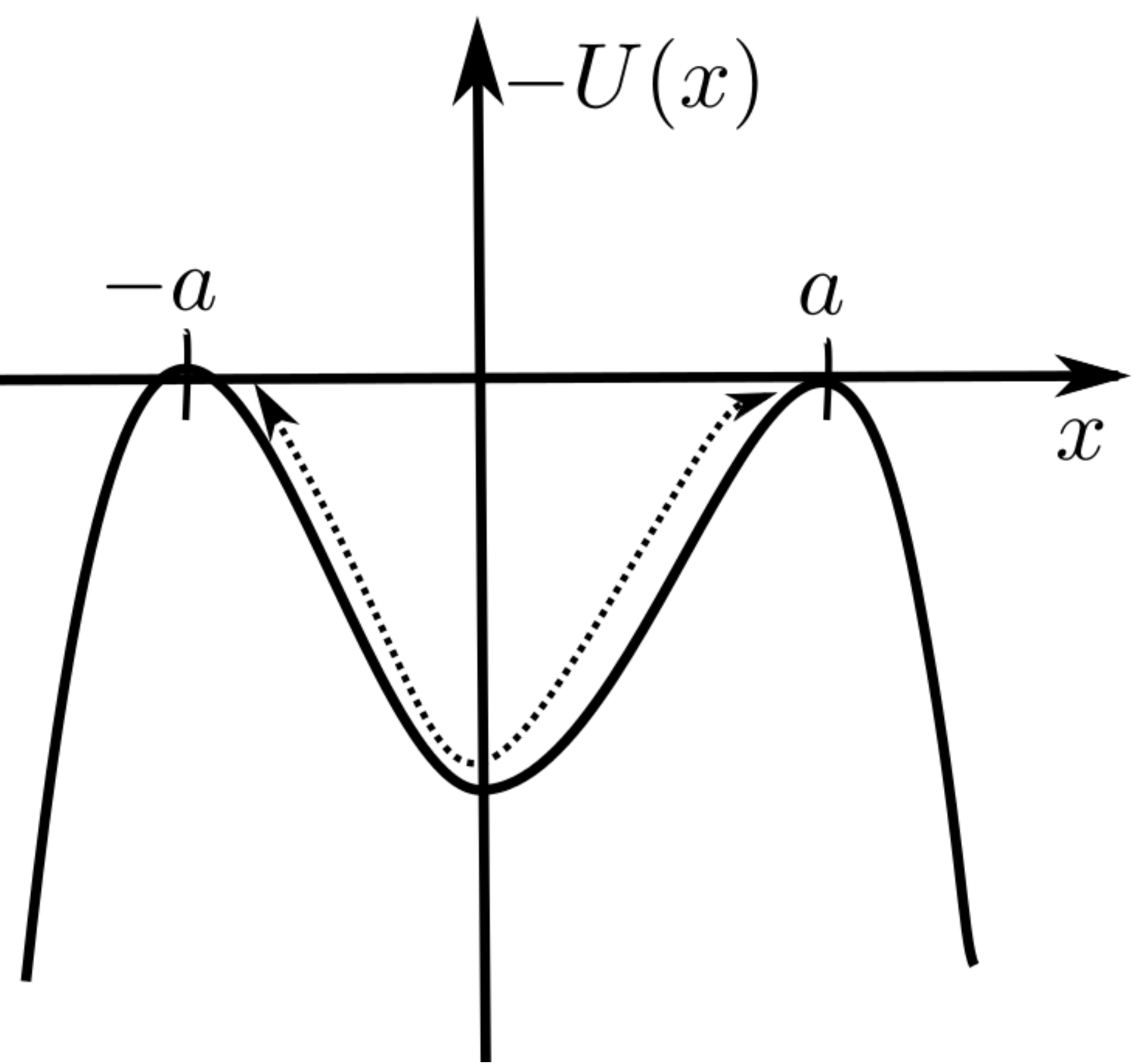}
\caption{A sketch of an inverted double well potential with two minima at $\pm a$. There are two trivial solutions corresponding to a fixed motion of the particle at the top of the left or right hill of the potential. Tunneling is achieved by a nontrivial solution in which the particle starts at the top of the left hill at $\tau\rightarrow -\infty$ and roll through the dashed line and emerges at the top of the right hill at $\tau\rightarrow +\infty$. Such a solution is called an instanton.}
\label{pot28}
\end{figure}

In many textbooks of quantum mechanics, tunneling (barrier penetration) is usually studied using the WKB method. In the tunneling region, the WKB exponent is imaginary, the wave function in the  becomes
\bea
\psi(x) \propto \frac{1}{\sqrt{|p|}}\exp\bigg[-\int_{-x_1}^{x_1}\frac{|p|}{\hbar}dx\bigg],
\eea
where $p=\sqrt{2m(U(x)-\mathcal E)}$ is the momentum of the particle, and $\pm x_1$ are the classical tunneling points $U(\pm x_1)=\mathcal{E}$. At  
the ground state, the energy splitting is given by \cite{ll1,ww}
\bea
\Delta = \frac{\hbar \omega}{\sqrt{e\pi}}\exp\bigg[-\int_{-a}^{a}\frac{|p|}{\hbar}dx\bigg],
\label{tunwkb}
\eea
where $\pm a$ are such that $U(\pm a)=\mathcal{E}_0$.
The instanton approach, however, uses the imaginary time formulation of path integral to find this ground state energy splitting. If we consider the classical equation of motion in imaginary time $\delta S_E=0$ we get:
\bea
m\ddot{x}= \frac{dU(x)}{dx}, \quad\text{where}\quad \ddot{x}\equiv \frac{d^2x}{d\tau^2}
\label{parin1},
\eea
which is the equation of motion with $-U(x)$. In other words, it describes  the motion of a particle in an inverted potential as shown in Fig.\eqref{pot28}.  Upon integration, one finds that the analog of the total ``energy'' is conserved:
\bea
\mathcal{E}=\frac{1}{2} m\lb\frac{dx}{d\tau}\rb^2 - U(x).
\label{parin}
\eea
There are at least three possible solutions of this equation of motion. The first solution corresponds to a particle sitting on the top of the left hill $x=-a$ in Fig.\eqref{pot28} , and  the second solution corresponds to a particle sitting on the top of the right hill $x=a$.  These  are constant solutions which do not give any tunneling.  However, there is a third solution in which the particle starts at the left hill at $\tau\to -\infty$ rolls over through the dashed line, and finally arrives at the right hill at $\tau\to \infty$ . This solution corresponds exactly to the barrier penetration in the WKB method. Such trajectory mediates tunneling and it is  called an instanton. Quantum mechanically, the propagator for this instanton trajectory is given by
\bea
\mathcal{A}\lb-a,-\frac{\beta}{2};a,\frac{\beta}{2}\rb=\braket{a|e^{-\beta\hat H /\hbar }|-a}.
\eea
For instance, 
the potential could be taken to be
\bea
U(x)= \frac{\omega_0^2}{4}(x^2-a^2)^2,\label{posspot}
\eea
but it is actually not necessary to make a specific choice, just the general form pictured in Fig.\ref{pot28} needs to be satisfied.  
Tunneling between the two minima of $U(x)$ requires the computation of the transition amplitudes:
\bea
\braket{\pm a|e^{-\beta\hat H /\hbar }|-a}.
\eea
In order to calculate this amplitude one has to know the solution of the classical equation of motion that obeys the boundary condition of Eqn.\eqref{quapa} as $\beta\rightarrow \infty$. There are two trivial solutions corresponding to  no motion with the particle fixed at the top of the left or right hill of the potential. Tunneling is achieved by a nontrivial solution in which the particle starts at the top of the left hill at $\tau\rightarrow -\infty$, roll through the dashed line in Fig.\eqref{pot28}, and emerges at the top of the right hill at $\tau\rightarrow +\infty$. This nontrivial solution has zero ``energy'' $\mathcal{E}=0$ since initially it starts at the top of the hill at $-a$ where the potential is zero and its kinetic energy is zero.  The solution of Eqn.\eqref{parin}  corresponding to the explicit potential Eqn. \eqref{posspot}, is given by \cite{poly, mackin}
\bea
x(\tau)= a\tanh[\frac{\omega_0}{2}(\tau-\tau_0)], \quad \omega_0^2= \gamma a^2/m,
\label{pinta}
\eea
where $\tau_0$ is an integration constant which corresponds to the time at which the solution crosses $x=0$.

The  action for the solution is given by
\bea
B&=& \int_{-\beta/2}^{\beta/2}  d\tau  \left[\frac{1}{2}m\left(\frac{dx}{d\tau}\right)^2 +U(x)\right],\\
&=&\int_{-\beta/2}^{\beta/2} d\tau\sqrt{2mU(x)}\frac{dx}{d\tau}, \\
&=& \int_{-a}^{a}dx\sqrt{2mU(x)},\\
&=& \frac{ 2\sqrt{{2m}}}{3}\omega_0 a^2,
\eea
where ${\cal E}=0$ from Eqn.\eqref{parin} is used in the second line, and only in the last equation is the specific potential Eqn.\eqref{posspot} used. This action is exactly the WKB exponent in Eqn.\eqref{tunwkb}.  In the approximation of the method of steepest descent, the path integral, Eqn.\eqref{quapa} is dominated by the path  which passes through the configuration for which the action is stationary, {\it i.e.,} Eqn.\eqref{pinta}, and the integral is given by the Gaussian approximation about the stationary point. Then, the one instanton contribution to the transition amplitude is \cite{cole1, cole3}  

\bea
\braket{ a|e^{-\beta\hat H /\hbar }|-a}\propto e^{-B/\hbar}[1+O(\hbar)].
\eea
In fact, one must consider other critical points which correspond to a dilute instanton gas.  The justification of the dilute instanton gas approximation is beyond the purview of this review, we refer the reader to dedicated expositions of the subject, \cite{cole1, cole3}.  The upshot is that one must sum over all  sequences of one instanton followed by any number of anti-instanton/instanton pairs, the total number of instantons and anti-instantons is odd for the transition $-a\leftrightarrow a$ but even for the  transition $-a\rightarrow -a$ ($a\rightarrow a$
). The result of this summation yields \cite{cole3}

\begin{align}
\braket{\pm a|e^{-\beta\hat H /\hbar }|-a}&=\mathcal{N}\frac{1}{2}[\exp(\mathscr{D}\beta e^{-B/\hbar})\mp\exp(-\mathscr{D}\beta e^{-B/\hbar})],
\label{tranint}
\end{align}
where $\mathcal{N}$ is the overall normalization including the square root of the free determinant which is given by $Ne^{-\beta\mathcal{E}_0}$ where $\mathcal{E}_0=\frac{1}{2}\hbar\omega_0$ is the unperturbed ground state energy and $N$ is a constant from the ground state wave function. $\mathscr{D}$ is the ratio of the square root of the determinant of the operator governing the second order fluctuations  about the instanton excluding the time translation zero mode, and that of the free determinant.  It can in principle be calculated.  A zero mode, occurring because of time translation invariance, is not integrated over, and is taken into account by integrating over the Euclidean time position of the occurrence of the instanton giving rise to  the  factor of $\beta$.  The left hand side of Eqn.\eqref{tranint} can also be written as

\bea
\braket{\pm a|e^{-\beta\hat H /\hbar }|-a}= \sum_n\braket{\pm a|n}\braket{n|-a}e^{-\beta\mathcal{E}_n},
\label{tranintt}
\eea
where $\hat H\ket{n}=\mathcal{E}_n\ket{n}$. Taking the upper sign on both sides of Eqs.\eqref{tranint} and \eqref{tranintt} and comparing the terms, one finds that the non-perturbative energy splitting between the ground and the first excited states is given by
\bea
\Delta =\mathcal{E}_1-\mathcal{E}_0= 2\hbar \mathscr{D}e^{-B/\hbar}.
\eea
In a similar manner, by comparing the coefficients one obtains symmetric ground state 
\bea
\ket{\mathcal{E}_0} = \frac{1}{\sqrt{2}}\lb\ket{a}+ \ket{-a}\rb,
\eea
and an antisymmetric first excited state
\bea
\ket{\mathcal{E}_1} = \frac{1}{\sqrt{2}}\lb\ket{a}- \ket{-a}\rb.
\eea

The analysis in the first part of this review will be based on computing the instanton trajectory, its action, and the corresponding energy splitting for any given model that possesses  tunneling.
 \subsection{Spin coherent state path integral}
 \label{sec:spinco}
For a spin system, the basic idea of path integral formulation is retained, however, instead of the orthogonal position $\ket{x}$ and momentum  $\ket{p}$ basis, a basis of spin coherent states is used \cite{rad,pere,klau,ll4}. This basis is defined through the following construction.  Let $\ket{s,s}$ be the highest weight vector in a particular representation of the rotation group, taken as its simply connected covering group $SU(2)$. This state is an eigenstate of the operators ${\hat S}_z$ and $\bold{\hat S}$:
\bea
{\hat S}_z\ket{s,s}&=s\ket{s,s};\quad {\bold{\hat S}}^2\ket{s,s}=s(s+1)\ket{s,s}.
\eea
The spin operators $\hat{S}_i$, $i=x,y,z$ form an irreducible representation of the Lie algebra of $SU(2)$, 
 \bea
 [\hat{S}_i, \hat{S}_j]=i\epsilon_{ijk}\hat{S}_k,
 \label{comm}
 \eea
 where $\epsilon_{ijk}$ is the totally antisymmetric tensor symbol and summation over repeated indices is implied in Eqn.\eqref{comm}.
\begin{figure}[ht]
\includegraphics[width=2in]{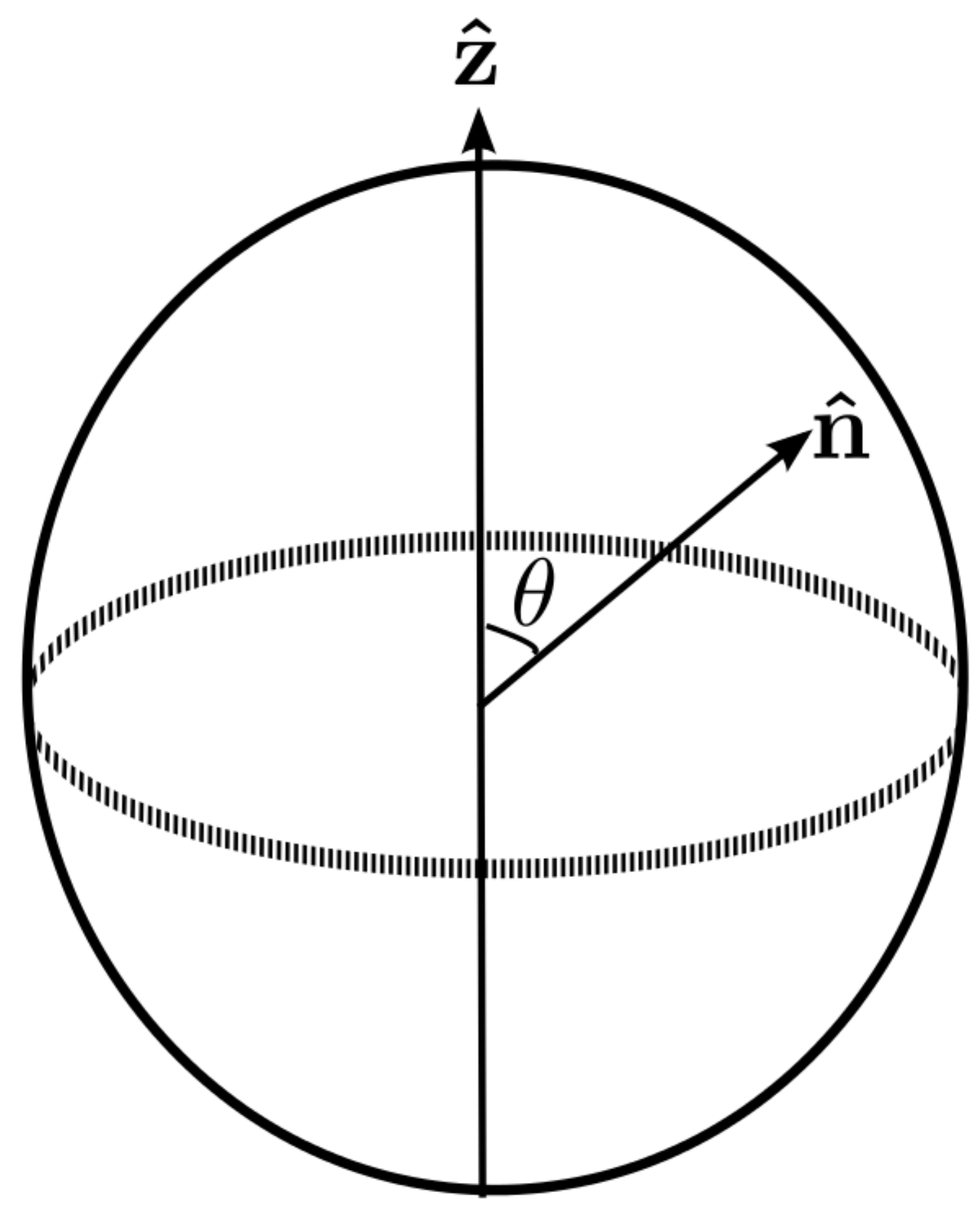}
\caption{The directions of the unit vectors $\bold{\bold{\hat{z}}}$ and $\bold{\hat n}$ on a two-sphere .}
\label{pot1}
\end{figure}
The coherent state, $\ket{\bold{\hat n}}$, an element of the $2s+1$ dimensional Hilbert (representation) space for the spin states,  is defined as \cite{ pere, ll4, klau, zfg, ed1, ed2}
\begin{equation}
\ket{\bold{\hat n}}= e^{i\theta\bold{\hat m}  \cdot \bold{ \hat S}}\ket{s,s} = \sum_{m=-s}^{s}\mathcal{M}^{s}(\bold{\hat n})_{ms}\ket{s,m},
\label{coh}
\end{equation}
where 
$\bold{\hat n} = (\cos\phi\sin\theta, \sin\phi\sin\theta, \cos\theta)$ is a unit vector 
ie. $\bold{\hat n}^2=1$ and  $\bold{\hat m}=(\bold{\hat n}\times \bold{\hat{z}})/|\bold{\hat n}\times \bold{\hat{z}}|$ is a unit vector orthogonal to $\bold{\hat n}$ and where  $\bold{\hat z}$ is the quantization axis pointing from the origin to the north pole of a unit sphere and $\bold{\hat n}\cdot  \bold{\hat z} = \cos\theta$  as shown in Fig.\eqref{pot1}.  Rotating the unit vector $\bold{\hat z}$ about the $\bold{\hat m}$ direction by the angle $\theta$ brings it exactly to the unit vector $\bold{\hat n}$.   $\ket{\bold{\hat n}}$ corresponds to a rotation of an eigenstate of $\hat{S}_z$, i.e $\ket{s,s}$,  to an eigenstate with a quantization axis along $\bold{\hat n}$ on a two-dimensional sphere $\mathcal{S}^2=SU(2)/U(1)$. The matrices $\mathcal{M}^{s}(\bold{\hat n})$ satisfy the relation
\bea
\mathcal{M}^{s}(\bold{\hat n}_1)\mathcal{M}^{s}(\bold{\hat n}_2)=\mathcal{M}^{s}(\bold{\hat n}_3)e^{i\mathcal{G}(\bold{\hat n}_1,\bold{\hat n}_2 ,\bold{\hat n}_3)\hat{S}_z},
\label{matr}
\eea
where $\mathcal{G}(\bold{\hat n}_1,\bold{\hat n}_2 ,\bold{\hat n}_3)$ is the area of a spherical triangle with vertices $\bold{\hat n}_1,\bold{\hat n}_2 ,\bold{\hat n}_3$. Note that Eqn.\eqref{matr} is not a group multiplication, thus the matrices $\mathcal{M}^{s}(\bold{\hat n})$ do not form a group representation.  Unlike the position and momentum eigenstates in Eqn.\eqref{prop1}, the inner product of two coherent states is not orthogonal:
\begin{equation}
\braket{\bold{\hat n}|\bold{\hat n}^{\prime}}= e^{is\mathcal{G}(\bold{\hat n},\bold{\hat n}^{\prime},\bold{\hat z})}[\frac{1}{2}(1+\bold{\hat n}\cdot\bold{\hat n}^{\prime})]^s.
\label{nonor}
\end{equation} 
It has the following property:
\bea
\bold{\hat n}\cdot \bold{\hat S}\ket{\bold{\hat n}}=s\ket{\bold{\hat n}}\Rightarrow\braket{\bold{\hat n}|\bold {\hat S}|\bold{\hat n}}=s\bold{\hat n}. \eea
 The resolution of identity is given by
\begin{equation}
\bold{\hat I}=\frac{2s+1}{4\pi}\int d^3 \bold{\hat n}\delta(\bold{\hat n}^2-1)\ket{\bold{\hat n}}\bra{\bold{\hat n}},
\label{id}
\end{equation}
where $\bold{\hat I}$ is a $(2s+1)\times(2s+1)$ identity matrix, and the delta function ensures that $\bold{\hat n}^2=1$.
The derivation of spin coherent state path integral now follows a similar fashion with Sec.\eqref{sec:spinpa}. Using  the expression in Eqn.\eqref{nonor} and Eqn.\eqref{id} one can express the imaginary time transition amplitude between $\ket{\bold{\hat n}_i}$ and $\ket{\bold{\hat n}_f}$ as a path integral. The analogous form of Eqn.\eqref{quapa} for spin system is given by \cite{ed2, zfg}
\begin{equation}
\braket{\bold{\hat n}_f|e^{-\beta \hat H(\bold{\hat S})}|\bold{\hat n}_i}= \int \mathcal{D}\bold{\hat n} \thinspace e^{-S_E[\bold{\hat n}]}, 
\label{pathint}
\end{equation}
where
\begin{equation}
S_E[\bold{\hat n}] = isS_{{WZ}}+ \int d\tau U(\bold{\hat n(\tau)}), \quad  U(\bold{\hat n(\tau)})=\braket{\bold{\hat n}| \hat H|\bold{\hat n}},
\label{act1}
\end{equation}
and $S_{WZ}$ arises because of the additional phase $e^{is\mathcal{G}(\bold{\hat n},\bold{\hat n}^{\prime},\bold{\hat z})}$ in Eqn.\eqref{nonor}.
We have set $\hbar=1$ in the path integral. The Wess-Zumino (WZ) action, $S_{WZ}$  is given by\footnote{An alternative way of deriving this equation can be found in \cite{petr}.} \cite{nov, witt, wesszu, ed1,ed2}
\begin{equation}
S_{WZ}=\int_{\frac{1}{2}\mathcal S^2} d\tau d \xi \thinspace \bold{\hat n}(\tau, \xi)\cdot[\partial_\tau \bold{\hat n}(\tau, \xi)\times\partial_ \xi\bold{\hat n}(\tau, \xi)],
\label{wz}
\end{equation}
where $\bold{\hat n(\tau)}$ has been extended over a topological half-sphere $\frac{1}{2}\mathcal{S}^2$ in the variables $\tau,\xi$.   In the topological half-sphere  we define  
$\bold{\hat n}$ with the boundary conditions 
\begin{equation}
\bold{\hat n}(\tau,0)=\bold{\hat n}(\tau),\,\,\, \bold{\hat n}(\tau,1)=\hat{\bold z},
\label{bo}
\end{equation}
so that the original configuration lies at the equator and the point $\xi=1$ is topologically compactified by the boundary condition.  This can be easily obtained by imagining that the original closed loop $\bold{\hat n(\tau)}$ at $\xi=0$ is simply pushed up to along the meridians to $\bold{\hat n(\tau)}=\hat {\bold z}$ at $\xi=1$.  The Wess-Zumino term originates from the non-orthogonality of spin coherent states in Eqn.\eqref{nonor}. Geometrically, it defines the area of the closed loop on the spin space, defined by the nominally periodic, original configuration $\bold{\hat n(\tau)}$.   It crucial to note that there is an ambiguity of modulo $4\pi$, since different ways of pushing the original configuration up can give different values for the area enclosed by the closed loop as one can imagine that the closed loop englobes the whole two sphere any integer number of times,  but this ambiguity has no physical significance since $e^{i4 N\pi s}=1$ for integer and half-odd integer $s$. The action, Eqn.\eqref{act1} is valid for a semiclassical spin system whose phase space  is $\mathcal{S}^2$. It is the starting point for studying macroscopic quantum spin tunneling  between the minima of the energy $U(\bold{\hat n})$.
\section{Macroscopic quantum tunneling of large spin systems}
\label{mactun}
\subsection{Coordinate dependent formalism}
Most often a coordinate dependent version of Eqn.\eqref{wz} is used in the condensed matter literature. It seems that most people find it difficult to study macroscopic quantum spin tunneling in the coordinate independent form. In this section, we will show how one can use any coordinate system of interest. In section \eqref{cif}, we will show that the coordinate independent form can reproduce all the known results in quantum spin tunneling. Since the spin particle lives on a two-sphere, the most convenient choice of coordinate are  spherical polar coordinates. Parametrizing the unit vector as $\bold{\hat n}(\tau, \xi)=(\cos\phi(\tau)\sin\theta_{\xi}(\tau), \sin\phi(\tau)\sin\theta_{\xi}(\tau), \cos\theta_{\xi}(\tau))$, with $\theta_{\xi}(\tau)=(1- \xi)\theta(\tau)$,  which satisfies the boundary conditions, Eqn.\eqref{bo} at $\xi=0$ and $\xi=1$.
Then
\beq
 \partial_\tau \bold{\hat n}=\bold{\hat{\bm\theta}} \dot\theta_\xi(\tau) +{\bm\hat{\bm\phi}} \sin\theta_{\xi}(\tau)\dot\phi(\tau),
\eeq
and
\beq
\partial_\xi\bold{\hat n}=\bold{\hat{\bm\theta}}( -\theta(\tau)),
\eeq
where $\bold{\hat{\bm\theta}}$ and ${\bm\hat{\bm\phi}}$ are the usual polar and azimuthal unit vectors which form an orthogonal triad with $\bold{\hat n}$ such that $\bold{\hat{\bm\theta}}\times{\bm\hat{\bm\phi}}=\bold{\hat n}$ (and cyclic permutations).  Thus we find the triple product  becomes 
 \begin{align}
 \bold{\hat n}(\tau, \xi)\cdot(\partial_\tau \bold{\hat n}(\tau, \xi)\times\partial_ \xi\bold{\hat n}(\tau, \xi))&= \dot{\phi}(\tau)\theta(\tau)\sin\theta_{\xi}(\tau).
  \end{align}
 Thus, the WZ term, Eqn.\eqref{wz} simplifies to \cite{pk,sm2}
\begin{align}
S_{WZ}&=\int d\tau \int_{0}^{1} d\xi \thinspace\dot{ \phi}(\tau)\theta(\tau)\sin\theta_{\xi}(\tau)= \int d\tau  \thinspace\dot{ \phi}(\tau) (1-\cos\theta(\tau)).
\label{wzcon}
\end{align}
This is the coordinate dependent form of WZ term or Berry phase \cite{berry84}, which is the expression found in most condensed matter literature.  It corresponds to the area of the unit two-sphere swept out by $\bold{\hat{n}}(\tau)$ as it forms a closed path  on $\mathcal{S}^2$.  To understand this explicitly, one can think of the integral in Eqn.\eqref{wzcon} as a line integral of a gauge field, which only has a $\phi$ component, integrated over a closed path on the two sphere, parametrized by $\tau$.  We denote the closed path as $\cal C$ and it is the boundary of a region $\cal S$, with evidently $\cal C=\partial S$, then
\beq
 \int d\tau  \thinspace\dot{ \phi}(\tau) (1-\cos\theta(\tau))=\oint _{\cal C}A_\phi d\phi.
\eeq
Then using Stokes theorem, we have
\beq
\oint _{\cal C}A_\phi d\phi=\int_{\cal S} d (A_\phi d\phi),
\eeq
written in the notation of differential forms.  However, the gauge field $\vec A =A_\phi {\bm\hat{\bm\phi}}=(1-\cos\theta){\bm\hat{\bm\phi}}$ corresponds exactly to the gauge field of a magnetic monopole located at the centre of the sphere.  Such a gauge field was first described by Dirac \cite{diracpole}, and gives rise to a constant radial magnetic field, apart from a string singularity located at the south pole, which is an unobservable gauge artefact if the magnetic charge is appropriately quantized.  The non observability of this string singularity in quantum mechanics was the seminal observation by Dirac if $s$, in Eqn.\eqref{act1}, is quantized to be a half integer.  Explicitly, the corresponding magnetic field is simply $d (A_\phi d\phi)=\partial_\theta A_\phi d\theta\wedge d\phi= \sin\theta d\theta\wedge d\phi$ which is the area element in spherical polar coordinates on the unit two sphere.   Thus $\oint _{\cal C}A_\phi d\phi=\int_{\cal S} d (A_\phi d\phi ) =\int_{\cal S}\sin\theta d\theta\wedge d\phi= {\rm area}\,(\cal S)$.

The general form of the Euclidean action in coordinate dependent formalism is then
\begin{align}
S_E = is\int d\tau  \thinspace \dot{ \phi}(\tau)  +S_0,
\label{act2}
\end{align}
where
\begin{equation}
S_0 = \int d\tau [ \thinspace -is\dot{ \phi}(\tau)\cos\theta(\tau) + U(\theta(\tau),\phi(\tau))].
\label{ac0}
\end{equation}
The first term in Eqn.\eqref{act2} is a boundary term,  which does not affect the classical equation of motion. It can be integrated out as
\bea
is\int_{-\frac{\beta}{2}}^{\frac{\beta}{2}} d\tau  \thinspace \dot{ \phi}(\tau) =is[\phi(\beta/2)-\phi(-\beta/2)+2\pi N],
\label{topo}
\eea
where $N$ is a winding number, that is the number of times $\phi(\tau)$ winds around the north pole of $\mathcal{S}^2$ as $\tau$ progresses from $-\beta/2$ to $\beta/2$.
 This term is insensitive to any continuous deformation of the field on $\mathcal{S}^2$, thus it is topological. Its effect on the transition amplitude will be studied later. \subsubsection{Easy $z$-axis uniaxial spin model in a magnetic field}
\label{unia}
\begin{figure}[ht]
\includegraphics[width=2.5in]{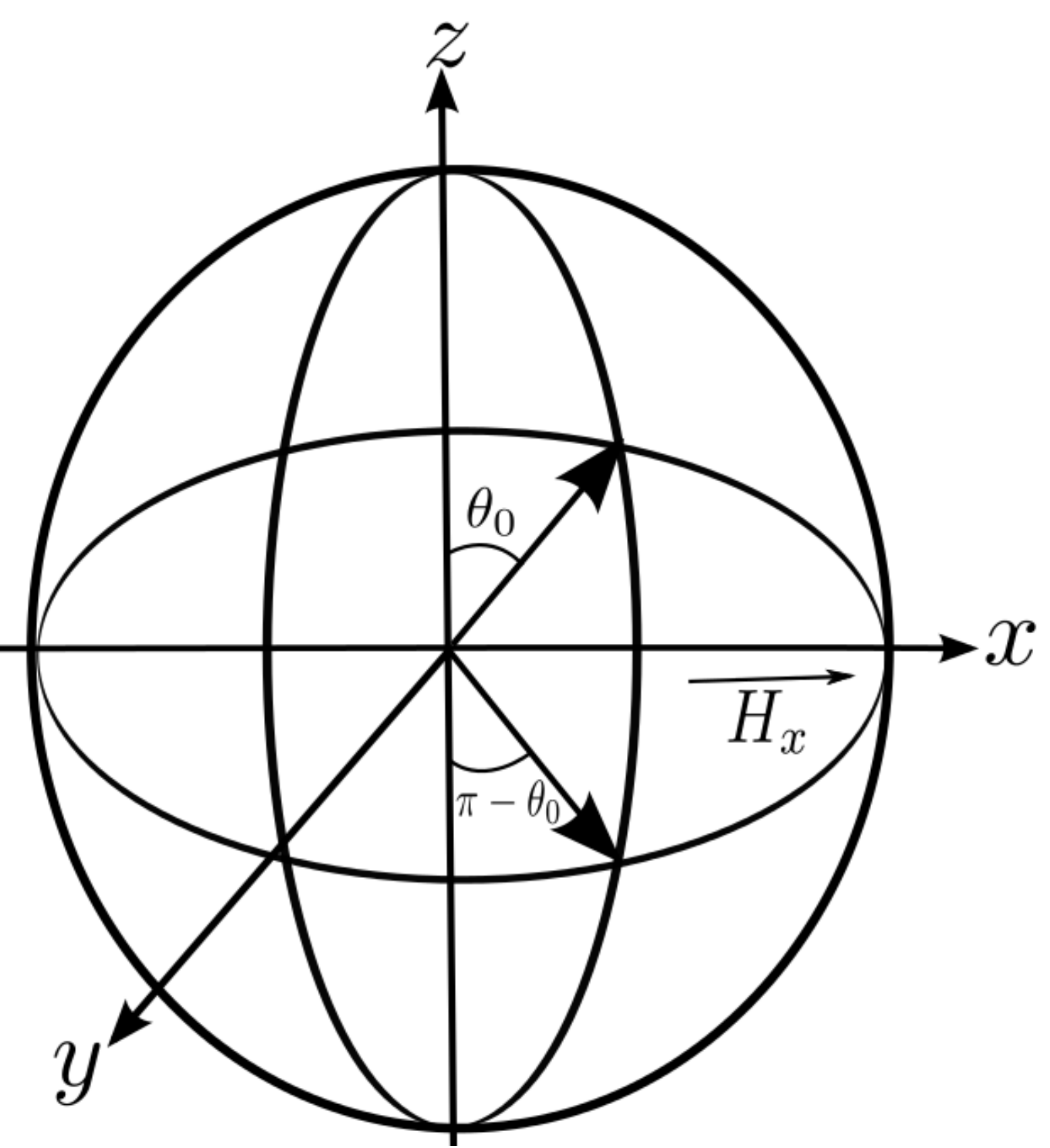}
\caption{The description of a classical spin (thick arrows) on a two-sphere with two classical ground states at $\phi=0$ . The magnetic field is applied parallel to the $x$-axis. The $x$-axis has been rotated on the right hand side for proper view.}
\label{pot26}
\end{figure}

Having derived the coordinate dependent action for a spin system, we will now turn to specific models where this formula can be implemented. Consider a uniaxial system with an easy $\bold{\hat{z}}$ axis (direction of minimum energy) and a magnetic field along the $\bold{\hat{x}}$ axis, the corresponding Hamiltonian is given by \cite{van,chud1}
\begin{align}
\hat H = -D\hat{S}_z^2 - H_x\hat{S}_x,
\label{hasn}
\end{align}
 where $D>0$ is the easy axis anisotropy  and $H_x=g\mu_Bh$, $h$ is the magnitude of the field, $g$ is the spin $g$-factor and $\mu_B$ is the Bohr magneton.  This model is a special case of the Lipkin-Meshkov-Glick model introduced in nuclear physics \cite{lmg}, which has been recently exactly solved \cite{rvm1,rvm2}.  This Hamiltonian is a good approximation for  Mn$_{12}$ acetate molecular magnet with a ground state of $s=10$ \cite{chud2,chud3,foss1, novs, papa}. An experimental review of this molecular magnet can be found in \cite{gs}. The description of the tunneling of spin in the quantum spin terminology is as follows.  For $H_x=0$, the Hamiltonian has a two fold degenerate ground state corresponding to the two ground states in the $\hat{S}_z$ representation, {\it i.e,} $\ket{\uparrow}$ and $\ket{\downarrow}$, where $\ket{\uparrow}\equiv\ket{s}$ and $\ket{\downarrow}\equiv\ket{-s}$. For  $H_x\neq0$, these two states are no longer degenerate since $\hat{S}_x=(\hat{S}_+ + \hat{S}_-)/2$ where $\hat{S}_+\ket{-s}\propto\ket{-s+1}$ and $\hat{S}_-\ket{s}\propto\ket{s-1}$.  In the limit of small magnetic field,  perturbation theory on the magnetic field term shows that the two degenerate ground states are split with an energy difference which is given by \cite{ga, zas4}
 
 \bea
 \Delta =\frac{4Ds^{3/2}}{\pi^{1/2}}\lb\frac{eh_x}{2}\rb^{2s}, \quad h_x = H_x/2Ds.
 \label{ga}
 \eea
The factor $h_x^{2s}$ signifies that the splitting arises from $2s^{\text{th}}$ order in degenerate perturbation theory. This implies that the two quantum states $\ket{\uparrow}$ and $\ket{\downarrow}$ can tunnel to each other through a magnetic energy barrier, a process called quantum spin tunneling \footnote{In the semi-classical description,  tunneling means the rotation of the two equivalent directions of the spin on a two-sphere as shown in Fig.\eqref{pot26}}. Thus, the ground and the first excited states become the symmetric and antisymmetric linear superposition of the degenerate states:
 \bea
 \ket{g}=\frac{1}{\sqrt{2}}(\ket{\uparrow}+\ket{\downarrow}); \quad \ket{e}=\frac{1}{\sqrt{2}}(\ket{\uparrow}-\ket{\downarrow}).
 \eea
 In the absence of the perturbative or splitting term, the energy splitting in Eqn.\eqref{ga} vanishes, which  directly implies that tunneling is only allowed when the Hamiltonian does not commute with  the quantization axis, in this case $\hat{S}_z$ . In the semi-classical analysis, the spin operator becomes a vector parametrized by spherical coordinate of length:
\beq
 S_x^2+S_y^2+S_z^2=s^2.
 \label{const}
  \eeq
  The corresponding classical energy of Eqn.\eqref{hasn} is given by
 \begin{align}
 U(\theta,\phi)= Ds^2\sin^2\theta-H_xs\sin\theta\cos\phi + H_x^2/4D,
 \label{unie}
 \end{align}
where an additional constants have been added to normalize the minimum of the potential to zero.
 The minimum energy requires
\begin{align}
\frac{\partial U}{\partial \theta}\bigg|_{\phi=0}=0\quad \text{and} \quad \frac{\partial^2 U}{\partial \theta^2}\bigg|_{\phi=0}>0,
\end{align}
which yields two classical degenerate minima at $(\phi,\theta)=(0, \theta_0)$ and $(\phi,\theta)=(0, \pi-\theta_0)$ with $\sin\theta_0=h_x=H_x/H_c$, provided $H_x<H_c=2Ds$. The maximum energy corresponds to $(\phi,\theta)=(0, \pi/2)$. 

These two classical minima correspond to the spin pointing in $\pm zx$ plane (see Fig.\eqref{pot26}), which are analogous to the  two quantum states $\ket{\uparrow}$ and $\ket{\downarrow}$. The barrier height is
\begin{align}
\Delta U= U_{\text{max}}-U_{\text{min}}= Ds^2(1-h_x)^2.
\end{align}

Due to tunneling the degeneracy of these ground states will be lifted and one finds that the true ground state is the linear superposition of the two unperturbed ground states. This tunneling is mediated by an instanton which is a solution of the classical equations of motion: 
\begin{align}
is\dot{\theta}\sin\theta &= \frac{\partial U}{\partial \phi},\label{cla1}\\ is\dot{\phi}\sin\theta & =- \frac{\partial U}{\partial \theta}.
\label{cla2}
\end{align}
 These equations are obtained from the least-action principle, whose solution gives the classical path for which the action, Eqn.\eqref{act2} is stationary $\delta S_E=0$.  Although one is usually interested in a real, physical trajectory, these equations are in fact, incompatible, unless one variable (either $\theta$ or $\phi$) becomes imaginary.  The energy along the trajectory has to vanish, since it is conserved by the dynamics, and normalized to zero at the starting point. This can be seen by  multiplying  Eqn.\eqref{cla1} by $\dot{\phi}$ and Eqn.\eqref{cla2} by $\dot{\theta}$ and subtracting the resulting equations which yields  
\begin{align}
\frac{\partial U}{\partial\phi}\dot\phi+\frac{\partial U}{\partial\theta}\dot\theta=0\quad\Rightarrow\quad U(\theta,\phi)= \text{const.} =0 .
\label{cons}
\end{align}
The transition amplitude, Eqn.\eqref{pathint}, in the coordinate dependent form can be written as
\begin{equation}
\braket{\theta_f,\phi_f|e^{-\beta \hat H }|\theta_i,\phi_i}= \int \mathcal{D} \phi\mathcal{D}(\cos\theta) \thinspace e^{-S_E },
\label{pathint1}
\end{equation}
which defines the transition from an initial state $\ket{\theta_i,\phi_i}$ at $\tau=-\beta/2$ to a final state $\ket{\theta_f,\phi_f}$ at $\tau=\beta/2$, subject to the boundary conditions $(\phi(-\beta/2),\theta(-\beta/2)) = (\phi_i,\theta_i)$ and $(\phi(\beta/2),\theta(\beta/2)) = (\phi_f,\theta_f)$. In most cases of physical interest, either $\phi_i=\phi_f$ or $\theta_i=\theta_f$. In the present problem $\phi_i=\phi_f=0$ while $\theta_i=\theta_0$ and $\theta_f=\pi-\theta_0$. Similar to the double well problem in Fig.\eqref{pot28}, the boundary conditions require that the real tunneling trajectory (either $\theta$ or $\phi$ not both) approaches the two minima of $U$  at $\tau=\pm \infty$.
Using Eqn.\eqref{unie} one obtains from Eqn.\eqref{cons}
\begin{align}
 \sin(\phi/2) = \pm i(\sin\theta-\sin\theta_0)/2\sqrt{\sin\theta\sin\theta_0}.
 \label{ima}
\end{align}
From Eqs.\eqref{unie}, \eqref{cla1} and \eqref{ima}, the classical trajectory (instanton)   is found to be \cite{anu1,chud1}
\begin{align}
\cos\theta(\tau) = -\cos\theta_0\tanh(\omega_h\tau), \quad \omega_h=Ds\cos\theta_0,
\label{unistan}
\end{align}
which interpolates from $\theta(\tau)=\theta_0$ at $\tau=-\infty$ to $\theta(\tau)=\pi-\theta_0$ at $\tau=\infty$.
Since the energy remains constant (which is normalized to zero) along the instanton trajectory, the action for this trajectory is determined only by the WZ term in Eqn.\eqref{act2}. It is found to be \cite{anu1}
\begin{align}
B = 2s\bigg[\frac{1}{2}\ln\lb\frac{1+\cos\theta_0}{1-\cos\theta_0}\rb-\cos\theta_0\bigg].
\label{anu4}
\end{align}
Absence of tunneling when $h_x=0$ corresponds to $B=\infty$. The energy splitting in the dilute instanton gas approximation is given by \cite{anu1,anu4}
\begin{equation}
\Delta = \frac{8Ds^{3/2}\cos^{5/2}\theta_0}{\pi^{1/2}\sin\theta_0}\lb\frac{1-\cos\theta_0}{1+\cos\theta_0}\rb^{s+\frac{1}{2}}e^{2s\cos\theta_0}.
\label{spp}
\end{equation}
In the perturbative limit, that is for a very small magnetic field, $\theta_0\rightarrow 0$,  the splitting, Eqn.\eqref{spp} reduces to
\begin{equation}
\Delta = \frac{8Ds^{3/2}(1-h_x^2/2)^{5/2}e^{2s(1-h_x^2/2)}}{\pi^{1/2}(4-h_x^2)^{s+\frac{1}{2}}}h_x^{2s}.
\label{spp1}
\end{equation}
The factor $h_x^{2s}$ reproduces the correct order of perturbation theory result as given in Eqn.\eqref{ga}.\subsubsection{Biaxial spin model and quantum phase inteference}
\label{bint}
Let us consider the biaxial spin model in the absence of an external magnetic field \cite{chud1, loss1,em} 
\begin{equation}
\hat H = D_1\hat{S}_z^2 + D_2\hat{S}_x^2; \quad D_1>D_2>0.
\label{bia}
\end{equation}

In the classical terminology, this model possesses an $XOY$-easy-plane anisotropy with an easy-axis along the $y$-direction, hard-axis along the $z$-direction and medium axis along the $x$-direction.   Quantum mechanically, the easy axis corresponds to the quantization axis, since the Casimir operator $\bold{\hat S}^2= \hat{S}_x^2+\hat{S}_y^2+\hat{S}_z^2= s(s+1)$, can be used to rewrite Eqn.\eqref{bia}  as 
\bea\hat H=-D_2\hat{S}_y^2+(D_1-D_2)\hat{S}_z^2 + \text{const.}
\label{bia1}
\eea
 The first term is the unperturbed term while the second term is the transverse or splitting term which does not commute with the unperturbed term.   Thus, the minimum energy of this Hamiltonian requires a representation in which $\hat{S}_y$ is diagonal.  This means that different representations of a biaxial spin Hamiltonian in the absence of an external magnetic field \footnote{\label{note3} In the presence of a magnetic field, different representation of a biaxial spin models can also be related by the anisotropy constants or rotation of axes} can be related to each other by redefining the anisotropy constants. For instance Eqn.\eqref{bia} is related to $\hat H=-A\hat{S}_x^2+B\hat{S}_z^2$ \cite{em} by $D_2=A$, $D_1=A+B$ . Thus, it suffices to consider just Eqn.\eqref{bia}. Semiclassically, the corresponding classical energy is
\begin{equation}
U(\theta,\phi) = D_1s^2\cos^2\theta +D_2s^2\sin^2\theta\cos^2\phi.
\label{biau}
\end{equation}
The minimum energy corresponds to $(\phi,\theta)=(\pm\pi/2, \pi/2)$,  which are located at $\pm\bold{\hat y}$ as shown in Fig.\eqref{pot27}, and the maximum is located at $(\phi,\theta)=(0, \pi/2)$. From the conservation of energy Eqn.\eqref{cons} one obtains
\begin{equation}
\cos\theta = \pm i\frac{\sqrt{\lambda}\cos\phi}{\sqrt{1-\lambda\cos^2\phi}}, \quad \lambda = D_2/D_1.
\end{equation}
Taking into account that the deviation of the spin away from the easy plane is very small, an alternative method to eliminate $\theta$ from the equation of motion is to integrate out $\cos\theta$ in Eqn.\eqref{pathint1}\cite{zha,em, chud7}. In this case the resulting action has a quadratic first order derivative term, a coordinate ($\phi$) dependent mass and   a potential . 
 Integration of the classical equation of motion Eqn.\eqref{cla2} yields \cite{chud1,em,zha}
\begin{equation}
\sin{\phi}(\tau) =\frac{\sqrt{1-\lambda}\tanh(\omega\tau)}{\sqrt{1-\lambda\tanh^2(\omega\tau)}}, \quad \omega = 2s\sqrt{D_1D_2},
\label{eze2}
\end{equation}
which corresponds to the tunneling of the spin from $\phi=\pi/2$ at $\tau=\infty$ to $\phi=-\pi/2$ at $\tau=-\infty$.
The instanton action for this trajectory is 
\begin{equation}
S_{c} =  is\int_{-\frac{\pi}{2}}^{\frac{\pi}{2}}d  {\phi} + B,
\label{act3}
\end{equation}
where $B$ is given by 
\begin{equation}
B=s\sqrt{\lambda} \int_{-\frac{\pi}{2}}^{\frac{\pi}{2}}d {\phi}\thinspace\frac{\cos {\phi}}{\sqrt{1-\lambda\cos^2 {\phi}}}= \ln\left(\frac{1+\sqrt{\lambda}}{1-\sqrt{\lambda}}\right)^{s}.
\label{act4}
\end{equation}

 Now, consider for example the path $({\phi}(\tau), \thinspace  {\theta}(\tau))$ connecting the two anisotropy minima at $(\phi,\theta)=(\pm\pi/2, \pi/2)$, then owing to the symmetry of the action $S_0$, Eqn.\eqref{ac0} (that is excluding the total derivative term), the path $(- {\phi}(\tau), \thinspace \pi- {\theta}(\tau))$ will also solve the classical equations of motion and $B$ will be the same for both paths but the total derivative term will be reversed:  $  is\int_{\mp\frac{\pi}{2}}^{\pm\frac{\pi}{2}}d {\phi}  =\pm is\pi$. Since the path integral in Eqn.\eqref{pathint1} contains all paths, in the semiclassical (small $\hbar$) approximation \cite{ww,cole1,cole3}, the contributions of these two paths can be combined to give
\begin{equation}
e^{i\pi s}e^{-B } + e^{-i\pi s}e^{-B }=2\cos(\pi s)e^{-B }.
\label{8.2}
\end{equation}
More appropriately, to obtain the tunneling rate one has to use the dilute-instanton gas  approximation  that is by summing over a sequences of one instanton followed by any number of anti-instanton/instanton pairs, with an odd number of instantons and anti-instantons (see Sec.\eqref{dwp}). The transition amplitude becomes \cite{loss1,hd}
\begin{align}
\braket{\frac{\pi}{2}|e^{-\beta \hat{H} }|-\frac{\pi}{2}} =\mathcal{N}\sinh\left[2\mathscr{D}\beta\cos(\pi s)e^{-B }\right],
\label{8.3}
\end{align}
where $\mathscr{D}$ is the fluctuation determinant \cite{cole1,cole2,cole3}. The computation of $\mathscr{D}$ can be done explicitly. $\mathcal{N} $ is a normalization constant and $B$ is the action for the instanton. The tunneling rate (energy splitting) from Eqn.\eqref{8.3} gives \cite{loss1}
\begin{equation}
\Delta  = 4\mathscr{D}\lvert\cos(\pi s)\rvert e^{-B },
\label{8.4}
\end{equation}
The factor $\cos(\pi s)$  is responsible for interference effect and it has markedly different consequences for  integer and half-odd integer spins. For integer spins (bosons), the interference is constructive $\cos(\pi s)= (-1)^s$, and the tunneling rate is non-zero, however, for half-odd-integer spins (fermions), the interference is destructive $\cos(\pi s)= 0$ and the tunneling rate vanishes. This suppression of tunneling for half-odd-integer spins in this model can be related to Kramers degeneracy \cite{kram, mess} due to the time reversal invariance of Eqn.\eqref{bia}. This directly implies that the  ground state is at least two-fold degeneracy in the semi-classical picture. This semi-classical degeneracy  sometimes implies that the two  degenerate quantum  ground states of the unperturbed term, $\ket{\uparrow}$ and $\ket{\downarrow}$ are exact ground states of the quantum Hamiltonian for half-odd integer spin \cite{hd}. 
\begin{figure}[ht]
\includegraphics[width=3.5in]{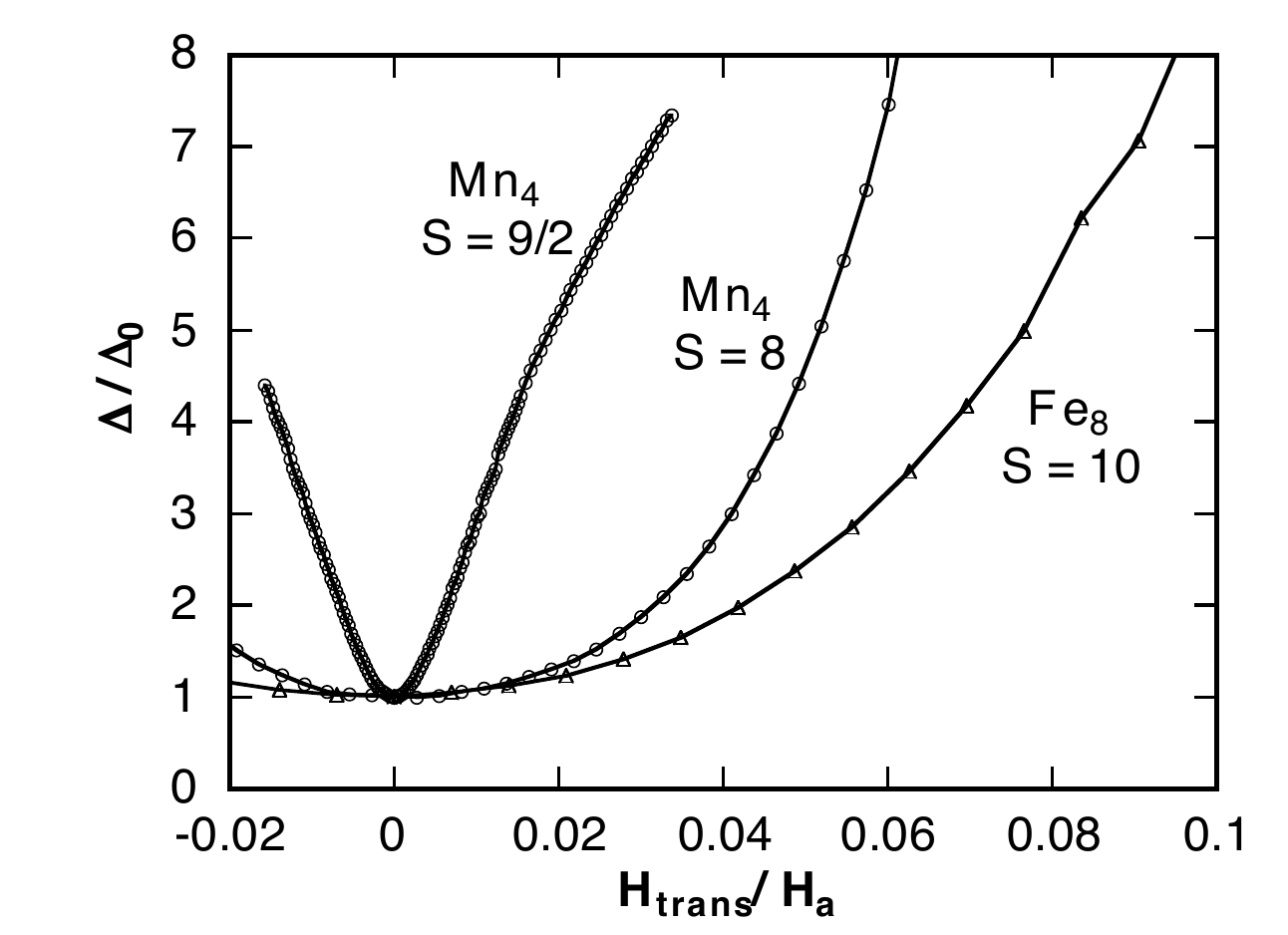}
\caption{Measured tunnel splittings obtained by the Landau-Zener method as a function of transverse field for all three SMMs. The tunnel splitting increases gradually for an integer spin, whereas it increases rapidly for a half-integer spin. Adapted with permission from \onlinecite{wern3}}
\label{pot25}
\end{figure}

In this biaxial model we have just reviewed, the quantum phase interference appeared naturally from the topological term in the action, Eqn.\eqref{topo}  since the instanton trajectory is in the $\phi$ variable. If we had considered the $z$-easy axis model such as 
\begin{equation}
\hat H = -k_z\hat{S}_z^2 + k_y\hat{S}_y^2, \quad k_z, k_y>0,
\label{zeasy}
\end{equation}
then the situation would have been different. 
This Hamiltonian is related to Eqn.\eqref{bia}
by $k_z= D_2$, $k_y=D_1-D_2$ or by rotation of axis $\hat{S}_z\leftrightarrow \hat{S_y}$. Suppose we wish to solve Eqn.\eqref{zeasy} as it is, then the corresponding classical energy is

\begin{equation}
U(\theta,\phi) = (k_z+k_y\sin^2\phi)s^2\sin^2\theta ,
\end{equation}
One finds from the conservation of energy that $\phi(\tau)$ is an imaginary constant and $\theta(\tau)$ is the real tunneling trajectory which is given by \cite{sm1}
\bea
\thinspace \theta\lb \tau\rb =  2 \arctan [ \exp(\omega (\tau-\tau_0))],
\eea
where $\omega = 2s\sqrt{k_z(k_y+k_z)}$, and $\theta(\tau)\rightarrow 0, \pi$ as $\tau\rightarrow\mp \infty$. The fact that $\phi(\tau)$, although imaginary,  is just a constant simply implies that the topological term in Eqn.\eqref{topo} which is responsible for the phase interference vanishes.  The transition amplitude arises from the necessity to translate $\phi$ from some fiducial value, taken without loss of generality to be zero, to the complex constant value before the instanton trajectory in $\theta$ and then followed by the translation of $\phi$ back to its fiducial value after the instanton trajectory.   It was explicitly shown, that translation of $\phi$ in the complex plane yields the transition amplitude and the corresponding  energy splitting is of the form: \cite{sm1,sm2}
 \bea
\Delta = 2\mathscr{D}(1+\cos(2\pi s))e^{-B },
\label{owep} 
\eea
where
 \begin{align}
B&=
  \begin{cases}
  s\ln\lb\frac{4k_z}{k_y}\rb    & \text{if } k_y \ll k_z,\\  2s\lb {k_z}/{k_y}\rb^{1/2} & \text{if } k_y \gg k_z.
  \end{cases}
  \label{21a}
 \end{align}
The fluctuation determinant is calculated to be $\mathscr{D}= 8\sqrt{2}k_zs^{3/2}/\pi^{1/2}$ for $k_y \ll k_z$ and $\mathscr{D}= 8(sk_zk_y)^{3/2}/\pi^{1/2}$ for $k_y \ll k_z$\cite{anu1}. Thus, we recover that tunneling is restricted  for half-odd integer spins. For integer spin and the semiclassical limit $s\gg 1$, simple operatorial quantum mechanical perturbation theory in the splitting term for $k_y\ll k_z$  gives \cite{ga}
\bea
 \Delta =\frac{8k_zs^{3/2}}{\pi^{1/2}}\lb\frac{k_y}{4k_z}\rb^{s},
 \label{ga1}
 \eea
which is consistent with Eqn.\eqref{owep} for integer spin $s$.
  The experimental confirmation of this spin-parity effect (i.e suppression of tunneling for half-odd integer spin) in spin systems was reported by \textcite{wern3}. They studied three SMMs in the presence of a transverse field using Landau-Zener method  to measure the tunnel splitting as a function of transverse field. They established the spin-parity effect by comparing the dependence of the tunneling splitting on the transverse field for integer and half-odd integer spin systems. Observation  showed that an integer spin system is insensitive to
small transverse fields whereas a half-odd integer spin system is much more sensitive as shown in Fig.\eqref{pot25}. This observation is analogous to the fact that half-odd integer spin does not tunnel.
\subsubsection{Biaxial spin model with an external magnetic field}
\label{solo}
The quantum phase interference (quenching of tunneling splitting) we saw in the previous section is a zero magnetic field effect. In the presence of a magnetic field complete destructive  interference for half-odd integer spins does not occur instead oscillation occurs. Consider the biaxial spin model with an external magnetic field applied along the hard-axis \cite{anu2, anu5, anu6}
\begin{equation}
\hat H = D_1\hat{S}_z^2 + D_2\hat{S}_x^2-h_z\hat{S}_z, 
\label{biam}
\end{equation}
where $h_z=g\mu_Bh$, $h$ is the magnitude of applied field and $g$ is the spin $g$-factor and $\mu_B$ is the Bohr magneton. This Hamiltonian can also be written as
\bea
\hat H=-D_2\hat{S}_y^2+(D_1-D_2)\hat{S}_z^2 -h_z\hat{S}_z+ \text{const.}
\label{biamax}
\eea
Thus, we see explicitly that the easy (quantization) axis is along the $y$-direction.
Unlike the previous model this Hamiltonian is no longer time reversal invariant due the presence of the magnetic field, so Kramers theorem is no longer applicable.   This Hamiltonian has been studied experimentally for Fe$_8$ molecular cluster \cite{wern,sess,san}. There are $2s+1$ energy level spectra where $s=10$ and a quantum number $m=-10,-9,\cdots, 10$. At very low temperature $(T<0.36K)$ only the lowest states $m=\pm10$ are occupied which can tunnel macroscopically. In the semi-classical analysis,  the classical energy up to an additional constant is 

\begin{equation}
U(\theta,\phi) = D_1s^2(\cos\theta-\alpha)^2+D_2s^2\sin^2\theta\cos^2\phi,
\label{biaem}
\end{equation}
with $\alpha = h_z/h_c$, $h_c=2D_1s$ being the coercive field. 
\begin{figure}[ht]
\includegraphics[width=2.5in]{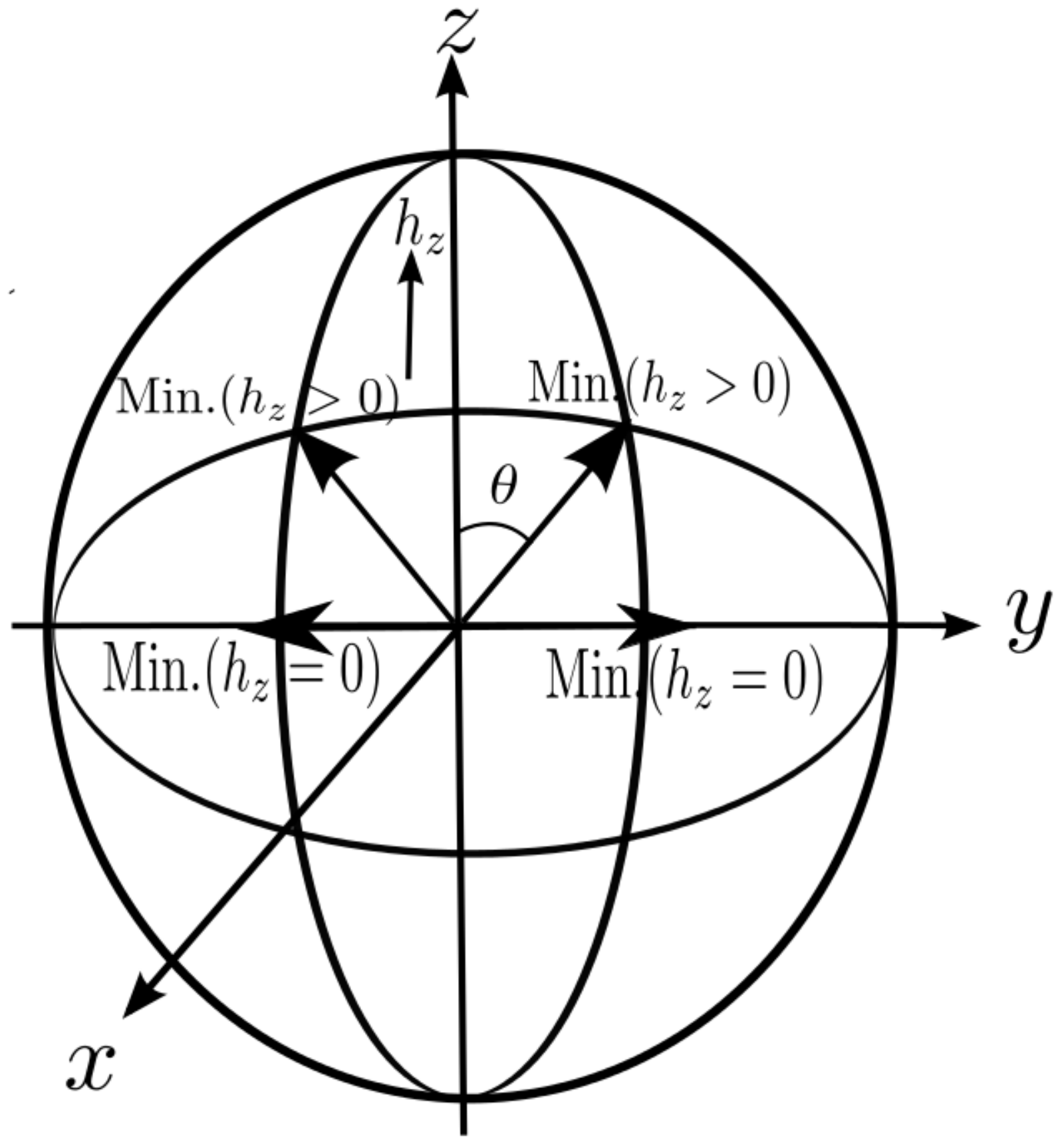}
\caption{The description of a classical spin (thick arrows) on a two-sphere with two classical ground states. For $h_z=0$, $\theta=\pm\pi/2$, the two classical ground states lie in the $\pm y$ directions which are joined by two tunneling paths in the equator.  For $h_z>0$, $\theta=\pm\arccos\alpha$, the two classical ground states lie in the $yz$ plane.  Reproduced from \textcite{qtm}}
\label{pot27}
\end{figure}

There are two classical degenerate minima located at $\cos\theta =\alpha,\thinspace \phi= -\pi/2$ and $\cos\theta =\alpha,\thinspace \phi =\pi/2$ provided $h_z<h_c$. These ground states  lie in the $xz$ and  $yz$ planes at an angle $\theta=\pm\arccos\alpha$ as shown in Fig.\eqref{pot27}.  From energy conservation , Eqn.\eqref{cons} the expression for $\cos{\theta}$ in terms of $\phi$  yields
\begin{equation}
\cos{\theta} = \frac{\alpha + i \lambda^{1/2}\cos{\phi}(1-\alpha^2 -\lambda\cos^2{\phi})^{1/2}}{1-\lambda\cos^2{\phi}},
\label{8.8}
\end{equation}We have chosen the positive solution in Eqn.\eqref{8.8} for convenience. Using this equation  and Eqn.\eqref{cla2}, one obtains the instanton solution: 
\begin{equation}
\sin{\phi}(\tau) =\frac{\sqrt{1-\lambda_H}\tanh(\omega_H\tau)}{\sqrt{1-\lambda_H\tanh^2(\omega_H\tau)}},
\label{eze1}
\end{equation}

\begin{figure}[ht]
\includegraphics[width=3.5in]{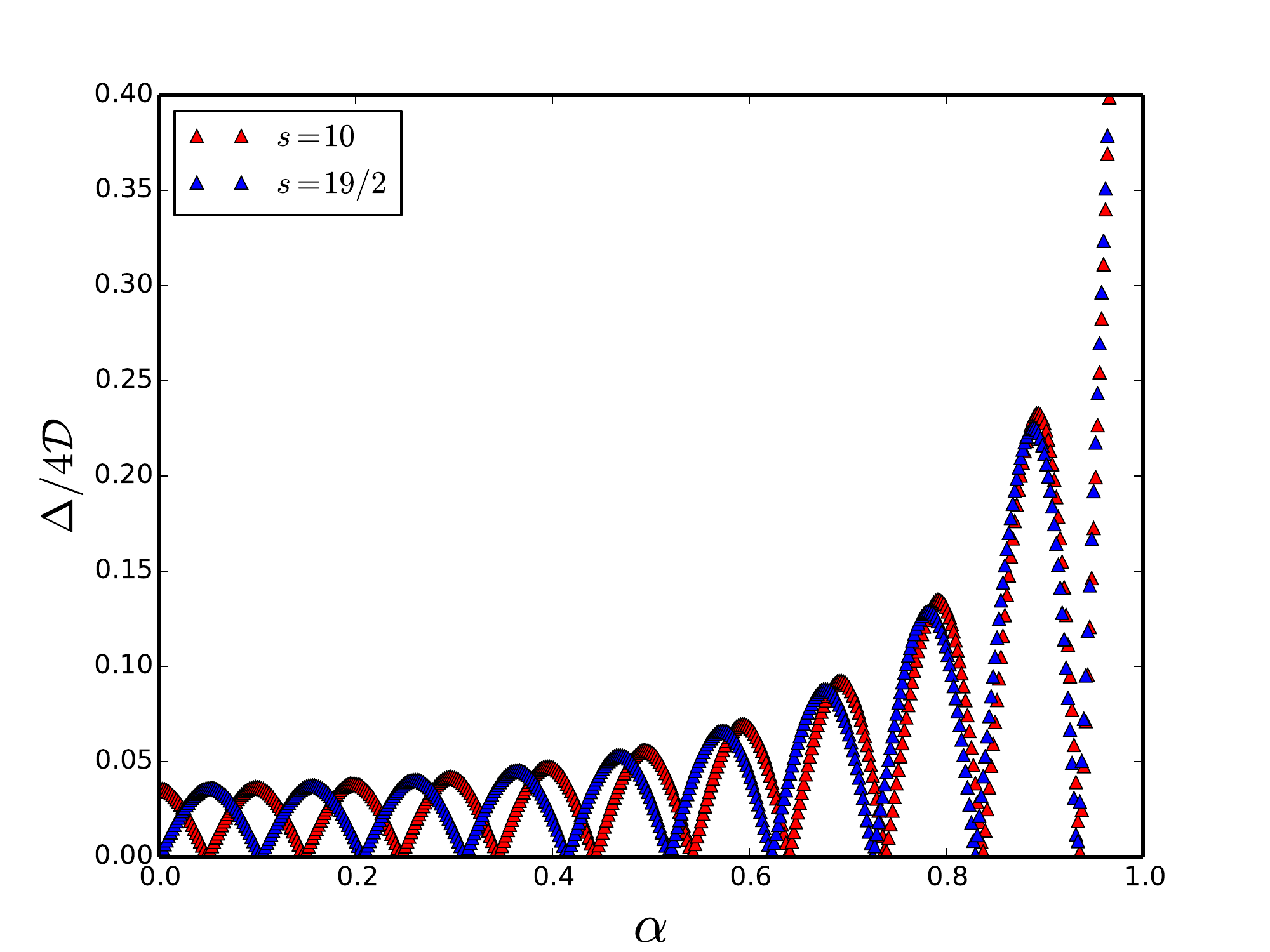}
\caption{ Color online: Oscillation of the tunneling splitting as a function of the magnetic field parameter $\alpha$. Solid line is for interger spins while dotted line is for half-odd integer spins.}
\label{pot}
\end{figure}

where $\omega_H = 2s\sqrt{D_1D_2(1-\alpha^2)}$ and $\lambda_H =  \lambda/(1-\alpha^2)$.
The classical action for this instanton path is
\begin{equation}
S_{c} = i\pi\Theta +B,
\label{9.1}
\end{equation}
where 
\begin{align}
\Theta &=\frac{s}{2\pi}\lb\mathscr{S}_+-\mathscr{S}_-\rb,\label{eze3}
\end{align}
and $\mathscr{S}_+-\mathscr{S}_-$ is the area  enclosed by the two tunneling paths on a 2-sphere as shown in Fig.\eqref{pot27}, which is given by 
\begin{align}
\mathscr{S}_{\pm}=\int_{\mp\frac{\pi}{2}}^{\pm\frac{\pi}{2}}d\phi\lb 1-\frac{\alpha}{1-\lambda\cos^2\phi}\rb=\pm\pi\left( 1-\frac{\alpha}{\sqrt{1-\lambda}}\right).
\end{align}
The instanton action is given by
\begin{align}
B&=s\ln \left(\frac{\sqrt{1-\alpha^2}+\sqrt{\lambda}}{\sqrt{1-\alpha^2}-\sqrt{\lambda}}\right)-\frac{2s\alpha}{\sqrt{1-\lambda}}\ln \left(\frac{\sqrt{(1-\alpha^2)(1-\lambda)}+\alpha\sqrt{\lambda}}{\sqrt{(1-\alpha^2)(1-\lambda)}-\alpha\sqrt{\lambda}}\right).
\end{align}
In this problem the imaginary path of the instanton action, Eqn.\eqref{9.1} has acquired an additional term due to the presence of the magnetic field. In the dilute instanton gas approximation, one obtains that the tunneling rate  is then given by
\begin{equation}
\Delta  =\Delta_0\lvert\cos(\pi \Theta)\rvert,\quad \Delta_0 =  4\mathscr{D} e^{-B},
\label{9.4}
\end{equation}
which clearly reduces to Eqn.\eqref{8.4} in the limit of zero magnetic field. Now, the tunneling splitting is no longer suppressed for half-odd integer spin but rather oscillates with the magnetic field (see Fig.\eqref{pot}) with a period of oscillation of
   \begin{equation}
 \Delta h =\frac{2D\sqrt{1-\lambda}}{g\mu_B},
 \label{ppp}
 \end{equation}
 only vanishes at  
  \begin{equation}
 \Theta = (n+1/2)\quad \text{or}\quad \alpha=\sqrt{1-\lambda}\left(s-n-1/2\right)\slash s,
 \label{eze4}
 \end{equation}
 where $n$ is an integer. It is crucial to note that the quenching of tunneling at a critical field  only occurs for biaxial spin system with a magnetic applied along the hard anisotropy axis.

\subsubsection{Landau Zener effect}
The uniaxial and the biaxial models we have studied so far can be mapped to a two-level pseudospin $\frac{1}{2}$ particle system \cite{owerre, chud12,chud11}. Let us consider a two-level system which is described by an unperturbed Hamiltonian $\hat{H}_0(\eta)$ that depends explicitly on a parameter $\eta$. Suppose that the eigenstates of this Hamiltonian are $\ket{m}$ and $\ket{m^{\prime}}$, then the eigenvalue equation yields
\begin{align}
\hat{H}_0(\eta)\ket{m}&=\zeta_1(\eta)\ket{m},\\
\hat{H}_0(\eta)\ket{m^{\prime}}&=\zeta_2(\eta)\ket{m^{\prime}},
\end{align}
where $\zeta_{1,2}(\eta)$ are the corresponding eigenenergies. It is assumed that the eigenstates $\ket{m}$ and $\ket{m^{\prime}}$ are independent of the parameter $\eta$, and that at some value of $\eta$, $\hat{H}_0(\eta)$ possesses a symmetry which allows level crossing (degeneracy) of the two eigenvalues $\zeta_{1,2}(\eta)$. The parameter $\eta$ could be an applied magnetic field\cite{wern6}. In the presence of a perturbative term $\hat V$, the total Hamiltonian can be written as
\bea
\hat H = \hat{H}_0+\hat V.
\eea
The Hamiltonian can be diagonalized in the basis $\big\{ \ket{m},\ket{m^{\prime}}\big\}$, the corresponding matrix is given by
\begin{equation}
\hat{H}(\eta) =
 \begin{pmatrix}
  \varepsilon_1(\eta) &  \Delta  \\
  \Delta^* &  \varepsilon_2(\eta)
 \end{pmatrix},
 \label{zene7}
\end{equation}
where
\begin{align}
 \varepsilon_1&= \zeta_1(\eta)+\braket{m|\hat V|m},\label{zene1}\\
  \varepsilon_2&= \zeta_2(\eta)+\braket{m^{\prime}|\hat V|m^{\prime}},\label{zene2}\\
   \Delta  &= 2|\braket{m|\hat V|m^{\prime}}|\Rightarrow \braket{m|\hat V|m^{\prime}}=\frac{1}{2}\Delta e^{-i\phi}.
 \label{zene3}
\end{align}
Diagonalizing Eqn.\eqref{zene7}, one obtains the eigenvalues:
\begin{align}
 \varepsilon_+ &= \frac{1}{2}\big[ ( \varepsilon_1+ \varepsilon_2)+\lb \varepsilon(\eta)^2+4| \Delta |\rb^{1/2}\big],\\
 \varepsilon_- &= \frac{1}{2}\big[ ( \varepsilon_1+ \varepsilon_2)-\lb \varepsilon(\eta)^2+4| \Delta |\rb^{1/2}\big],
\end{align}

where $\varepsilon(\eta)=\varepsilon_1- \varepsilon_2$. If both the unperturbed energies are degenerate at some critical value $\eta_c$ where $\varepsilon(\eta_c)=0$, we see that the two levels $\varepsilon_{\pm}$ never cross each other unless the avoided crossing term $\Delta$ vanishes. Let us consider the time-dependent Schr\"odinger equation:
\begin{align}
\hat H\ket{\psi(t)}=i\frac{\partial \ket{\psi(t)}}{\partial t}.
\end{align}
The wave function can be taken as a linear combination of the unperturbed states:
\begin{align}
\ket{\psi(t)}=C_1(t)e^{-i\int \varepsilon_1 dt}\ket{m}+C_2(t)e^{-i\int \varepsilon_2 dt}\ket{m^{\prime}}.
\end{align}
Using Eqn.\eqref{zene1}--\eqref{zene3}, the time-dependent Schr\"odinger equation can be written as
\begin{align}
i\dot{C}_1&=\Delta e^{-i\int_0^t\varepsilon(t^{\prime}) dt^{\prime}}C_2,
\label{zene4}\\
i\dot{C}_2&=\Delta^*e^{i\int_0^t\varepsilon(t^{\prime}) dt^{\prime}}C_1.
\label{zene5}
\end{align}
These two differential equations must be solved with the boundary conditions:
\begin{align}
C_1(-\infty)=0, \quad |C_2(-\infty)|=1.
\end{align}
Using the fact that $\Delta$ is time-independent, differentiating Eqn.\eqref{zene4} and substituting Eqn.\eqref{zene5} into the resulting equation yields
\begin{align}
\ddot{C}_1-i\varepsilon(t)\dot{C}_1 +|\Delta|^2{C}_1=0.
\label{zene6}
\end{align}
Writing $\varepsilon(t)=\alpha t$, $f=\Delta e^{i\phi}$ and
\begin{align}
C_1=ye^{i\frac{1}{2}\int_0^t\varepsilon(t^{\prime}) dt^{\prime}}.
\end{align}
Eqn.\eqref{zene6} transforms into the form:
\begin{align}
\ddot{y}+\lb f^2 +i\frac{\alpha}{2} +\frac{\alpha^2}{4}t^2\rb y=0,
\end{align}
which transforms into the Weber equation:\cite{whitta}
 by setting $n=-if^2/\alpha$ and $z=\sqrt{\alpha}e^{i\pi/4 }t$ :
\bea
\frac{d^2y}{dz^2}+\lb n+\frac{1}{2}-\frac{1}{4}z^2\rb y.
\eea
\begin{figure}[ht]
\includegraphics[width=3.5in]{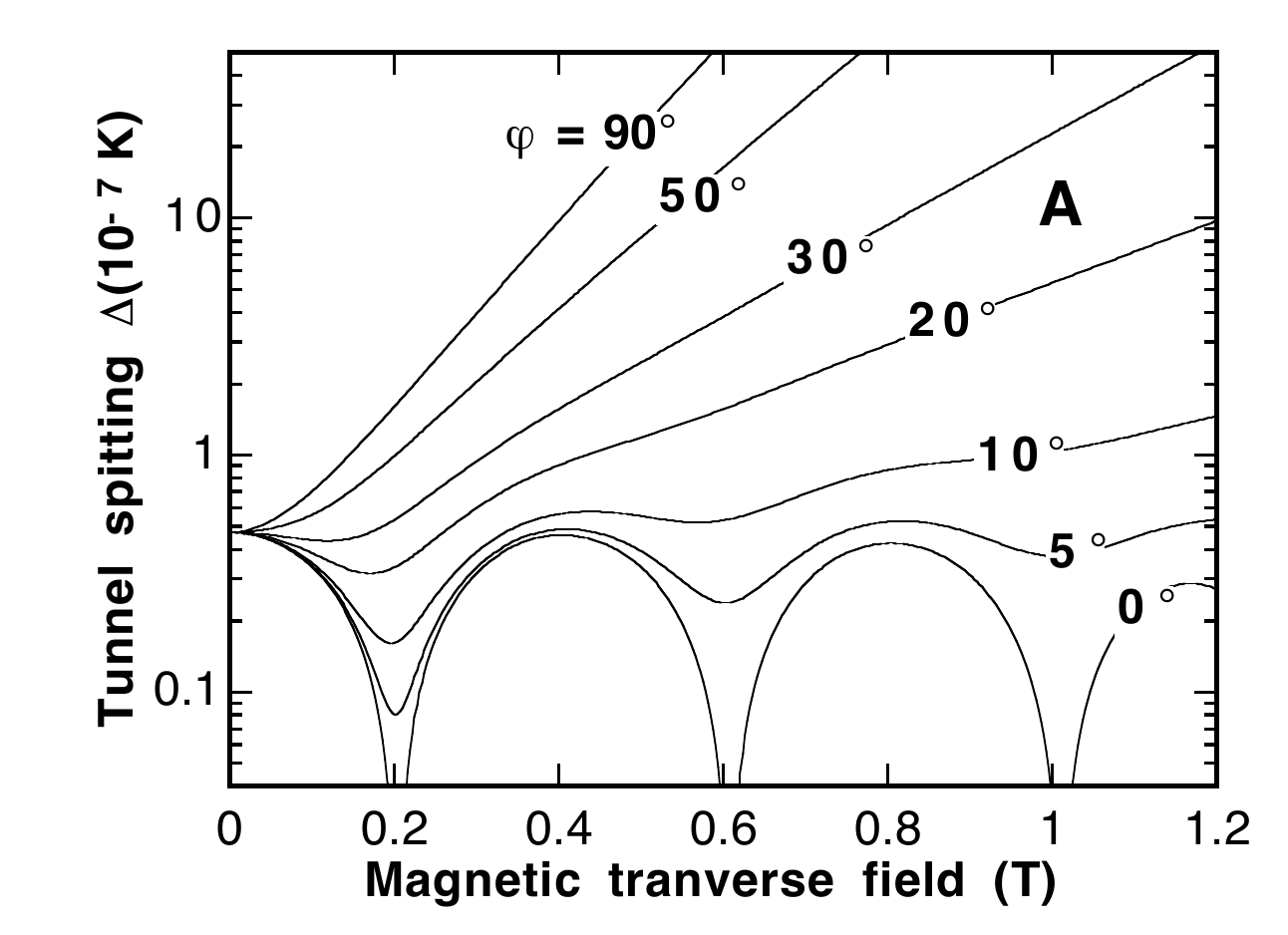}
\includegraphics[width=3.5in]{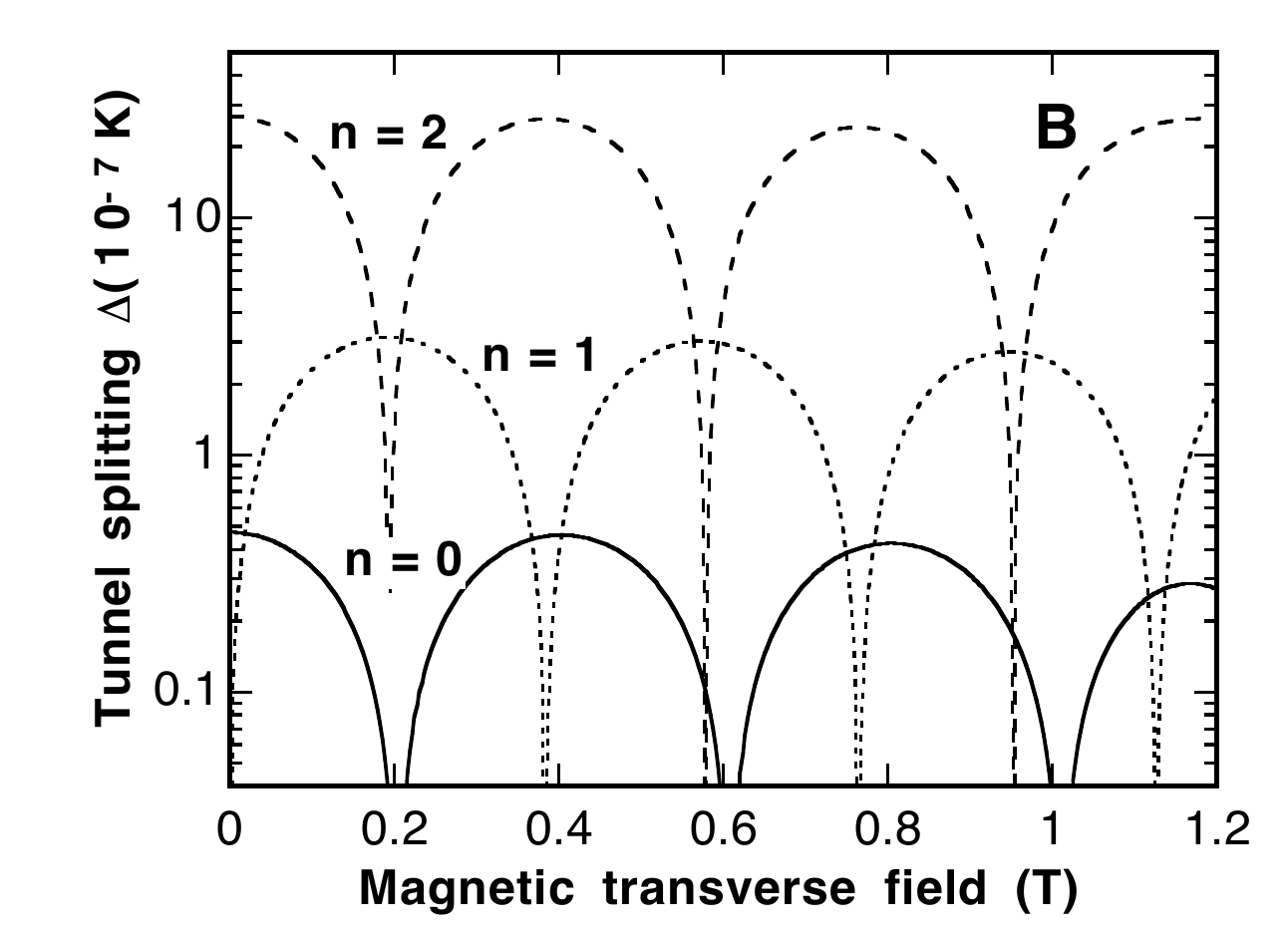}
\caption{ 
 Calculated tunneling splitting as a function of the applied field using Landau-Zener method for the Hamiltonian $\hat H = -AS_z^2+B(S_x^2-S_y^2)+ C(S^{4}_++S^{4}_-)-g\mu_BhS_x$. For  $C=0$, it is related to that of Eqn.\eqref{biam} by $D_1=A+B$ and $D_2=A-B$. $(\bold{A})$ is the quantum transition between $m=\pm10$ for several values of the azimuth angles $\phi$. $(\bold{B})$ is the quantum transition between $m=-10$ and $m=10-n$ at  $\phi=0$, where $n=0,1,2,\cdots$, $m=-s,\cdots, s$, and $s=10$ $A=0.275 K$, $B=0.046 K$ and $C=-2.9\times 10^{-5} K$ for Fe$_8$ molecular cluster. Adapted with permission from \onlinecite{wern}} 
\label{lz}
\end{figure}

The solutions of this differential equation are parabolic cylinder functions. The general solution of Eqn.\eqref{zene6} has the form\cite{zener}
\begin{align}
C_1(t)=\bigg[ aD_{-\nu-1}(-i\sqrt{\alpha}e^{i\pi/4 }t)+bD_{\nu}(\sqrt{\alpha}e^{i\pi/4 }t)\bigg]e^{i\varepsilon(t)/4},
\end{align}
where  $a$ and $b$ are constants determined by the initial conditions. In the limit $t\to\infty$, the asymptotic form of the excitation probability is found to be\cite{zener,ll1,llz, jan}
\bea
\mathcal P= 1-|C_1(\infty)|^2= 1-\exp\bigg[-\frac{2\pi|\Delta|^2}{\frac{d\varepsilon}{dt}}\bigg],
\eea
which is the famous Landau-Zener formula. The theoretical prediction of the oscillation of tunneling splitting of the model in Sec.\eqref{solo} has been observed experimentally in Fe$_8$ molecular cluster and Mn$_{12}$ SMMs using this Landau Zener technique \cite{wern0,wern,wern2}.  In Fig.\eqref{lz} we have shown the experimental confirmation of this theoretical prediction. It  explicitly shows the oscillations of the tunnel splittings as a function of the magnetic field applied along the hard anisotropy axis.  This field is responsible for the periodic change in the avoided level crossing $\Delta$, which we found from the semiclassical analysis as a destructive or constructive quantum interference, with the period of oscillation given in Eqn.\eqref{ppp}.  The tunneling probability from the Landau Zener formula is given by \cite{wern}
\bea
\mathcal P= 1-\exp\bigg[-\frac{\pi|\Delta|^2}{4s\hbar g\mu_B\frac{dH}{dt}}\bigg],
\eea
where $\frac{dH}{dt}$ is the constant field sweeping rate and $g\approx 2$.

 The value of the period of oscillation, Eqn.\eqref{ppp} using the anisotropy parameters for Fe$_8$ molecular cluster in Fig.\eqref{lz} with $D_1=A+B$ and $D_2=A-B$  is $\Delta h=0.26 T$. The value is very small compare to its experimental measured value $0.41 T$. In order to fix this discrepancy an additional  fourth order anisotropy of the form $C(S^{4}_++S^{4}_-)$  is required in Eqn.\eqref{biam} \cite{wern, wern0}.  The inclusion of this term involves a tedious theoretical analysis. There is no exact instanton solution but some approximate schemes have been developed to tackle this problem \cite{chg,foss,kim3}.

\subsubsection{Antiferromagnetic exchange coupled dimer model}
\label{sec:anti}
We have considered only the tunneling phenomenon of   single molecule magnets (SMMs) . In many cases of physical interest, interactions between two large spins are taken into account. These interactions can be either ferromagnetic, which aligns the neighbouring spins or antiferromagnetic, which anti-aligns the neighbouring spins. One physical system in which these interactions occur is the dimerized molecular magnet [Mn$_4$]$_2$. It comprises two Mn$_4$ SMMs of equal spins $s_1=s_2=9/2$, which are coupled antiferromagnetically. The phenomenon of quantum tunneling of spins in this system has been be studied both numerically and experimentally \cite{tiron, hill}.
For this system, the simplest form of the Hamiltonian in the absence of an external magnetic field can be written as
\begin{align}
\hat{H}&= -D(\hat{S}_{1,z}^2+\hat{S}_{2,z}^2)  + J \hat{\bold{S}}_{1}\cdot \hat{\bold{S}}_{2},  \label{ant}
\end{align}
where $J>0$ is the antiferromagnetic exchange coupling.   and $D \gg  J>0$ is the easy-axis anisotropy constant, $S_{i,z}, i=1,2$ is the projection of the component of the spin along the $z$ easy-axis. In this model the exchange term acts as a field bias on its neighbour. We will report here on the analysis of this model by \cite{sm2}, however the nature of the ground states was first proposed by \cite{bab} and the energy splitting was obtained by \cite{kim2} and the quantum operator perturbation theoretical analysis is given in \cite {chud9,chud10}.  \textcite{parkma} demonstrated using density-functional theory that this simple model can reproduce experimental results in  [Mn$_4$]$_2$ dimer with $D=0.58 K$ and $J=0.27 K$.  It also plays a crucial role in quantum CNOT gates and SWAP gates  for spin $1/2$ \cite{loss2}. 

The total $z$-component of the spins $\hat{S}_{z}=\hat{S}_{1,z}+\hat{S}_{2,z}$ is a conserved quantity. However, the individual $z$-component spins $\hat{S}_{1,z}, \hat{S}_{2,z}$ and the staggered configuration $\hat{S}_{1,z}-\hat{S}_{2,z}$ are not conserved.  The Hilbert space of this system is the tensor product of the two spaces $\mathscr{H}=\mathscr{H}_1 \otimes \mathscr{H}_2$ with dim$(\mathscr{H})$= $(2s_1+1)\otimes (2s_2+1)$. The basis of $S_{j}^z$ in this product space is given by  $\ket{s_1,\sigma_1}\otimes\ket{s_2, \sigma_2} \equiv\ket{\sigma_1,\sigma_2}$.  We immediately specialize to the case $s_1=s_2=s$.    In the absence of the exchange interaction, the ground state of the Hamiltonian is four-fold degenerate corresponding to the states where the individual spins are in their highest weight or lowest weight states, $|\hskip-1 mm\uparrow, \uparrow\rangle, |\hskip-1 mm\downarrow, \downarrow\rangle, |\hskip-1 mm\uparrow, \downarrow\rangle, |\hskip-1 mm\downarrow, \uparrow\rangle$, where $\ket{\uparrow, \downarrow}=\ket{\uparrow}\otimes\ket{\downarrow}\equiv \ket{s,-s}$ etc, with the exchange interaction term $J$, the two ferromagnetic states $\ket{\uparrow, \uparrow}$ and $\ket{\downarrow, \downarrow}$  are still degenerate, exact eigenstate of the Hamiltonian,  but the antiferromagnetic states  $\ket{\uparrow, \downarrow}$ and $\ket{\downarrow, \uparrow}$ are not. These two antiferromagnetic states  link with  each other at $2s^{\text{th}}$ order in degenerate perturbation  theory in the exchange transverse term, that is at order $J^{2s}$
 \cite{kim2, sm2}. Thus, the exchange interaction plays the same role as the splitting terms in the uniaxial and biaxial models considered previously. This is completely understandable since tunneling requires a term that does not commute with the quantization axis. However, in this model we will see that both integer and half-odd integer spins can  tunnel\footnote{It is crucial to note that Kramers degeneracy only applies to a system with an odd total number of half-odd integer spin.} but their ground and first excited states are different. Up to an additional constant, the classical energy corresponds to
\begin{align}
U&= J s^2\left(\sin \theta_1\sin \theta_2 \cos(\phi_1-\phi_2)+\cos \theta_1\cos \theta_2+1\right)+Ds^2(\sin^2\theta_1+\sin^2\theta_2).
\end{align}
The minimum energy corresponds to $\phi_1-\phi_2=\pi$: $\theta_1=0$, $\theta_2=\pi$, $\phi_1-\phi_2=\pi$: $\theta_1=\pi$, $\theta_2=0$ and the maximum at $\phi_1-\phi_2=\pi$: $\theta_1=\pi/2$, $\theta_2=\pi/2$. There are four classical equations of motion but we already have the constraint that the total $z$-component spins is conserved, that is $\cos \theta_1+\cos \theta_2=0\Rightarrow \theta_2=\pi-\theta_1=\pi-\theta$. Introducing the variables $\phi=\phi_1-\phi_2$ and  $\Phi=\phi_1+\phi_2$ (which is cyclic), one finds that the two spin problem reduces to an effective single spin problem which is described by the Lagrangian:
\begin{align}
\mathcal{L}_E=is \dot{\phi}(1- \cos\theta)+ U(\theta,\phi),
\end{align}
where the effective energy is
\begin{equation}
U(\theta,\phi)=2Ds^2\sin^2\theta\lb 1+ \frac{\lambda}{2}  (1+\cos\phi)\rb,
\end{equation}
and  $\lambda= J/D\ll1$ . Since $\sin^2\theta\neq0$ as $\theta$ varies as the tunneling progresses,  energy conservation requires
\begin{equation}
\cos\phi=-\lb\frac{2}{\lambda}+1\rb .
\label{aka1}  
\end{equation}
Thus,  $|\cos\phi|>1$ as $\lambda\ll1$.   Therefore there is no real solution for $\phi$ as expected.  It was  shown that the proper choice of $\phi$ for antiferromagnetic coupling is $\phi=\pi+i\phi_I$, where $\phi_I$ is real \cite{sm2}. Plugging this into Eqn.\eqref{aka1} we obtain $\phi_I \approx \ln(4/\lambda)$.

From the classical equation of motion Eqn.\eqref{cla1} one finds that the classical trajectory has the form  
\begin{equation}
\theta\lb \tau\rb =  2 \arctan \lb e^{\omega (\tau-\tau_0)}\rb,
\end{equation}
where $\omega= Js\sinh\phi_I=2Ds\sqrt{1+\kappa}$, $\kappa=J/D$ and at $\tau=\tau_0$ we have $\theta(\tau)= \pi/2$.  Thus $\theta(\tau)$ interpolates  from 0 to $\pi$ as $\tau=-\infty\rightarrow\infty$ for the instanton and from $\pi$ to 0 for an anti-instanton. The action for this trajectory is found to be
\begin{equation}
S_0 = -i2s\pi +2s\phi_I.
\label{aka}
\end{equation}

\begin{figure}[ht]
\includegraphics[width=3.5in]{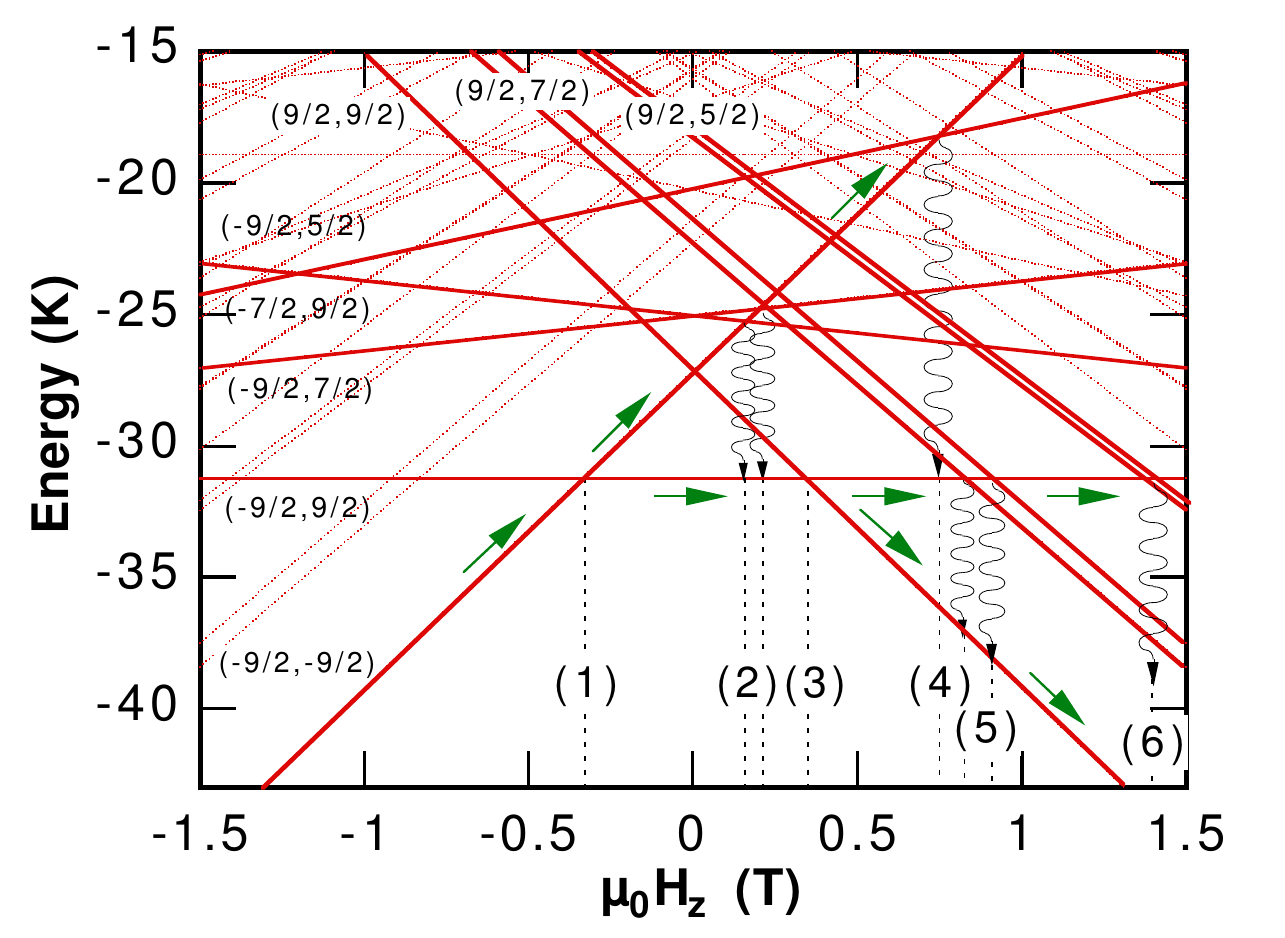}
\caption{ Color online:  
The so-called exact numerical diagonalization of the dimer model plotted as a function of applied magnetic field with the parameters $D=0.77K$, $J=0.13K$. Each state is labeled by $\ket{m_1,m_2}$. Dotted lines, labeled 1 to 5, indicate the strongest tunnel resonances: 1: (-9/2,9/2) to (-9/2 ,9/2); 2: (-9/2,9/2) to (-9/2 ,7/2), followed by relaxation to  (-9/2,9/2); 3:  (-9/2,9/2) to (9/2 ,9/2); 4:  (-9/2,-9/2) to (-9/2 ,5/2), followed by re- laxation to  (-9/2,9/2); 5:  (-9/2,9/2) to (7/2 ,9/2), followed by relaxation to  (9/2,9/2). In order to get most of these transitions theoretical one needs to add term like $J(S_1^{+}S_2^{+}+S_1^{-}S_2^{-})$ in Eqn.\eqref{ant}. Adapted with permission from \onlinecite{tiron}} 
\label{tipic}
\end{figure}

The energy splitting between the ground and the first excited states is given by \cite{sm2, kim2}
\begin{equation}
\Delta  =2\mathscr{D}\lb\frac{J}{4 D }\rb^{2s}\cos(2\pi s).
\label{maa}
\end{equation}
For half-odd integer spin $\Delta<0$ the ground $\ket{g}$ and the first excited $\ket{e}$ states are\begin{align}
| g\rangle =\frac{1}{\sqrt 2}(|\hskip-1mm\downarrow,\uparrow \rangle -|\hskip-1mm\uparrow,\downarrow \rangle);\quad\quad | e\rangle =\frac{1}{\sqrt 2}(|\hskip-1mm\downarrow,\uparrow \rangle +|\hskip-1mm\uparrow,\downarrow \rangle);
\end{align}

while for integer spins  $\Delta>0$ we have
\begin{align}
| g\rangle =\frac{1}{\sqrt 2}(|\hskip-1mm\downarrow,\uparrow \rangle +|\hskip-1mm\uparrow,\downarrow \rangle);\quad\quad | e\rangle =\frac{1}{\sqrt 2}(|\hskip-1mm\downarrow,\uparrow \rangle -|\hskip-1mm\uparrow,\downarrow \rangle).
\end{align}
In this case there is no suppression of tunneling even at zero field, the phase term that arises from the   imaginary term in Eqn.\eqref{aka} switches the ground state from odd to even for half-odd integer and integer spins respectively. This shows that for half-odd integer spins, the ground state is the state with $s=0$. This result has been experimentally  shown that  [Mn$_4$]$_2$ represents an unequivocal and unprecedented example of quantum tunneling in a monodisperse antiferromagnet with no uncompensated spin $ (s=0)$ in the ground state \cite{wern4}. In the presence of an external magnetic field applied along the easy axis, there are $(2s+1)^2 \times (2s+1)^2 = 100\times100$ matrices which are sparsely populated giving rise to an exact numerical diagonalization of 100 non-zero energy states as shown in Fig.\eqref{tipic} , \cite{tiron,hill, wern4, hu, tironb}. The values of the anisotropy parameters that were used to fit experimental data for this dimer are $J=0.13K$, $D=0.77K$\cite{tiron, hill}.
An analogous two spin problem is that of a biaxial antiferromagnetic particle of two collinear ferromagnetic sublattices with a small non-compensation $s=s_1-s_2\neq 0$. The corresponding Hamiltonian \cite{mull0, chud5, anu3}
is
\begin{align}
\hat H&= \sum_{a=1,2}(k_1\hat{S}_a^{z2}+k_2\hat{S}_a^{y2}-h\hat{S}_a^{z})+J \hat{\bold{S}}_{1}\cdot \hat{\bold{S}}_{2},
\end{align}
where $k_1\gg k_2>0$ are the anisotropy constants. It possesses an easy $x$-axis and $xy$ easy plane, and the magnetic field $h$ is applied along the hard $z$-axis. The two spins are unequal unlike the dimer model considered above so one is interested in the sublattice rotation of the N\'eel vector \cite{bab}. The classical energy is of the form:
\begin{align}
U&= J s_1s_2\left(\sin \theta_1\sin \theta_2 \cos(\phi_1-\phi_2)+\cos \theta_1\cos \theta_2\right)+\sum_{a=1,2}(k_1s_a^2\cos^2\theta_a+k_2s_a^2\sin^2\theta_a\sin^2\phi_a-hs_a\cos\theta_a)
\end{align}
The full action contains two WZ terms thus, there are four equations of motion in general. There is no operator that commutes with this Hamiltonian therefore there is no constraint.
In order to get  an effective single spin problem, several approximations have to be made. Firstly, we have to assume that the two spins $s_1$ and $s_2$ are almost antiparallel. Therefore, one can replace $\theta_2$ and $\phi_2$ by $\theta_2=\pi-\theta_1-\epsilon_\theta$ and $\phi_2=\pi+\phi_1+\epsilon_\phi$ where $\epsilon_\theta, \epsilon_\phi \ll 1$ are small fluctuations. Replacing $\theta_2$ and $\phi_2$ in the action and setting $s_1=s_2=s_0$ except for the terms containing $s_1-s_2=s$, and integrating out the fluctuations  $\epsilon_\theta, \epsilon_\phi$ from the path integral one obtains an effective single spin model, which can then be solved using the procedures outline above. However, unlike the dimer model, one finds in this case that in the absence of the magnetic field, tunneling of hampered when $s$ is half-odd integer \cite{chud5} while in the presence of the magnetic field, tunneling splitting oscillates with the field only vanishes at a certain critical value \cite{mull0}
\subsection{Coordinate independent formalism}
\label{cif}
\subsubsection{Equation of motion and Wess-Zumino action}
The coordinate dependent formalism we have just reviewed in the previous section is widely used in most condensed matter literature, but not much seems to be written about the solutions of these models in a coordinate independent form. The solution of a physical problem should  be independent of the coordinate system.  Having solutions only in a coordinate dependent form leaves a slight but persistent, irritating doubt that somehow the results may have some coordinate dependent artefacts, which of course should not be there.  In section \eqref{sec:spinpa} we derived the classical action for the spin system without the use of coordinates. In this section we will show that one can solve the spin models we have considered so far in totally coordinate independent way and also recover the quantum phase interference exactly as before. First of all, we need to know the classical path that minimizes the coordinate independent action Eqn.\eqref{act1}: 
\begin{equation}
S_E[\bold{\hat n}] = isS_{{WZ}}+ \int d\tau U(\bold{\hat n(\tau)}), \quad  U(\bold{\hat n(\tau)})=\braket{\bold{\hat n}| \hat H|\bold{\hat n}}.
\label{act1again}
\end{equation}
The variation of coordinate independent WZ term, Eqn.\eqref{wz} due to small variation  of $\bold{\hat n}$  gives
\begin{align}
\delta S_{WZ}&=\int d\tau\int d\xi\thinspace\partial_{\tau}[\bold{\hat n}\cdot(\delta\bold{\hat n}\times\partial_\xi\bold{\hat n})]+ \int d\tau\int d\xi\thinspace\partial_{\xi}[\bold{\hat n}\cdot( \partial_\tau\bold{\hat n}\times \delta\bold{\hat n})].
\label{varia}
\end{align}
To obtain this variation we must remember that   $0=\delta (\bold{\hat n}\cdot \bold{\hat n}) =2 \bold{\hat n}\cdot \delta\bold{\hat n}$, and  $0=\partial_{\tau,\xi} (\bold{\hat n}\cdot \bold{\hat n}) =2\bold{\hat n}\cdot \partial_{\tau,\xi} \bold{\hat n}$, since  $\bold{\hat n}$ is a unit vector.  Consequently, the volume defined by the parallelepiped traced out by the three vectors, the variation and the two derivatives, must vanish,  $\delta\bold{\hat n}\cdot(\partial_\tau\bold{\hat n}\times\partial_\xi\bold{\hat n})=0$ since any three vectors orthogonal to a given vector $\bold{\hat n}$, lie in the same plane.
The first term in Eqn.\eqref{varia} vanishes by virtue of the boundary conditions Eqn.\eqref{bo} and the second term yields
\begin{align}
\delta S_{WZ}&=-\int d\tau \thinspace \delta\bold{\hat n(\tau)}\cdot[\bold{\hat n}(\tau)\times\partial_\tau\bold{\hat n}(\tau)] .
\label{varia1}
\end{align}
As $\delta\bold{\hat n(\tau)}$ is still a constrained variation, necessarily orthogonal to $\bold{\hat n}$, therefore
\beq
\frac{\delta S_{WZ}}{\delta\bold{\hat n(\tau)}}\ne[\bold{\hat n}(\tau)\times\partial_\tau\bold{\hat n}(\tau)] .
\eeq
What we may conclude is that the part  of $[\bold{\hat n}(\tau)\times\partial_\tau\bold{\hat n}(\tau)]$ which is orthogonal to $\bold{\hat n}$ will contribute to the equation of motion.  The way to implement this, is to take the vector product with $\bold{\hat n}$, which implements the projection to the appropriate orthogonal directions.  
Then, using the fact that $\bold{\hat n}(\tau)\times[\bold{\hat n}(\tau)\times\partial_\tau\bold{\hat n}(\tau)]=-\partial_\tau\bold{\hat n}(\tau)$, the variation of the total action gives the equation of motion:
\begin{align}
is\partial_\tau\bold{\hat n}(\tau)=-\bold{\hat n}(\tau)\times\frac{\partial U(\bold{\hat n}(\tau))}{\partial\bold{\hat n}(\tau)}.
\label{ll}
\end{align}
This is the imaginary-time equivalent for the equation for Larmor precession in the effective magnetic field ${\delta U(\bold{\hat n}(\tau))}/{\delta\bold{\hat n}(\tau)}$, often called the Landau-Lifshitz equation \cite{ll,ll0}. Taking the cross product of Eqn.\eqref{ll} with $\partial_\tau\bold{\hat n}(\tau)$, and subsequently the dot product with $\bold{\hat n}(\tau)$, one finds  immediately the equation of  energy  conservation:
\bea
 U(\bold{\hat n}(\tau)) =\text{const.}
\label{corcon}
\eea
 Having obtained the equation of motion as a function of the trajectory $\bold{\hat n}(\tau)$, we wish need to write the WZ action, Eqn.\eqref{wz} as a function of $\tau$ alone, as done in the coordinate dependent formulation as in Eqn.\eqref{wzcon}, in order to compute the instanton action for the trajectory $\bold{\hat n}(\tau)$. This can only be achieved  if the integration over $\xi$ can be done leaving us with the integration over $\tau$ in terms of the unit vector $\bold{\hat n}(\tau)$. This integration can indeed  be done. Let us express the unit vector $\bold{\hat n}(\tau, \xi)$ as
\bea
\bold{\hat n}(\tau, \xi)= f(\tau,\xi)n_z(\tau)\bold{\hat z} + g(\tau,\xi)[n_x(\tau) \bold{\hat{x}}+n_y(\tau)\bold{\hat y}],
\label{3.14}
\eea
with the boundary conditions given in Eqn.\eqref{bo}. From Eqn.\eqref{3.14} and $\bold{\hat n}\cdot \bold{\hat n}=1$ one obtains immediately
\bea
g^2=\frac{1-f^2n_z^2}{1-n_z^2}.
\label{eze6}
\eea
Owing to the boundary conditions in Eqn.\eqref{bo}, these functions must obey 
 \begin{align}
& f(\tau,\xi=0)=1;\quad f(\tau,\xi=1)=\frac{1}{n_z(\tau)};\quad  g(\tau,\xi=0)=1;\quad g(\tau,\xi=1)=0.
\end{align}
A long but straightforward calculation \cite{sm1}  shows that
\begin{align}
\bold{\hat n}(\tau, \xi)\cdot (\partial_\tau\bold{\hat n}(\tau, \xi)\times\partial_\xi\bold{\hat n}(\tau, \xi))  = \frac{n_z\partial_\xi f}{1-n_z}(n_x\dot{n}_y-n_y\dot{n}_x).
\end{align}
The WZ term becomes\footnote{A similar expression is given in \cite{klau,petr,spg, anu0}}\cite{sm1}
\bea
S_{WZ} =is\int d\tau \frac{(n_x\dot{n}_y-n_y\dot{n}_x 
)}{1+n_z}.
\label{3.13a}
\eea
This expression defines the WZ term in the coordinate independent form as a function of time alone. By spherical parameterization
one can easily recover the coordinate dependent  form given by Eqn. \eqref{wzcon}. Further simplification of Eqn. \eqref{3.13a} yields
\begin{align}
S_{WZ} & = is\int \frac{d(n_y/n_x)}{1+(n_y/n_x)^2}(1-n_z) =is\int d[\arctan(n_y/n_x)](1-n_z).
\label{wzconin}
\end{align}
\subsubsection{Coordinate independent uniaxial spin model in a magnetic field}
Now let us consider the uniaxial model in
section\eqref{unia}. The corresponding classical energy in coordinate independent form is
\bea
U(\bold{\hat n})  = -Ds^2(\bold{\hat n}\cdot \bold{\hat z})^2 -H_xs  \bold{ \hat n}\cdot \bold{\hat x}.
\eea
From Eqn.\eqref{ll} we obtain the equation of motion
\beq
is\partial_\tau\bold{\hat n}-2Ds^2(\bold{\hat n}\cdot\bold{\hat{z}})(\bold{\hat n}\times\bold{\hat{z}}) -H_xs (\bold{\hat n}\times\bold{\hat{x}})=0. \label{ea}           
\eeq
 Taking the cross product of this equation with $\partial_\tau\bold{\hat n}$ and using the fact that $\bold{\hat n}(\tau)\cdot\partial_\tau\bold{\hat n}(\tau)=0$ we obtain  the conservation of energy 
\begin{align}
 Ds^2((\bold{\hat n}\cdot \bold{\hat{z}})^2-1) +H_xs \bold{\hat n}\cdot \bold{\hat{x}} -H_x^2/4D
=0,
\end{align}
where an additional constants have been added for convenience. Using this expression together with the constraint $\bold{\hat n}\cdot \bold{\hat n}=1$ we find the relations
\begin{align}
\bold{\hat n}\cdot\bold{\hat{x}}&= \frac{1}{2h_x}
(1+h_x^2-(\bold{\hat n}\cdot\bold{\hat{z}})^2);
\quad
\bold{\hat n}\cdot\bold{\hat{y}}= \pm\frac{i}{2h_x}(1-h_x^2-(\bold{\hat n}\cdot\bold{\hat{z}})^2).\label{hxhy1}
\end{align}
The ratio of these two expressions give
\begin{equation}
\frac{\bold{\hat n}\cdot\bold{\hat{y}}}{\bold{\hat n}\cdot\bold{\hat{x}}}=\pm i\frac{ 1-h_x^2-(\bold{\hat n}\cdot\bold{\hat{z}})^2}{1+h_x^2-(\bold{\hat n}\cdot\bold{\hat{z}})^2}=\tan\chi,
\label{ratio}
\end{equation}
which is imaginary.
Taking the scalar product of Eqn. \eqref{ea} with $\bold{\hat{z}}$ and using Eqn.\eqref{hxhy1} we obtain
\beq
  is\partial_\tau(\bold{\hat n}\cdot\bold{\hat{z}}) \pm iDs^2 (1-h_x^2-(\bold{\hat n}\cdot\bold{\hat{z}})^2)=0.
\eeq
The above equation integrates as
\beq
\bold{\hat n}\cdot \bold{\hat{z}} = \pm \sqrt{1-h_x^2}\tanh(\omega_h\tau), \quad \omega_h = Ds\sqrt{1-h_x^2},
\eeq
which is the same as Eqn.\eqref{unistan}.
 To determine the action for this trajectory we use Eqn.\eqref{wzconin}, that is
\begin{align}
B &=is\int \frac{d(n_y/n_x)}{1+(n_y/n_x)^2}(1-n_z).
\end{align}

From Eqn.\eqref{ratio} we find:

\begin{align}
B &=\pm s\int_{\mp\sqrt{1-h_x^2}}^{\pm\sqrt{1-h_x^2}} \frac{n_zdn_z}{1-n_z^2}(1-n_z) = 2s\bigg[\frac{1}{2}\ln\lb\frac{1+\sqrt{1-h_x^2}}{1-\sqrt{1-h_x^2}}\rb-\sqrt{1-h_x^2}\bigg], 
\end{align}
which is exactly the coordinate dependent result in  Eqn.\eqref{anu4}.
\subsubsection{Coordinate independent biaxial model and suppression of tunneling}
In section\eqref{bint}, we reviewed the  suppression of tunneling for half-odd integer spin for a biaxial single molecule magnet a particular choice of coordinate. In this section we will show that these results can be recovered in terms of the unit vector $\bold{\hat n}(\tau)$. Thus, the suppression of tunneling for half-odd integer spin is independent of the choice of coordinate. In the coordinate independent form, the classical energy of the Hamiltonian, Eqn.\eqref{bia} can be written as
\bea
U = D_1s^2(\bold{\hat n}\cdot\bold{\hat{z}})^2 +D_2s^2(\bold{\hat n}\cdot\bold{\hat{x}})^2,
\eea

The classical equation of motion, Eqn.\eqref{ll} yields
\beq
is\partial_\tau\bold{\hat n}+2D_1s^2(\bold{\hat n}\cdot\bold{\hat{z}})(\bold{\hat n}\times\bold{\hat{z}}) +2D_2s^2  (\bold{\hat n}\cdot\bold{\hat{x}})(\bold{\hat n}\times\bold{\hat{x}})=0. \label{eomnh}           
\eeq
From the conservation of energy and the fact that $\bold{\hat n}\cdot \bold{\hat n}=1$, it follows that

\begin{align}
\bold{\hat n}\cdot\bold{\hat{z}}&=\pm i\sqrt{\frac{D_2}{D_1}}\bold{\hat n}\cdot\bold{\hat{x}}=\pm i\sqrt{\frac{D_2}{D_1-D_2}(1-(\bold{\hat n}\cdot\bold{\hat{y}})^2)};\quad \bold{\hat n}\cdot\bold{\hat{x}}=\pm\sqrt{\frac{D_1}{D_1-D_2}(1-(\bold{\hat n}\cdot\bold{\hat{y}})^2)}.
\label{hxhy}
\end{align}
Then
\beq
\frac{\bold{\hat n}\cdot\bold{\hat{y}}}{\bold{\hat n}\cdot\bold{\hat{x}}}= \pm\frac{\bold{\hat n}\cdot\bold{\hat{y}}}{\sqrt{\frac{D_1}{D_1-D_2}(1-(\bold{\hat n}\cdot\bold{\hat{y}})^2)}}=\tan\chi.
\label{phi}
\eeq
Taking the scalar product of Eqn. \eqref{eomnh} with $\bold{\hat{x}}$ and using Eqn.\eqref{hxhy} yields
 \beq
is\partial_\tau(\bold{\hat n}\cdot\bold{\hat{y}}) - i2s^2\sqrt{D_1D_2}(1-(\bold{\hat n}\cdot\bold{\hat{y}})^2)=0.
\eeq
Upon  integration we obtain the instanton:
\beq
\bold{\hat n}\cdot\bold{\hat{y}}= n_y=\tanh \lb \omega(\tau-\tau_0)\rb,
\eeq
where $\omega= 2s\sqrt{D_1D_2}$. The instanton interpolates from $n_y=1$ to $n_y=-1$ as $\tau\rightarrow\pm\infty$. Thus, $\arctan(\bold{\hat n}\cdot\bold{\hat{y}}/{\bold{\hat n}\cdot\bold{\hat{y}}})\rightarrow \pm\pi/2$ as $\tau\rightarrow \pm\infty$. Since the energy remains constant along the instanton trajectory, the action is determined only from the WZ term:\begin{align}
S_c&  = is\int_{-\frac{\pi}{2}}^{\frac{\pi}{2}} d[\arctan(n_y/n_x)](1-n_z).\end{align}
From Eqn.\eqref{hxhy} and Eqn.\eqref{phi} we find
\bea
\bold{\hat n}\cdot\bold{\hat{z}}=n_z= \pm\frac{i\sqrt{\lambda}}{\sqrt{1-\lambda +\lb\frac{n_y}{n_x}\rb^2}}, \quad \lambda = D_2/D_1.
\eea
Thus, we recover the  action in Eqn.\eqref{act3}
\begin{align}
S_c &=is\pi + \ln\lb\frac{1+\sqrt{\lambda}}{1-\sqrt{\lambda}}\rb^s.\end{align}
The calculation of the energy splitting follows directly from section\eqref{bint}. Thus, one recovers the spin-parity effect in a coordinate independent manner.  This simply means that the spin-parity effect is independent of the choice of coordinate.

\section{Effective potential (EP) method}
\label{epmth}

As we mentioned earlier, the spin coherent state path integral formalism is valid in the large $s$ limit, in other words if one imposes the commutator relation $[\phi, p]=i\hbar$, where $p=s\cos\theta$, then the spin commutator relation $[\hat{S}_i,\hat{S}_j]=i\epsilon_{ijk}\hat{S}_k$ is only recovered in the large $s$ limit\footnote{The proof of this is given in \cite{mull1}, Appendix A}. On the other hand, the effective potential method uses an exact mapping \cite{swh,zas1,zas2}.  In this method, one introduces the spin wave function  using the $\hat{S}_{z}$ eigenstates, and the resulting eigenvalue equation $\hat H \ket{\psi}=\mathcal{E}\ket{\psi}$ is then transformed to a differential equation, which is further reduced to a Schr\"{o}dinger equation with an effective potential and a constant or coordinate dependent mass. The energy spectrum of the spin system now coincides with the $2s+1$ energy levels for the particle moving in a potential field. The limitations of the method are as follow:
\begin{enumerate}[1).]

\item{ In the effective potential method, the WZ term (Berry phase) does not appear in the corresponding particle action, the quantum phase interference effect seems to disappear, however, in some special cases with a magnetic field one can recover the quenching of tunneling at the critical field from the periodicity of the particle wave function.}
 \item{The effective potential method of higher order anisotropy spin models such as $\hat H = -D\hat{S}_z^2 -B\hat{S}_z^4 + C(\hat{S}_+^4 + \hat{S}_-^4)-H_x\hat{S}_x$ and $\hat H = D_1\hat{S}_z^2+D_2\hat{S}_x^2 + C(\hat{S}_+^4 + \hat{S}_-^4)$ are very cumbersome to map onto a particle problem. In fact there is no effective potential method for such systems.  Therefore the effective potential method is only efficient for large spin systems that are quadratic in the spin operators.}
\end{enumerate}
\subsection{Effective method for a uniaxial spin model with a transverse magnetic field}
\label{ep1}
In this section we will consider the effective potential method of the uniaxial model we studied in section\eqref{unia}. The Hamiltonian of this system is given by
\begin{align}
\hat H = -D\hat{S}_z^2 - H_x\hat{S}_x.
\label{u1}
\end{align}
Consider the the problem of finding the exact eigenstates of this Hamiltonian. The eigenvalue equation is 
\bea
\hat H \ket{\psi}=\mathcal{E}\ket{\psi},
\label{u2}
\eea
where the spin wave function in the $\hat{S}_z$ representation is given by \cite{swh}
\beq
\ket{\psi} =\sum_{\sigma=-s}^{s}\binom{2s}{s+\sigma}^{-1/2}  c_\sigma\ket{ s,\sigma}.
\label{u3}
\eeq
Using the fact that $\hat{S}_x=\frac{1}{2}\lb\hat{S}_+ + \hat{S}_-\rb$ and
\bea
\hat{S}_{\pm}\ket{s,\sigma}=\sqrt{(s\mp \sigma)(s\pm \sigma+1)}\ket{s,\sigma \pm 1}.
\label{raislow}
\eea
A straightforward calculation using Eqns.\eqref{u1},\eqref{raislow}, and \eqref{u3} in Eqn.\eqref{u2}  gives:
\begin{align}
&-D\sigma^2c_\sigma-\frac{1}{2}H_x[(s-\sigma+1)c_{\sigma-1}+(s+\sigma+1)c_{\sigma+1}]=\mathcal{E}c_\sigma,
\label{u4}
\end{align}
where $\sigma=-s,-s+1,\cdots, s$, and $ c_\sigma=0$ for $|\sigma|>s$.
Introducing a generating function of the form:
\bea
\mathcal{G}(x)= \sum_{\sigma=-s}^{s}c_\sigma e^{\sigma x},
\label{gn}
\eea
the eigenvalue equation, that is Eqn.\eqref{u4} transforms to a second-order differential equation of the form:
\bea
b_1\frac{d^2\mathcal{G}}{d^2x}+b_2\frac{d\mathcal{G}}{dx} -b_3\mathcal{G}=\mathcal{E}\mathcal{G},
\label{pipi}
\eea
where
\begin{align}
b_1&=-D;\quad b_2=H_x\sinh x;\quad b_3=H_xs\cosh x.
\end{align}

The spin-particle correspondence follows from a special transformation of the form\footnote{\label{note10} Substituting Eqn.\eqref{u8} into Eqn.\eqref{pipi} gives $b_1 \Psi^{\prime\prime} +\lb 2b_1y^{\prime}+b_2\rb\Psi^{\prime} + [b_2y^{\prime} + b_3+b_1\lb y^{\prime\prime} +y^{\prime 2}\rb]\Psi=\mathcal E\Psi$. The function $y(x)$ is determined by  demanding the coefficient of $\Psi^{\prime}$ vanishes.} 
\bea
\Psi(x)= e^{-y(x)}\mathcal{G}(x),
\label{u8}
\eea
where $y(x)=\tilde{s}h_x\cosh(x)$, $h_x= H_x/2D\tilde{s}<1$, and $\tilde{s}=(s+\frac{1}{2})$ is a quantum renormalization. This transformation in Eqn.\eqref{u8} is regarded as the coordinate or particle wave function since $\Psi(x)\rightarrow  0$ as $x\rightarrow \pm \infty$. Plugging this transformation into Eqn.\eqref{pipi} removes the first derivative term yielding the Schr\"odinger equation\cite{swh,zas2,zas1}:
\begin{align}
\hat H \Psi(x)=\mathcal{E} \Psi(x);\quad\hat H = -\frac{1}{2m}\frac{d^2}{dx^2}+U(x),
\end{align}
where 
\begin{align}
U(x)= D\tilde{s}^2(h_x\cosh x-1)^2; \quad m=\frac{1}{2D}.
\label{u6}
\end{align}
As before we have added a constant to normalize the potential to zero at the minimum $\cosh x= 1/h_x$. In Eqn.\eqref{u8}, the generating function contains a real exponential function. This choice is usually a matter of convenience. In most cases it is convenient to use an imaginary exponential function to avoid some technical issues, as we will see in the next section. The minimum of the potential is now at $x_{\text{mim}}=\pm\arccosh(1/h_x)$ and the maximum is at $x_{\text{max}}=0$ with the height of the barrier given by
\begin{equation}
\Delta U = D\tilde{s}^2(1-h_x)^2.
\label{hg}
\end{equation}
It is possible to analytically solve the Schr\"odinger equation and find the energy levels of the particle in the potential Eqn.\eqref{u6}, such solution has been reported \cite{raz}. This potential is of the form of a double well we saw in Sec.\eqref{sec:spinpa} with $\pm a = \pm\arccosh(1/h_x)$. The instanton solution of such a problem follows the same approach \cite{cole3}. The Euclidean Lagrangian corresponds to Eqn.\eqref{eucl} with the mass and the potential given by Eqn.\eqref{u6}. The solution of the Euclidean classical equation of motion, Eqn.\eqref{parin} yields the instanton trajectory \cite{zas1, zas2}
\begin{equation}
x(\tau)= \pm 2\arctanh \bigg[\sqrt{\frac{1-h_x}{1+h_x}}\tanh(\omega\tau)\bigg],
\end{equation}
where $\omega= D\tilde{s}\sqrt{{1-h_x^2}}$. This nontrivial solution corresponds to the motion of the spin particle at the top of the left hill at  $\tau\rightarrow -\infty$, $x(\tau)\rightarrow -a$ and roll through the dashed line in Fig.\eqref{pot28} and emerges at the top of the right hill at  $\tau\rightarrow \infty$, $x(\tau)\rightarrow a$. The corresponding action for this trajectory is
\begin{equation}
B= 2\tilde{s}\bigg[\frac{1}{2}\ln\lb \frac{1+\sqrt{1-h_x^2}}{h_x}\rb-\sqrt{1-h_x^2}\bigg].
\label{chud10}
\end{equation}
The computation of the ground state energy splitting yields \cite{zas1,chud2}
\begin{equation}
\Delta = \frac{8D\tilde{s}^{3/2}(1-h_x^2)^{5/2}}{\pi^{1/2}}\lb \frac{e^{\sqrt{1-h_x^2}}}{1+\sqrt{1-h_x^2}}\rb^{2\tilde{s}}h_x^{2s},
\label{u7}
\end{equation}
which recovers the factor $h_x^{2s}$ we saw previously in the spin coherent state path integral formalism. In the presence of a longitudinal magnetic field i.e along $z$-axis, the two degenerate minima of the potential become biased, one with lower energy and the other with higher energy. The problem becomes that of a quantum decay of a metastable state \cite{zas3}.
\subsection{Effective method for  biaxial spin models}
\label{rmp1}
\subsubsection{Biaxial ferromagnetic spin with hard axis magnetic field}
The biaxial spin model also possesses a particle mapping via the EP method. Consider the biaxial system studied in sec.\eqref{solo}
\begin{equation}
\hat H = D_1\hat{S}_z^2 + D_2\hat{S}_x^2-h_z\hat{S}_z.
\label{biaf}
\end{equation}
A convenient way to  map this system to particle Hamiltonian is by introducing a non-normalized spin coherent state \cite{rad,anu0,pere,erg}:
\begin{align}
\ket{z} &= e^{zS^{-}}\ket{s,s}=\sum_{\sigma=-s}^{s}\binom{2s}{s+\sigma}^{1/2} z^{s-\sigma}\ket{ s,\sigma}=e^{is\phi}\sum_{\sigma=-s}^{s}\binom{2s}{s+\sigma}^{1/2} e^{-i\sigma \phi}\ket{ s,\sigma}.
\label{non1}
\end{align}
The last equality sign follows by restricting the complex variable on a unit circle, i.e $z=e^{i\phi}$. Acting from the left  by $e^{-is\phi}\bra{\psi}$ and subsequently taking the complex conjugate
we obtain\begin{align}
\braket{z|\psi}=e^{is\phi}\sum_{\sigma=-s}^{s}\binom{2s}{s+\sigma}^{1/2}c_{\sigma} e^{i\sigma \phi}\equiv e^{is\phi}\Phi(\phi),
 \label{u11}
 \end{align}
where $c_\sigma= \braket{s,\sigma|\psi}$ and $\Phi(\phi)$ is the generating function\footnote{It is convenient to use the generating function for $x$ or $y$ easy axis models while Eqn.\eqref{gn} is convenient for $z$ easy axis model. In that way one avoids the problem of a negative mass particle.},
with periodic boundary condition $\Phi(\phi+2\pi)$= $e^{2i\pi s}\Phi(\phi)$.
From Eqn.\eqref{non1} we have
\begin{align}
\braket{z|\hat{S}_z|\psi}&=e^{is\phi}\sum_{\sigma=-s}^{s}\binom{2s}{s+\sigma} ^{1/2}\sigma c_{\sigma}e^{i\sigma \phi}=-ie^{is\phi}\frac{d\Phi(\phi)}{d\phi}.
\end{align}
Similar expressions can be derived for $\braket{z|\hat{S}_x|\psi}$ and $\braket{z|\hat{S}_y|\psi}$. Thus, the action of the spin operators on this function yields the following expressions \cite{zas1,zas2}:
\begin{align}
& {\hat{S}}_z =  -i\frac{d}{d\phi}; \quad {\hat{S}}_x = s\cos\phi-\sin\phi\frac{d}{d\phi}; \quad {\hat{S}}_y = s\sin\phi+\cos\phi\frac{d}{d\phi}.
\label{u9}
\end{align}
The Schr\"{o}dinger equation can then be written as
  \bea
  \hat{H}\Phi(\phi) = \mathcal{E}\Phi(\phi).
  \label{u10}
  \eea
 From Eqn.\eqref{biaf} and Eqn.\eqref{u9}  one obtains the differential \cite{zas1,mull1}:
 \begin{equation}
\begin{split}
& -{D_1}(1-\lambda\sin^2\phi)\frac{d^2\Phi}{d\phi}-D_2(s-\frac{1}{2})\sin 2\phi \frac{d\Phi}{d\phi} +ih_z\frac{d\Phi}{d\phi}+(D_2s^2\cos^2\phi + D_2s\sin^2\phi )\Phi= \mathcal E\Phi .
\label{3.4b} 
\end{split}
\end{equation} 
A convenient way to obtain a Schr\"{o}dinger equation with a constant is by introducing an incomplete elliptic integral of first kind \cite{by, abra} and the particle wave function:
 \begin{align}
 x &= F(\phi,\kappa)= \int_{0}^{\phi}d\varphi  \frac{1}{\sqrt{1-\kappa^2\sin^2\varphi}};\quad
 \Psi(x)= e^{-iu(x)}[\dn(x)]^{-s }\Phi(\phi(x)),
  \label{eze5}
  \end{align}
 with amplitude $\phi$ and modulus $\kappa^2=\lambda$. The trigonometric functions are related to the Jacobi elliptic functions by $\sn(x)=\sin\phi$, $\cn(x)=\cos\phi$ and $\dn(x)=\sqrt{1-\kappa^2\sn^2(x)}$. The function $u(x)$ is defined by
 \bea
 \frac{du}{dx}=\frac{\alpha s}{\dn(x)},\quad \alpha = h_z/2D_2s .
 \label{der}
 \eea
 The imaginary phase is a topological shift in the wave function which is related to Aharonov Bohm effect \cite{abom}.
In this new variable, Eqn.\eqref{3.4b} transforms into a  Schr\"{o}dinger equation  
with
\beq
H =  \frac{1}{2m}\bigg[-i\frac{d}{dx}+A(x)\bigg]  + U(x); \quad m= \frac{1}{2 {D_1} } .
\label{3.18}
\eeq
The effective potential and the gauge field are given by
\bea
U(x) = \eta \cd(x)^2;  \quad \cd(x)=\frac{\cn(x)}{\dn(x)};
\label{eze7}
\eea
\bea
A(x)=-\frac{(2s+1)\alpha}{\dn(x)};
\eea
where $\eta = D_2s(s+1)+\frac{\lambda \alpha^2}{4(1-\lambda)}$. The potential has a period of $2\mathcal{K}(\kappa)$, where $\mathcal{K}(\kappa) $ is the complete elliptic function of first kind that is $\phi=\pi/2$ in the upper limit of Eqn.\eqref{eze5}. Using Eqn.\eqref{der} one finds that the wave function obeys the periodic boundary condition\cite{mull3, ssd}
\begin{equation}
\Psi(x+4\mathcal{K}(\kappa))=e^{i2\pi s(1-\alpha/\sqrt{1-\lambda})}\Psi(x) .
\label{wavef}
\end{equation}
The corresponding Euclidean Lagrangian of this particle Hamiltonian is
\begin{equation}
L_E = \frac{1}{2}m\dot{x}^2+iA(x)\dot{x}+U(x).
\end{equation}
The second term of this equation drops out from the classical equation of motion, however, it is responsible for the suppression of tunneling splitting just like the WZ term (Berry phase) in the spin coherent state path integral formalism. Thus one finds that
the exact instanton solution is 
\bea
\sn[x(\tau)]=\tanh(\omega\tau), \quad \omega^2= 4s(s+1) {D_1D_2},
\eea
which interpolates from $x_i= -\mathcal{K}(\kappa)$ $(\phi=-\pi/2)$ at $\tau=-\infty$ to $x_f= \mathcal{K}(\kappa)$ $(\phi=\pi/2)$ at $\tau=\infty$.   The action for this trajectory is found to be
\begin{equation}
S_c = -i(2s+1) b + B,
\end{equation}
where $b= \pi \alpha/\sqrt{1-\lambda}$ and $B$
is  given by
\begin{equation}
B = \sqrt{\frac{\eta}{D_2}}\ln \lb \frac{1+ \sqrt{\lambda}}{1- \sqrt{\lambda}}\rb.
\end{equation}
By summing over instantons and anti-instantons configurations, it was shown that the energy splitting is given by\cite{mull1}
\begin{equation}
\Delta  =\Delta_0\lvert\cos(s\pi +b)\rvert,\quad \Delta_0 =  4\mathscr{D} e^{-B}.
\end{equation}
Thus one recovers the suppression of tunneling as before.

 As an alternative approach of recovering the quenching of tunneling splitting, consider the transition from $x=0$ to $x= 2\mathcal{K}(\kappa)$ and $x=0$ to $x=-2\mathcal{K}(\kappa)$. The former is counterclockwise transition while the latter is clockwise transition, thus the total transition amplitude vanishes:
\begin{equation}
\mathcal{A}(2\mathcal{K}(\kappa),t;0,0)+\mathcal{A}(-2\mathcal{K}(\kappa),t;0,0)=0,
\label{propa}
\end{equation}
where $\mathcal{A}$ represent the Feynman propagator given in Eqn.\eqref{qua}. In terms of the wave function the propagator can be written as\cite{cole1,cole2,cole3,fey}
\begin{equation}
\mathcal{A}(x_f,t;x_i,0)=\sum_{l}\Psi_l(x_f)\Psi_l^*(x_i)e^{-i\mathcal{E}_lt}.
\label{propa1}
\end{equation}
Then from Eqs.\eqref{propa} and \eqref{propa1} one obtains the relation
\bea
\Psi_l(2\mathcal{K}(\kappa))=-\Psi_l(-2\mathcal{K}(\kappa)),
\eea
which yields from Eqn.\eqref{wavef}
\begin{equation}
e^{i2\pi s(1-\alpha/\sqrt{1-\lambda})}=-1,
\label{wavefq}
\end{equation}
for any quantum number $l$.
From this equation one obtains the condition for suppression of tunneling \cite{ssd}
\begin{equation}
 \alpha=\sqrt{1-\lambda}\left(s-n-1/2\right)\slash s,
 \end{equation}
just as Eqn.\eqref{eze4}.
\subsubsection{Biaxial ferromagnetic spin with medium axis magnetic field}
\label{med}
 Suppose we apply a magnetic field in the medium $x$-axis corresponding to the Hamiltonian:
\begin{equation}
\hat H = D_1\hat{S}_z^2 + D_2\hat{S}_x^2-H_x\hat{S}_x.
\label{eze10}
\end{equation}
As we pointed out in sec.\eqref{solo}, the quenching of tunneling at the critical field is only seen with biaxial spin models with magnetic field along the hard-axis, thus this model does not possess such effect. At zero magnetic field, there are two classical degenerate ground states corresponding to the minima of the energy located at $\pm \bold{\hat y}$, these ground states remain degenerate for $h_x\neq0$ in the easy $XY$ plane. 
The particle Hamiltonian is
\beq
H =  -\frac{1}{2m}\frac{d^2}{dx^2} + U(x) , \quad m= \frac{1}{2 {D_1} },
\eeq
with the effective potential and the wave function given by \cite{sm3}
\begin{align}
U(x) &= \frac{D_2\tilde{s}^2[\cn(x)-\alpha_x]^2}{\dn^2(x)};\quad
\Psi(x)=\frac{\Phi(\phi(x))}{ [\dn(x)]^{s }} \exp\bigg[-\arccot\lb \sqrt{\frac{\lambda}{(1-\lambda)}}\cn(x)\rb\bigg],
\label{eze8} 
\end{align}
where $\tilde{s}=(s+\frac{1}{2})$ and $\alpha_x = H_x/2 {D_2}\tilde{s}$. In order to arrive at this potential we have used the approximation $s(s+1)\sim \tilde{s}^2$ and shifted the minimum energy to zero by  adding a constant  of the form $D_2 \tilde{s}^2\alpha_x^2$.  The potential, Eqn.\eqref{eze8} has minima at $x_0=4n\mathcal{K(\kappa)}\pm\cn^{-1}(\alpha_x)$ and maxima at $x_{sb}=\pm4n\mathcal{K(\kappa)}$ for small barrier and at $x_{lb}=\pm2(2n+1)\mathcal{K(\kappa)}$ for large barrier. The heights of the potential for small and large barriers are given by \cite{sm3,mull2}
\begin{align}
\Delta U_{sb} =D_2\tilde{s}^2(1-\alpha_x)^2;\quad
\Delta U_{lb} = D_2\tilde{s}^2(1+\alpha_x)^2,
\label{ba}
\end{align}
\begin{figure}[ht]
\centering
\includegraphics[scale=0.4]{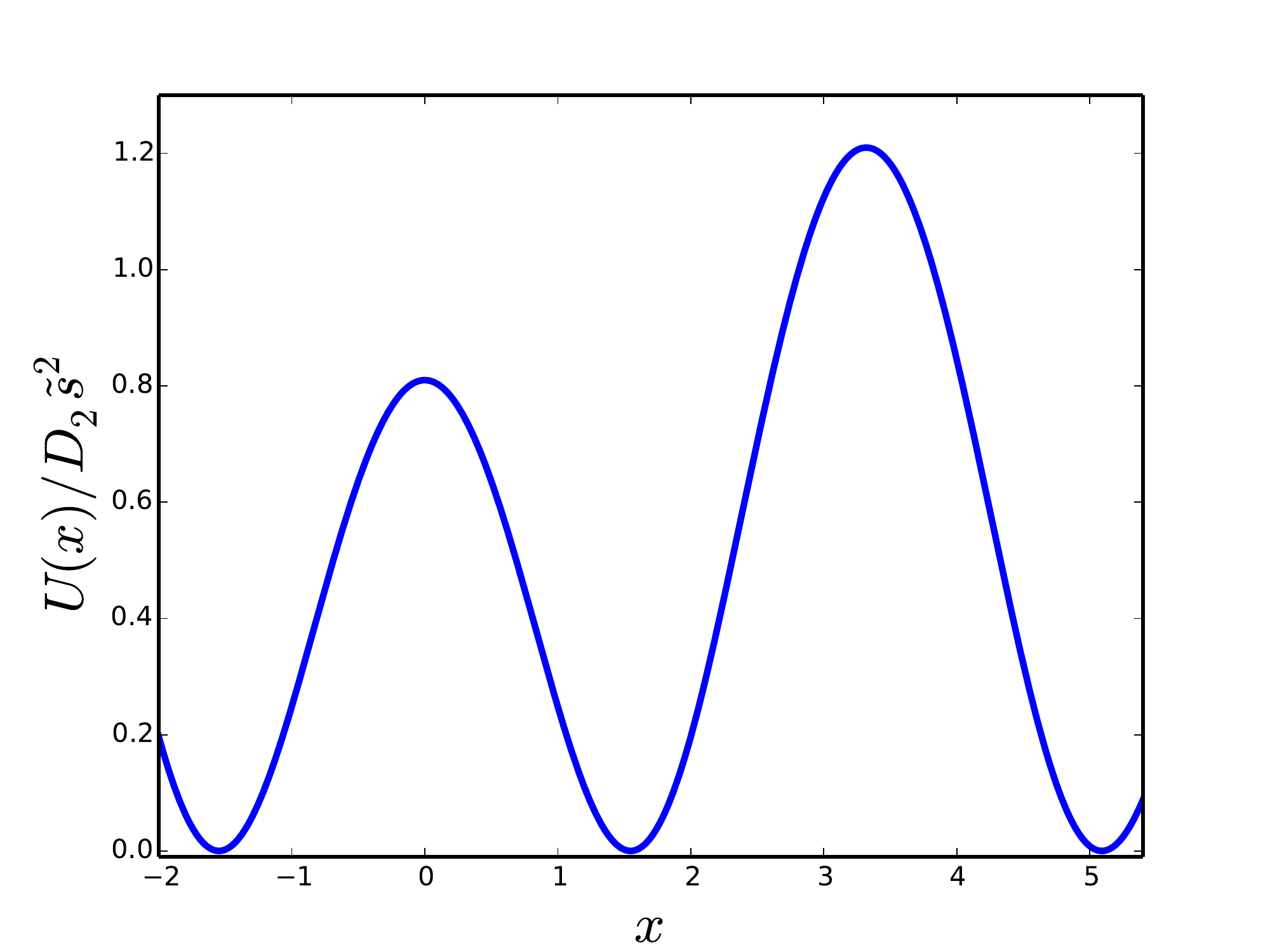}
\caption{ Color online: The plot of the effective potential in Eqn.\eqref{eze8} for $\alpha_x=0.1$, $\kappa=0.2$.}
\label{pot6}
\end{figure}

The classical trajectory yields
\bea
 \sn[x(\tau)] =\pm \frac{2 \sqrt{\frac{1-\alpha_x}{1+\alpha_x}}\tanh(\omega\tau)}{[1+\frac{1-\alpha_x}{1+\alpha_x}\tanh^2(\omega\tau)]},
\label{int}
\eea
and the corresponding  action is \cite{zas1,sm3}
\begin{align}
B&= \tilde{s}\bigg[ \ln\lb\frac{1+\sqrt{\lambda(1-\alpha_x^2)}}{1-\sqrt{\lambda(1-\alpha_x^2)}}\rb\pm 2\alpha_x \sqrt{\frac{\lambda}{1-\lambda}}\arctan\lb \frac{\sqrt{(1-\lambda)(1-\alpha_x^2)}}{\alpha_x}\rb\bigg],
\label{eq4}
\end{align}
where the upper and lower signs are for tunneling in large and small barriers respectively. The tunneling splitting can be found in the usual way by summing over instanton and anti-instanton configurations.
During our discussion of phase transition in the next section, we will return to this concept of large and small barriers in detail. In this section we have specifically chosen biaxial spin models that possess an exact instanton solution. The transformations in Eqns.\eqref{u9}---\eqref{u11} are derived by restricting the analysis on a unit circle parameterize by the angle $\phi$. In these two models, the variable $\phi$ and then $x$ correspond exactly to the azimuthal angle $\phi$ in the spin coherent state path integral. In other representations of a biaxial spin system, this is not true and the EP method gives a very complicated effective potential, one can neither find the exact instanton solution nor the suppression of tunneling. However, without computing the explicit instanton trajectory, the action at the bottom of the potential well can be found in some cases by another elegant approach as we will see in the next section.

\section{Quantum-classical phase transitions of the escape rate in large spin systems}
\label{pt}
\subsection{Methods for studying quantum-classical phase transitions of the escape rate }
In the preceding sections, we have reviewed quantum tunneling in spin systems which is dominated by instanton trajectory at zero temperature. As we mentioned in Section\eqref{sec:intro}, transitions at finite temperature can be either first or second-order. In this section we will now discuss the phase transition of the escape rate from thermal to quantum regime at nonzero temperature.  The escape rate of a particle through a potential barrier in the semiclassical approximation is obtained by taking the Boltzmann average over tunneling probabilities \cite{aff, chud2}:
\begin{equation}
\Gamma = \int_{U_{\text{min}}}^{U_{\text{max}}} d \mathcal{E} \mathscr{P}(\mathcal{E})e^{-\beta(\mathcal{E}-U_{\text{min}})},
\label{decay}
\end{equation}
where $\beta^{-1}=T$ is the temperature of the system, which is much less than the height of the potential barrier. This defines the temperature assisted tunneling rate, and  $\mathscr{P}(\mathcal{E})$ is an imaginary time transition amplitude from excited states at an energy $\mathcal{E}$. The integration limits $U_{\text{max}}$ and $U_{\text{min}}$ are the  top and bottom  of the potential energy respectively. The transition amplitude is defined as
\begin{equation}
\mathscr{P}(\mathcal{E})=\mathscr{A} e^{-S(\mathcal{E})},
\label{trans}
\end{equation}
where $\mathscr{A}$ is a prefactor independent of $\mathcal{E}$. The Euclidean action is of the form:
\bea
S(\mathcal{E}) = 2\int_{x_1(\mathcal{E}) }^{x_2(\mathcal{E})} dx \sqrt{2m(x)(U(x)-\mathcal{E})},
\label{eneact}
\eea
where $x_{1,2}(\mathcal{E})$ are the roots of the  integrand in Eqn.\eqref{eneact}, which are the classical turning points ($U(x_{1,2})=\mathcal{E}$)  of a particle with energy $-\mathcal{E}$ in the inverted potential $-U(x)$ as depicted in Fig.\eqref{pot23}. The mass $m(x)$ is  coordinate dependent in general.   The factor of $2$ in Eqn.\eqref{eneact} corresponds to the back and forth oscillatory motion of the particle in the inverted potential (see Fig.\ref{pot23}). In other words, the particle crosses the barrier  twice.  
\subsubsection{Phase transition with thermon action}
\label{pt1}
The escape rate can as well be written as
\begin{equation}
\Gamma=\mathscr A \int_{U_{\text{min}}}^{U_{\text{max}}} d \mathcal{E} e^{-\mathcal{S}_p},
\label{decay0}
\end{equation}where 
\bea
\mathcal{S}_p= S({\mathcal{E}}) + \beta (\mathcal{E}-U_{\text{min}}),
\label{thermon}
\eea
  is the thermon action \cite{chud6}. In the method of steepest decent (for small temperatures $T< \hbar \omega_0$, $\omega_0$ is the frequency at the bottom of the potential), one can introduce fluctuations around the classical path that minimizes this thermon action, i.e $\frac{d \mathcal{S}_p}{d\mathcal{E}}=0$. The escape rate, Eqn.\eqref{decay0} in this method   is thus written as \cite{chud3}
\begin{equation}
\Gamma \sim  e^{-\mathcal{S}_{\text{min}}(\mathcal{E})},
\label{decay2}
\end{equation}
and $\mathcal{S}_{\text{min}}(\mathcal{E})$ is the minimum of the thermon action in Eqn.\eqref{thermon} with respect to energy. 

In many cases of physical interest, when the energy is in the range $U_{\text{min}}<\mathcal{E}<U_{\text{max}}$, the Euclidean action $S(\mathcal{E})$  can be computed exactly or numerically in the whole range of energy for any given potential in terms of complete elliptic integrals and hence the thermon action $\mathcal{S}_p$. This corresponds to the action of the periodic instanton\cite{mull0} or thermon.  At the bottom of the potential $\mathcal{E}=U_{\text{min}}$, the minimum thermon action becomes the vacuum instanton action, that is \bea \mathcal{S}_{\text{min}}(U_{\text{min}})= {S}(U_{\text{min}}).\eea Thus, the vacuum instanton action of the previous sections becomes $ B=S(U_{\text{min}})/2$, since it corresponds to half of the period of oscillation.  Eqn.\eqref{decay2} becomes the transition amplitude formula for a pure quantum tunneling.  However, at the top of the barrier $\mathcal{E}= U_{\text{max}}$, the Euclidean action vanishes, $S(U_{\text{max}})=0$, the minimum thermon action (thermodynamic action) becomes
\bea
\mathcal{S}_{\text{min}}(U_{\text{max}})=\mathcal{S}_0 = \beta {\Delta U}.
\label{thermonda}
\eea
This corresponds to the action of a constant trajectory $x(\tau)=x_s$ at the bottom of the inverted potential \cite{chud6}. The escape rate Eqn.\eqref{decay2} becomes the Boltzmann formula for a pure thermal activation.
As we showed  in section\eqref{sec:intro}, the crossover temperature from thermal to quantum regimes (``first-order phase transition") occurs when the escape rate Eqn.\eqref{decay2} with $\mathcal{S}_{\text{min}}(U_{\text{min}})$ is equal to that with $\mathcal{S}_{\text{min}}(U_{\text{max}})$, which yields Eqn.\eqref{temp0}. At this temperature the thermon action $\mathcal{S}_p$ sharply intersects with the thermodynamic action $\mathcal{S}_0$ leading to a discontinuity in the first-derivative of the action $\mathcal{S}_p$ at $\beta_0^{(1)}$. For second-order phase transition the thermon action $\mathcal{S}_p$ smoothly joins the thermodynamic action $\mathcal{S}_0$ at $\beta=\beta_0^{(2)}$.
\subsubsection{Phase transition with thermon period of oscillation}
\label{pt2}
The dominant term in Eqn.\eqref{decay0} comes from the minimum of the thermon action Eqn.\eqref{thermon}, which is given by

\begin{align}
\beta(\mathcal{E}) &=-\frac{dS(\mathcal{E})}{d\mathcal{E}}= \int_{x_1(\mathcal{E})}^{x_2(\mathcal{E})}dx\sqrt{\frac{2m(x)}{U(x)-\mathcal{E}}} \equiv \tau(\mathcal E).
\label{pper}
\end{align}
This is the period of oscillation of a particle with energy $-\mathcal{E}$ in the inverted potential $-U(x)$. At the bottom of the potential $\mathcal{E}= U_{\text{min}}$, the period  $\beta(\mathcal{E})= \infty$ i.e $T= 0$ which corresponds to the vacuum instanton of section\eqref{epmth}, while at the top of the barrier $\mathcal{E}= U_{\text{max}}$,  $\beta(\mathcal{E}) \to \beta_0^{(2)}= 2\pi/\omega_b$ \cite{aff}. The first and second-order transitions can be studied  from the behaviour of $\beta(\mathcal{E})$ as a function of $\mathcal{E}$.
\begin{enumerate}[1).]
 \item {If $\beta(\mathcal{E})$ has a minimum  at some point $\mathcal{E}_0< U_\text{max}$, $\beta_\text{min}=\beta(\mathcal{E}_0)$ and then rises again, i.e non-monotonic, then first-order phase transition occurs \cite{chud6}. At a certain energy within the range $U_\text{min}<\mathcal{E}_1<\mathcal{E}_0$, the thermon action sharply intersects with the thermodynamic action, yielding the actual crossover temperature $\beta_0^{(1)}=\beta(\mathcal{E}_1)$.}
  \item { A monotonic decrease of $\beta(\mathcal{E})$ with increasing $\mathcal{E}$ from the bottom to the top of the barrier indicates the presence of second-order phase transition\cite{chud2,chud3, chud6}. In this case  the thermon action $\mathcal{S}_p$ smoothly intersects with the thermodynamic action $\mathcal{S}_0$, yielding the crossover temperature $\beta_0^{(2)}$ \cite{blatter, chud6}, which is exactly Eqn.\eqref{temp1}.}
\end{enumerate}

\subsubsection{Phase transition with free energy}
\label{pt3}
The semiclassical escape rate Eqn.\eqref{decay2} can be written in a slightly different form:
\begin{equation}
\Gamma \sim  e^{-\beta F_{\text{min}}},
\label{decay1}
\end{equation}
where $F_{\text{min}}=\beta^{-1}\mathcal{S}_{\text{min}}(\mathcal E)$ is the minimum of the effective free energy
\bea
F =\beta^{-1}\mathcal{S}_p= \mathcal{E}+ \beta^{-1} S(\mathcal{E})-U_{\text{min}},
\label{freen}
\eea
with respect to $\mathcal{E}$. 
The crossover from thermal to quantum regimes (first-order phase transition) occurs when two minima  in the $F$ vs. $\mathcal E$ curve have the same free energy.
All the interesting physics of phase transition in spin systems can also be captured when the energy is very close (but not equal) to the top of the potential barrier, $\mathcal{E}\rightarrow U_{\text{max}}$. In this case the free energy can  then be used to characterize first- and second-order phase transitions in analogy with Landau's theory of phase transition if one knows the exact expression of the action $S(\mathcal{E})$ for any given mass and potential. In most models with a magnetic field the action $S(\mathcal{E})$ cannot be obtained exactly, one has to study the free energy numerically.
\subsubsection{Phase transition with criterion formula}
\label{pt4}
An alternative method for determining the phase transition of the escape rate, as well as the phase boundary was considered by \textcite{mull4}. They  studied the Euclidean action near the top of the potential barrier, which had been considered earlier by \textcite{blatter}. For  the general case of a particle that possesses  a coordinate dependent mass, they found that near the top of the potential barrier  the expression that depends on the potential, which determines the type of phase transition is given by \cite{mull4}
\begin{align}
\mathscr{C}&= \bigg[U^{\prime\prime\prime}(x_s)\lb g_1+\frac{g_2}{2}\rb+\frac{1}{8}U^{\prime\prime\prime\prime}(x_s) +\omega^2m^{\prime}(x_s)g_2 +\omega^2m^{\prime}(x_s) \lb g_1+\frac{g_2}{2}\rb+\frac{1}{4}\omega^2m^{\prime\prime}(x_s)\bigg]_{\omega=\omega_b}, 
\label{mull}
\end{align}
where  

\begin{align}
&g_1 = -\frac{\omega^2m^{\prime}(x_s) +U^{\prime\prime\prime}(x_s)}{4U^{\prime\prime}(x_s)}, \label{3.2}\\&
g_2 = -\frac{ 3m^{\prime}(x_s)\omega^2 +U^{\prime\prime\prime}(x_s)}{4\left[ 4 m(x_s)\omega^2+U^{\prime\prime}(x_s)\right]},  \\&\omega_b^2 =-\frac{U^{\prime\prime}(x_s) }{m(x_s)}; \quad m^{\prime}\equiv \frac{dm(x)}{dx} \thinspace, \text{etc.}
\label{omega}
\end{align}
 The coordinate $x_s$ represents the position of the sphaleron\footnote{Sphalerons are static, unstable, finite-energy solutions of the
classical equations of motion.} at the bottom of the inverted potential  as shown in Fig.\eqref{pot23}. The criterion for first-order phase transition requires $\mathscr{C}<0$, while $\mathscr{C}>0$ implies a second-order transition, and the phase boundary  between the first- and the second-order phase transitions  is of course $\mathscr{C}=0$. The criterion formula in Eq.\eqref{mull} is quite general. It can be simplified in two special cases:
\begin{enumerate}[1).]
 \item{If the mass of the particle is a constant and the  potential energy is an  even function, Eq.\eqref{mull} reduces to
 \begin{align}
\mathscr{C}&= \frac{1}{8}U^{\prime\prime\prime\prime}(x_s).
\label{conmull1}
\end{align}
Thus, expanding the potential around $x_{s}$, the coefficient of the fourth-order term quickly determines the first- and the second-order phase transitions, as well as the phase boundary \cite{chud3}.}

\item{If mass is still a constant but the potential is an odd function, Eq.\eqref{mull} reduces to \begin{align}
\mathscr{C}&= -\frac{5}{24}\frac{[U^{\prime\prime\prime}(x_s)]^2}{U^{\prime\prime}(x_s)}+\frac{1}{8}U^{\prime\prime\prime\prime}(x_s).\label{conmull2}
\end{align}
}
\end{enumerate}
\subsection{Phase transition in uniaxial spin model in a magnetic field}
\label{phaseuni}
\subsubsection{Spin model Hamiltonian}
\begin{figure}[ht]
\centering
\includegraphics[scale=0.35]{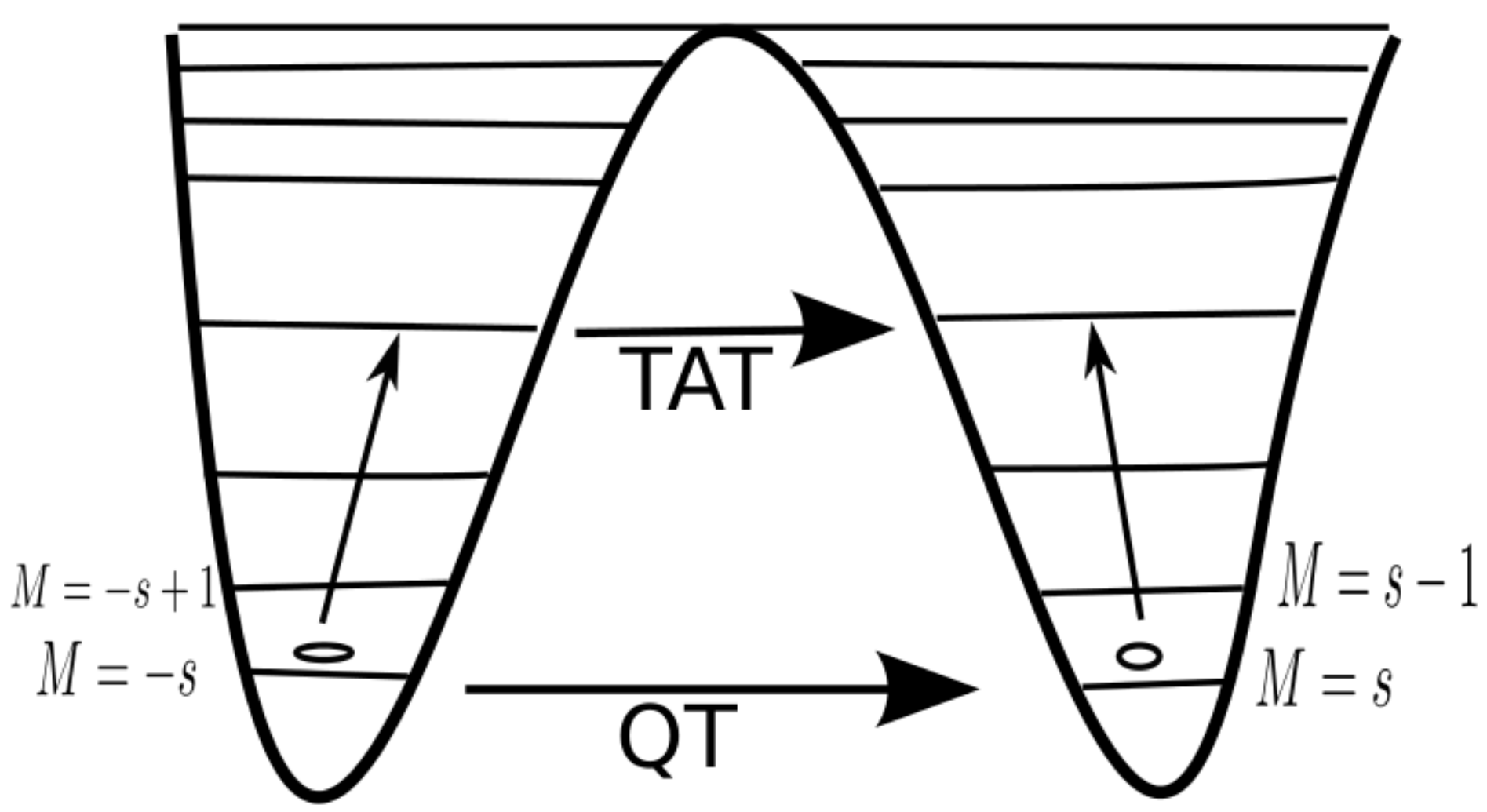}
\caption{A sketch of quantum tunneling(QT) and thermally assisted resonant tunneling (TAT) in Mn$_{12}$Ac molecular magnet.}
\label{pot31}
\end{figure}

We have written down all the necessary formulae for studying the phase transition of the escape rate for a uniaxial spin model in an applied field. An extensive analysis of this model can be found in \cite{chud2, chud8}. In this section we will briefly review the theoretical analysis and recent experimental development. For this system we saw that the spin Hamiltonian is given by
\begin{align}
\hat H = -D\hat{S}_z^2 - H_x\hat{S}_x.
\label{chud2}
\end{align}
As we mentioned before this system is a good approximation for Mn$_{12}$Ac, with a ground state of $s=10$  and  $21$ energy levels. Transition between these states can occur either by quantum tunneling (QT) or thermally assisted tunneling (TAT) as depicted in Fig.\eqref{pot31}. 
\subsubsection{Particle Hamiltonian}
\label{uniham}
As we explicitly showed in Sec.\eqref{ep1}, the spin Hamiltonian in Eqn.\eqref{chud2} corresponds to the particle potential and the mass:
\begin{align}
U(x)= D\tilde{s}^2(h_x\cosh x-1)^2, \quad m=\frac{1}{2D}.
\label{chud3}
\end{align}
Since the potential is an even function and the mass is constant, the quickest way to determine the regime where the first-order transition sets in, is  by considering where the coefficient of the fourth order term changes sign near $x_s=0$.  Expanding the potential around  $x_s$ we have
\begin{align}
U(x)&\approx U(0) + D\tilde{s}^2[-h_x(1-h_x)x^2 +\frac{h_x}{3}\lb h_x -\frac{1}{4}\rb x^4 + O(x^6)].
\label{expa}
\end{align}
The coefficient of $x^2$ in Eqn.\eqref{expa} is negative for $h_x<1$, which corresponds to nonvanishing of the potential barrier, Eqn.\eqref{hg}.
 The coefficient of $x^4$ is similar to $\mathscr C$ in Eq.\eqref{conmull1}, it is given by\cite{mull4}
 \bea
 \mathscr C=D\tilde{s}^2 h_x\lb h_x-\frac{1}{4}\rb.
 \eea
 \begin{figure}[ht]
\centering
\includegraphics[width=3.5in]{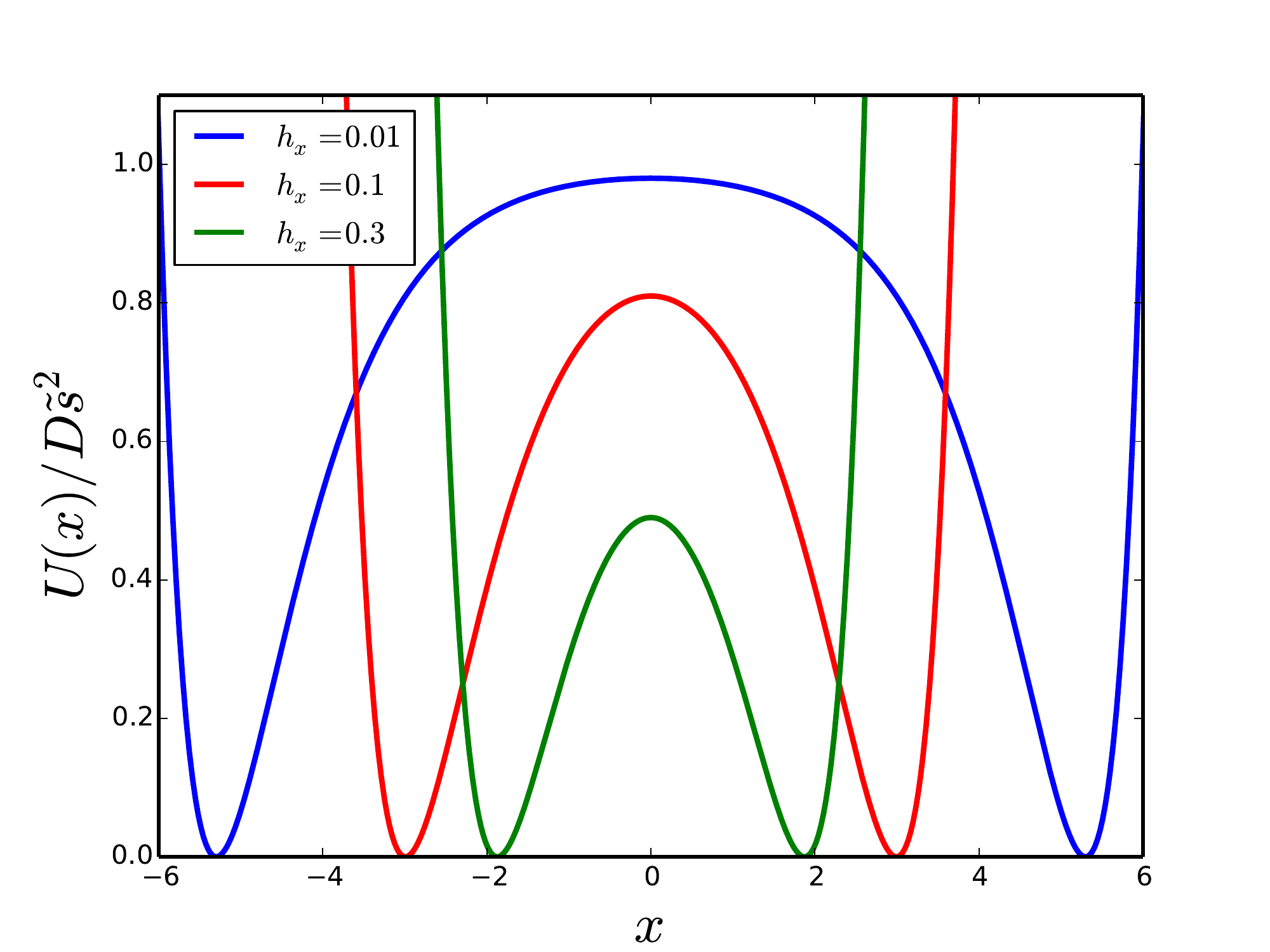}
\caption{Color online: The plot of the potential in Eq.\eqref{chud3} for several values of $h_x$.}
\label{chud_pont}
\end{figure}
Clearly, $\mathscr C$ changes sign for $h_x<\frac{1}{4}$, which corresponds to the regime of the first-order transition from thermal activation to quantum tunneling. It is positive for $h_x>\frac{1}{4}$, which is the regime of second-order phase transition, and of course vanishes at the phase boundary $h_{cx}=\frac{1}{4}$.  \subsubsection{Thermon or periodic instanton action}
\begin{figure}[ht]
\centering
\includegraphics[width=3.5in]{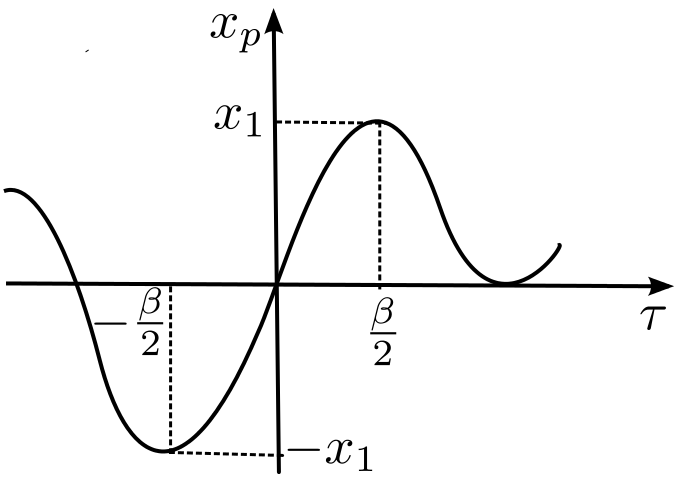}
\caption{ The plot of the periodic trajectory in Eq.\eqref{uniperiod}.}
\label{peri}
\end{figure}

 An alternative approach to investigate quantum-classical phase transitions of the escape rate is by computing the thermon action:
  \begin{align}
\mathcal{S}_p = 2\sqrt{2m}\int_{-x_1 }^{x_1} dx \sqrt{U(x)-\mathcal{E}} +\beta (\mathcal{E}-U_{\text{min}}),
\label{uniact}
\end{align}
 where $\pm x_{1}$ are the roots of the integrand which are the classical turning points. This action corresponds to the action of the periodic instanton trajectory\footnote{The thermon action in Eq.\eqref{uniact} only differs from the periodic instanton action by a factor of 2 in Eq.\eqref{uniact}. So we will use the two names interchangeably.} of Eq.\eqref{chud3}. That is the solution of the classical equation of motion:
 \bea
 \frac{1}{2}m\dot{x}^2 -U(x)=-\mathcal E.
 \eea
Integrating this equation using Eq.\eqref{chud3} one finds that the periodic instanton trajectory is given by \cite{zha3}
\begin{align}
x_p =\pm 2\tanh^{-1}\big[\xi_p\sn\lb \omega_p\tau, k\rb\big],
\label{uniperiod}
\end{align}
where
\begin{align}
k^2&=\frac{1-\lb\sqrt{\tilde{\mathcal{E}}} +h_x\rb^2}{1-\lb\sqrt{\tilde{\mathcal{E}}} -h_x\rb^2}, \quad \xi_p^2 =\frac{1- h_x\pm \sqrt{\tilde{\mathcal{E}}}}{1+ h_x\pm \sqrt{\tilde{\mathcal{E}}}},\\
\omega_p^2 &= (D\tilde{s})^2\big[1-(\sqrt{\tilde{\mathcal{E}}} -h_x)^2\big],\quad
\tilde{\mathcal{E}}= \mathcal{E}/D\tilde{s}^2.
\end{align}
It is required that as $\tau\to\pm \frac{\beta}{2}$, this trajectory must tend to the classical turning points $x_p\to\pm x_1$ as depicted in Fig\eqref{peri}.
   Let us define dimensionless energy quantity\cite{chud3,chud2}:
 \bea
 P= \frac{U_\text{max}-\mathcal E}{U_\text{max}-U_\text{min}}.
 \eea
 Clearly $P\to 0$ at the top of the barrier $\mathcal E\to U_\text{max}$, and $P\to 1$ at the bottom of the barrier $\mathcal E\to U_\text{min}$. By making a change of variable $y=\cosh x$,  Eq.\eqref{uniact} can be reduced to complete elliptic integrals \cite{abra,by} in the whole range of energy. It is found to be of the form \cite{chud2}:
 \begin{figure}[ht]
\centering
\includegraphics[width=3.5in]{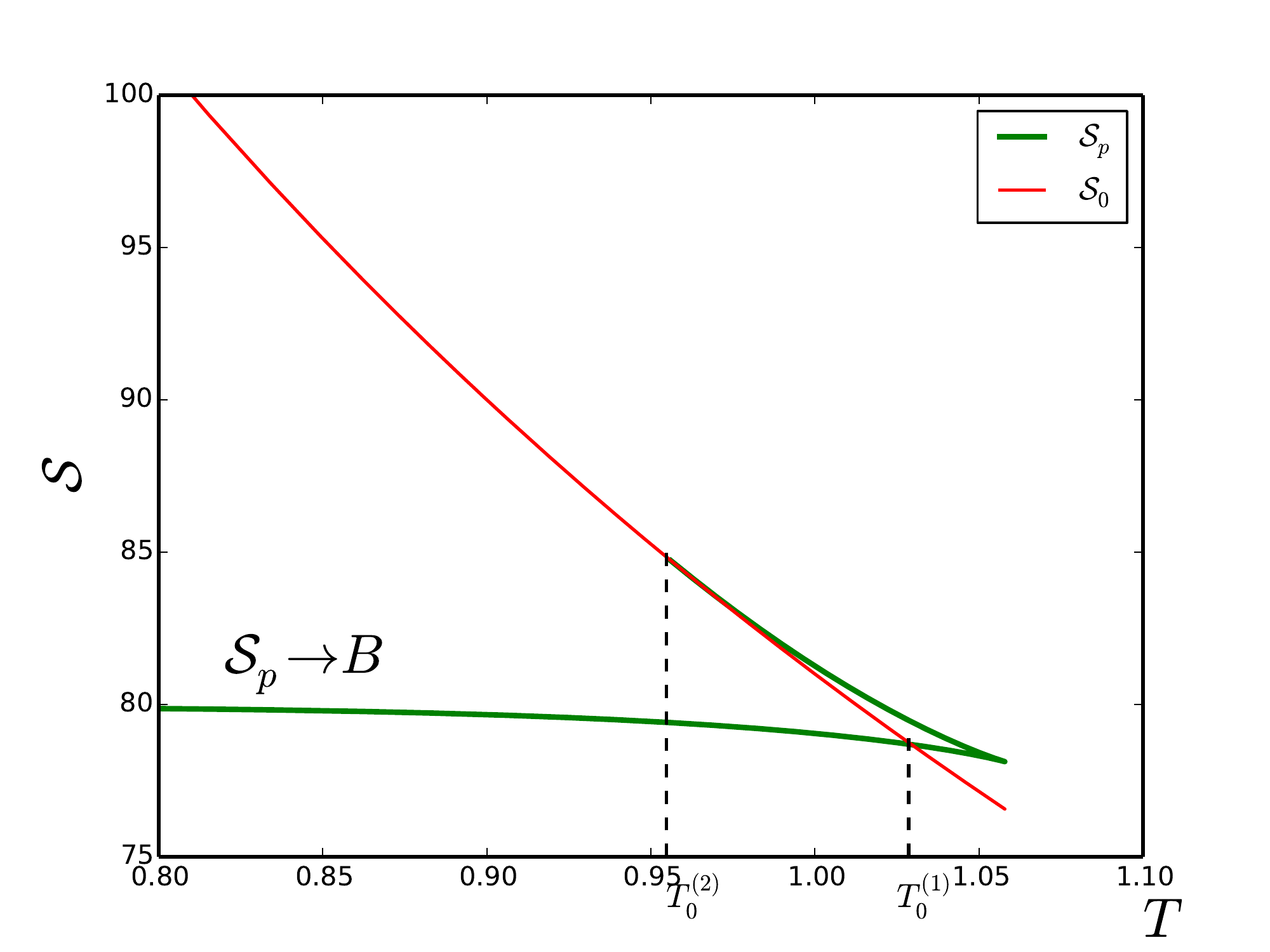}
\includegraphics[width=3.5in]{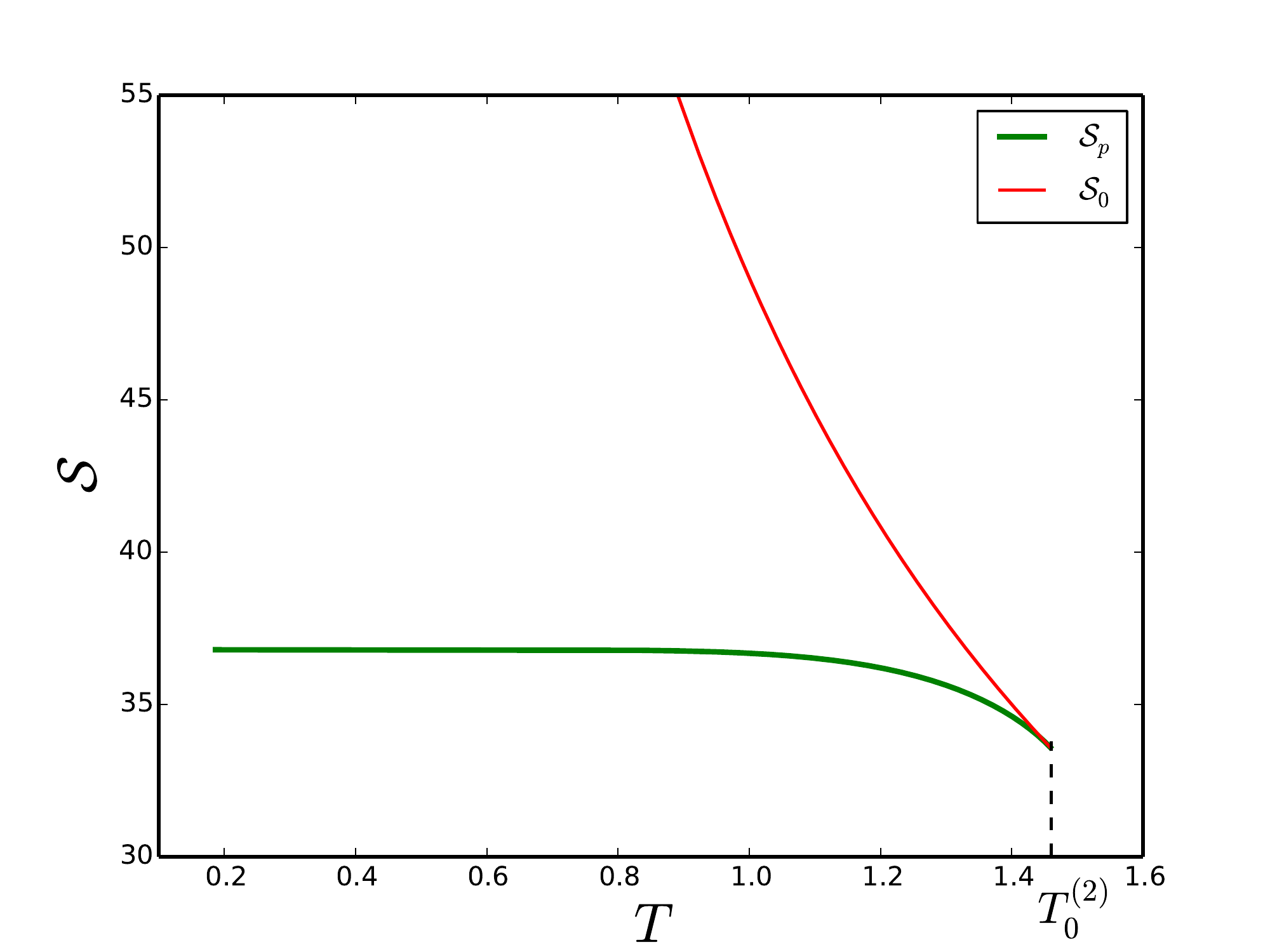}
\caption{ Color online: The plot of the thermon action Eqn.\eqref{thermon} and thermodynamic action Eqn.\eqref{thermonda} against temperature. Left:  $s=10$, $D=1$ $h_x=0.1$ (first-order transition), where $B$ is the vacuum instanton action. Right: $s=10$, $D=1$, $h_x=0.3$ (second-order transition).}
\label{unithermon}
\end{figure}
\begin{align}
\mathcal{S}_p=4\tilde{s} \sqrt{(1-h_x)g_+}\mathcal{I}(\alpha^2, k)+\beta \Delta U(1-P),
\label{chudthem}
\end{align}
 where
 \bea
 g_{\pm}= P+h_x\lb 1 \pm \sqrt{1-P}\rb^2,
 \eea
 \begin{align}
 \mathcal{I}(\alpha^2, k)&= (1+\alpha^2)\mathcal{K}(k)-E(k)+(\alpha^2-k^2/\alpha^2)(\Pi(\alpha^2,k)-\mathcal{K}(k));\quad\alpha^2= (1-h_x)P/g_+; \quad k^2=g_-/g_+;
 \end{align}
where $\mathcal{K}(k)$, $E(k)$ and $\Pi(\alpha^2,k)$ are the complete elliptic integral of the first, the second and the third kinds respectively. In Fig.\eqref{unithermon} we have shown the plot of the actions Eqn.\eqref{chudthem} and Eqn.\eqref{thermonda} as a function of temperature. Indeed one observes the sharp and smooth intersections corresponding to the first- and the second-order phase transitions temperatures respectively.

\subsubsection{Free energy function}

The free energy can also be used to study the quantum-classical phase transitions in this systems. It can be written down exactly from Eqn.\eqref{chudthem}. It is given by
\begin{align}
\frac{F}{\Delta U}= 1-P +\frac{4\theta\sqrt{h_xg_+}}{\pi\lb 1-h_x\rb^{3/2}}\mathcal{I}(\alpha^2, k),
\end{align}
where $\theta= T/T_0^{(2)}$, and $T_0^{(2)}$ is given by Eq.\eqref{unisecond}. 
Near the top of the barrier $P\ll1$, the free energy of this spin model yields \cite{chud2,chud3}
\begin{align}
F(P)/\Delta U &= 1 +(\theta-1)P +\frac{\theta}{8}\lb1-\frac{1}{4h_x}\rb P^2  +\frac{3\theta}{64}\lb 1-\frac{1}{3h_x} +\frac{1}{64h_x^3}\rb P^3 + O(P^4).
\end{align}
This free energy should be compared with the Landau's free energy function:
\bea
F= F_0+ a\psi^2 +b\psi^4 + c\psi^6.
\label{land}
\eea
\begin{figure}[ht]
\centering
\includegraphics[width=3.5in]{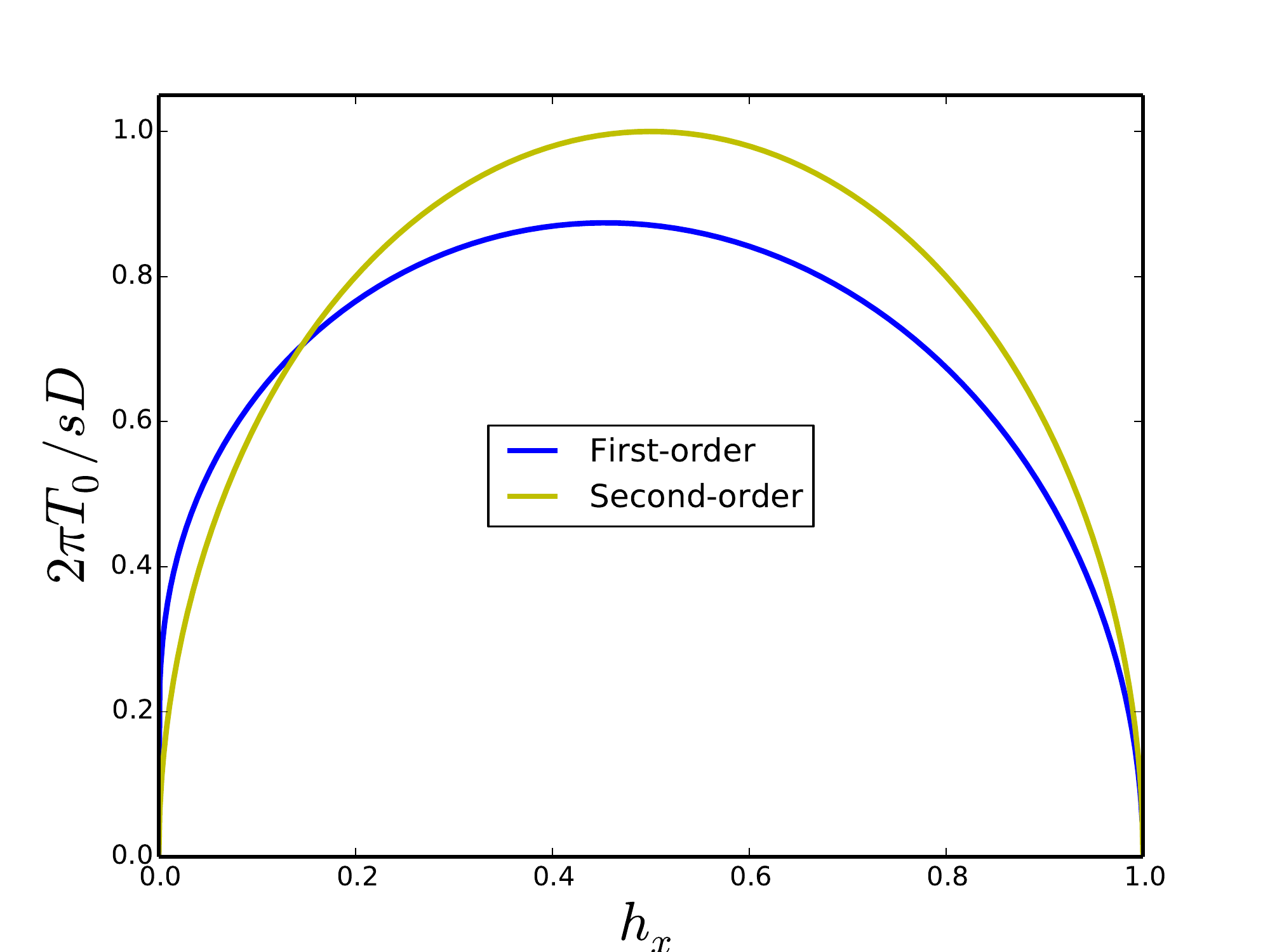}
\caption{ Color online: The plot of the of the first-and second-order cross over temperatures against $h_x$. Reproduced with permission from \onlinecite{chud2}.}
\label{pot7}
\end{figure}
\begin{figure}[ht]
\centering
\includegraphics[width=3.5in]{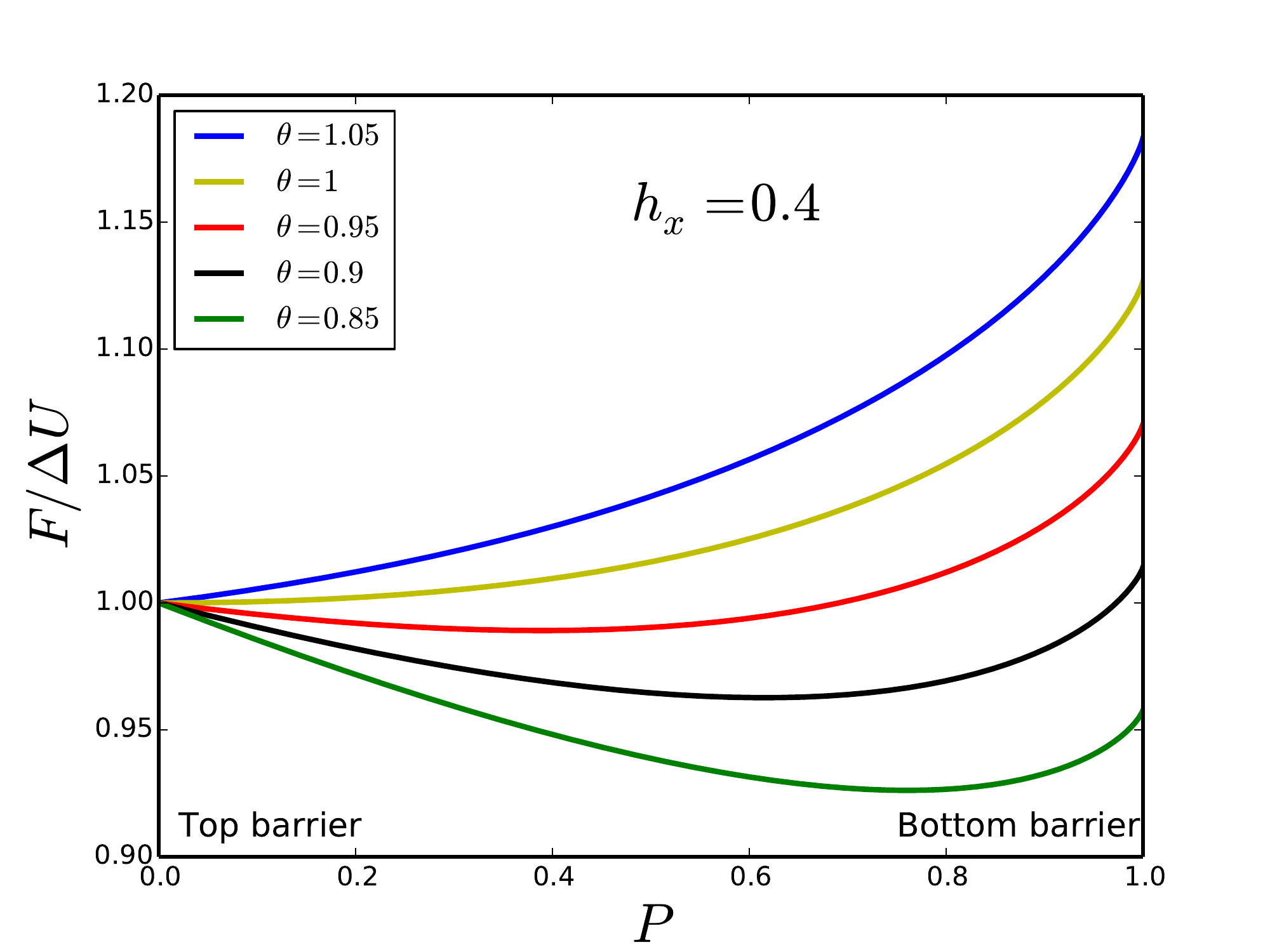}
\includegraphics[width=3.5in]{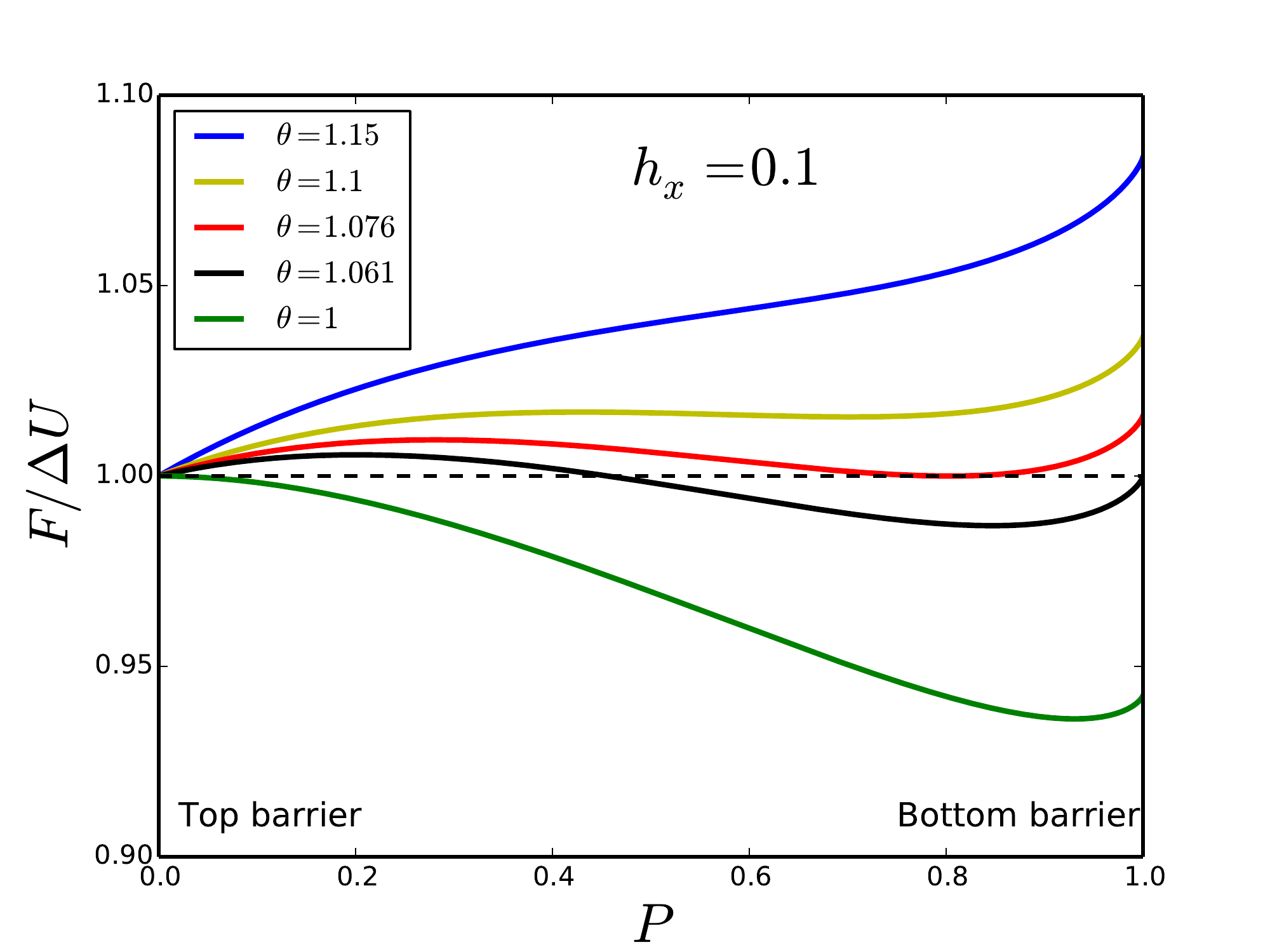}
\caption{ Color online: The plot of the of the free energy as a function of $P$ for $h_x=0.4$ second-order transition and $h_x=0.1$  first-order transition. Reproduced with permission from \onlinecite{chud3}.}
\label{pot8}
\end{figure}
The analogy between these two free energies comes from identifying the coefficient of $P^2$ as the Landau coefficient $b$, which determines the regime of first-order phase transition $b<0$, and that of second-order phase transition $b>0$. The boundary between the first- and  the second-order phase transition corresponds $b=0$. We see that these conditions recover the results in Sec.\eqref{uniham}.
The plot of the free energy in the whole range of energy is shown in Fig.\eqref{pot8}. The actual crossover temperature from thermal to quantum  regimes is determined when two minima of a curve have the free energy. For $h_x=0.1$, it is found to be at $T_0^{(1)}= 1.076T_0^{(2)}$.
This crossover temperature is approximately given by $T_0^{(1)}= \Delta U/S(\mathcal E \to 0)$, which can be obtained easily from Eqn.\eqref{hg} and Eqn.\eqref{chud10}. For the second order transition one finds that at $x_s=x_{\text{max}}=0$
\bea
T_0^{(2)}= \frac{1}{\beta_0^{(2)}}=\frac{D\tilde{s}}{\pi}\sqrt{h_x(1-h_x)}.
\label{unisecond}
\eea
In the limit $h_x\rightarrow 1$ one finds $T_0^{(1)}/T_0^{(2)}\approx 0.833$. The plot of these temperatures are shown in Fig.\eqref{pot7}.  It is crucial to note that the magnetic field plays a decisive role on the crossover temperatures for the first-and the second-order phase transitions. It is the  main parameter that drives tunneling in this system. Physically, the sharp first-order phase transition in this model occurs due to the flatness or wideness of the barrier top in the small magnetic field limit as shown in Fig.\eqref{chud_pont}. In the strong magnetic field limit, the top of the barrier is of the parabolic form (see Fig.\eqref{chud_pont}) which leads to suppression of first-order phase transition.  In the limit of zero magnetic field the Hamiltonian Eq.\eqref{chud2} commutes with the $z$-component of the spin, thus $\hat{S}_z$ is a constant of motion and the potential in Eq.\eqref{chud3} becomes a constant. Hence, there is no dynamics (tunneling) and no quantum-classical phase transitions\footnote{The basic understanding is that in the zero magnetic field limit,  the barrier becomes infinitely thick and tunneling cannot occur.}.
\subsubsection{Experimental results}

\begin{figure}
\centering
\includegraphics[scale=1]{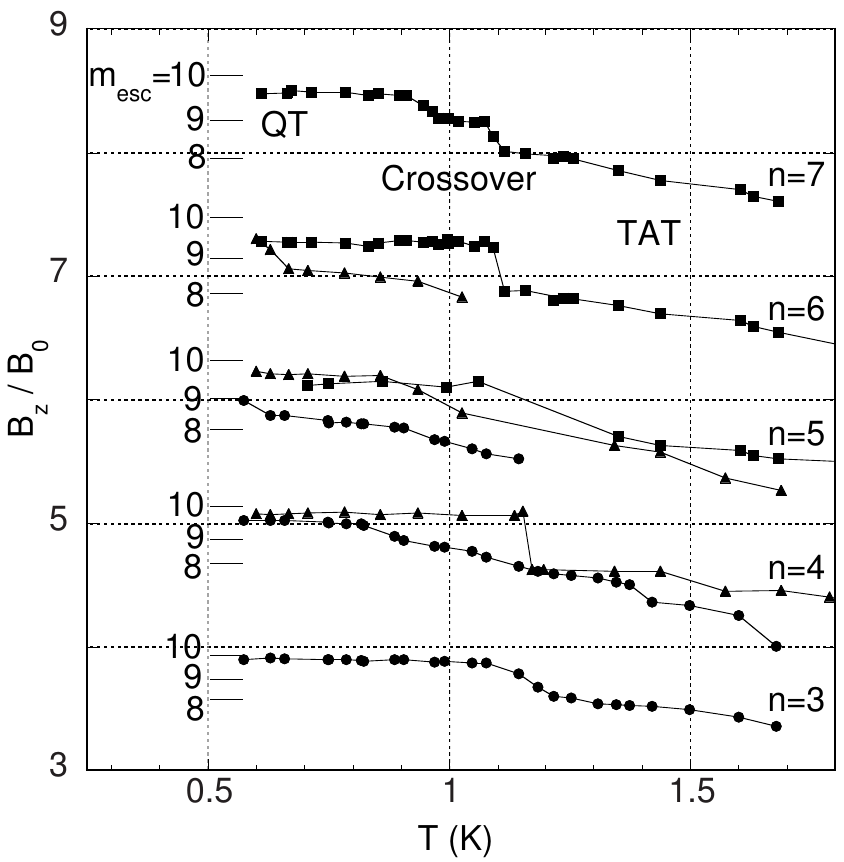}
\caption{ Peak positions of the first derivative of the magnetization plotted against the temperature in Mn$_{12}$Ac molecular magnet. At the ground state $M=s=10$, the peak is independent of temperature (QT) while for excited states $M<s$ transition occurs by TAT. Adapted with permission from \onlinecite{bkm}  }
\label{pot32}
\end{figure}
\begin{figure}
\centering
\includegraphics[scale=0.6]{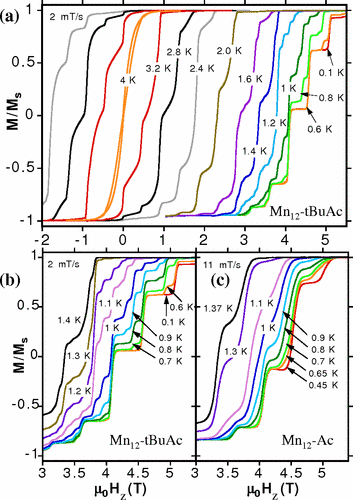}
\caption{ Color online: Temperature dependence of hysteresis loops  of (b) Mn$_{12}$-tBuAc and (c) Mn$_{12}$Ac SMMs at different temperatures and a constant field sweep rate as indicated in the figure. With decreasing temperature  , the hysteresis increases due to a decrease in the transition rate of thermal assisted tunneling. Adapted with permission from \onlinecite{wern5}  }
\label{pot33}
\end{figure}
 Recently, experiments have been conducted to measure these crossover temperatures.  Experimental result for Mn$_{12}$Ac molecular magnet with the model in Eqn.\eqref{wern5} has confirmed the existence of an abrupt and gradual  crossover temperature $(T\sim 1.1 K)$ between thermally assisted and pure quantum tunneling \cite{bkm, ll2000, kent2000, ga1} as shown in Fig.\eqref{pot32}. Below the crossover temperature the magnetization relaxation becomes temperature independent, which indicates that transition occur by QT between the $m_{esc}= \pm s$ states. Above the crossover temperature transition favours the excited states with $m_{esc}< s$  (TAT).  Quite recently, a similar result was observed in  Mn$_{12}$-tBuAc molecular nanomagnet with a spin ground state of $s=10$. This molecular nanomagnet  has the same magnetic anisotropy as Mn$_{12}$Ac but the molecules are very isolated and the crystals have less disorder and a higher symmetry\cite{wern5}. The Hamiltonian for this system has the form:
\bea
\hat H = -D\hat{S}_z^2 - B\hat{S}_z^4-h_z\hat{S}_z  + \hat {H}_{\perp},
\label{wern5}
\eea
where $h_z =g_z\mu_0\mu_BH_z$ and $\hat {H}_{\perp}$ is the splitting term which is comprised of $\hat{S}_x$ and $\hat{S}_y$. In the absence of $\hat {H}_{\perp}$, the  $21$ energy levels of Eqn.\eqref{wern5} can be found by the so-called exact numerical diagonalization in the $\hat{S}_z$ representation. The inclusion of a small perturbation $\hat {H}_{\perp}$ leads to an avoided level crossings in the degenerate energy subspace.  The crossover temperature for the compound occurs at $T\sim 0.6 K$. The hysteresis loops in Fig.\eqref{pot33} show a temperature independent quantum tunneling at the lowest energy levels  below $0.6 K$, while the temperature dependent thermal assisted tunneling at the excited states occurs above $0.6K$.

\subsection{Phase transition in biaxial spin model}
\subsubsection{Model Hamiltonian and spin coherent state path integral}
The phase transition in biaxial spin systems follow a similar trend to that of uniaxial spin model in a magnetic field. The first work on this system was begun by \textcite{mull00}. They studied the model:
\bea \hat H = K_1\hat{S}_z^2+K_2\hat{S}_y^2, \quad K_2/K_1=\lambda <1,
\eea
 by spin coherent state path integral and periodic instanton method. This Hamiltonian is related to that of Eqn.\eqref{bia} and Eqn.\eqref{biau} by $K_1=D_1$, $K_2=D_2$ and $\phi\rightarrow \pi/2-\phi$. It can also be related to any biaxial spin model in the absence of a magnetic field.
The effective Lagrangian of this system can be obtained by  integrating out $\cos\theta$ from the spin coherent state path integral, Eqn.\eqref{pathint1}, one finds that the effective classical Euclidean Lagrangian is
\begin{equation}
L_E =is\dot{\phi} +\frac{1}{2}m(\phi)\dot{\phi}^2+ U(\phi),
\end{equation}
where
\begin{align}
m(\phi)= \frac{1}{2K_1(1-\lambda\sin^2\phi)},\quad U(\phi)= K_2s^2\sin^2\phi.
\label{cohere}
\end{align}
The potential barrier height is located at $\phi_s=\frac{\pi}{2}$, and the minimum energy is located at $\phi=0$.
\subsubsection{Periodic instanton}
The periodic instanton trajectory of this model can be computed from the classical equation of motion:
\bea
m({\phi}_p)\ddot{{\phi}}_p+\frac{1}{2}\frac{dm({\phi}_p)}{d{\phi}_p}\dot{{\phi}}_p^2-\frac{dU}{d{\phi}_p}=0.
\eea
\begin{figure}[ht]
\centering
\includegraphics[width=3.5in]{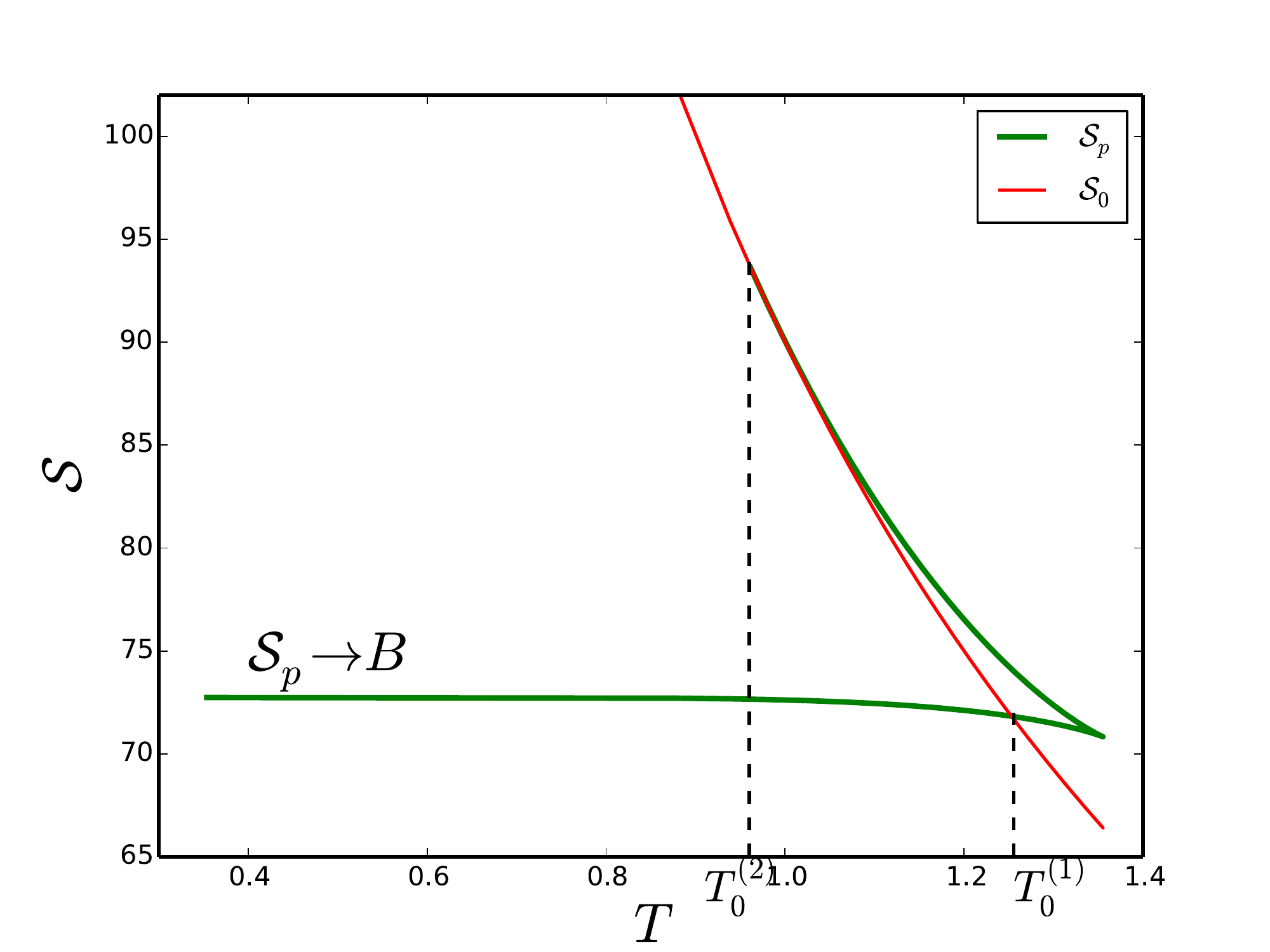}
\includegraphics[width=3.5in]{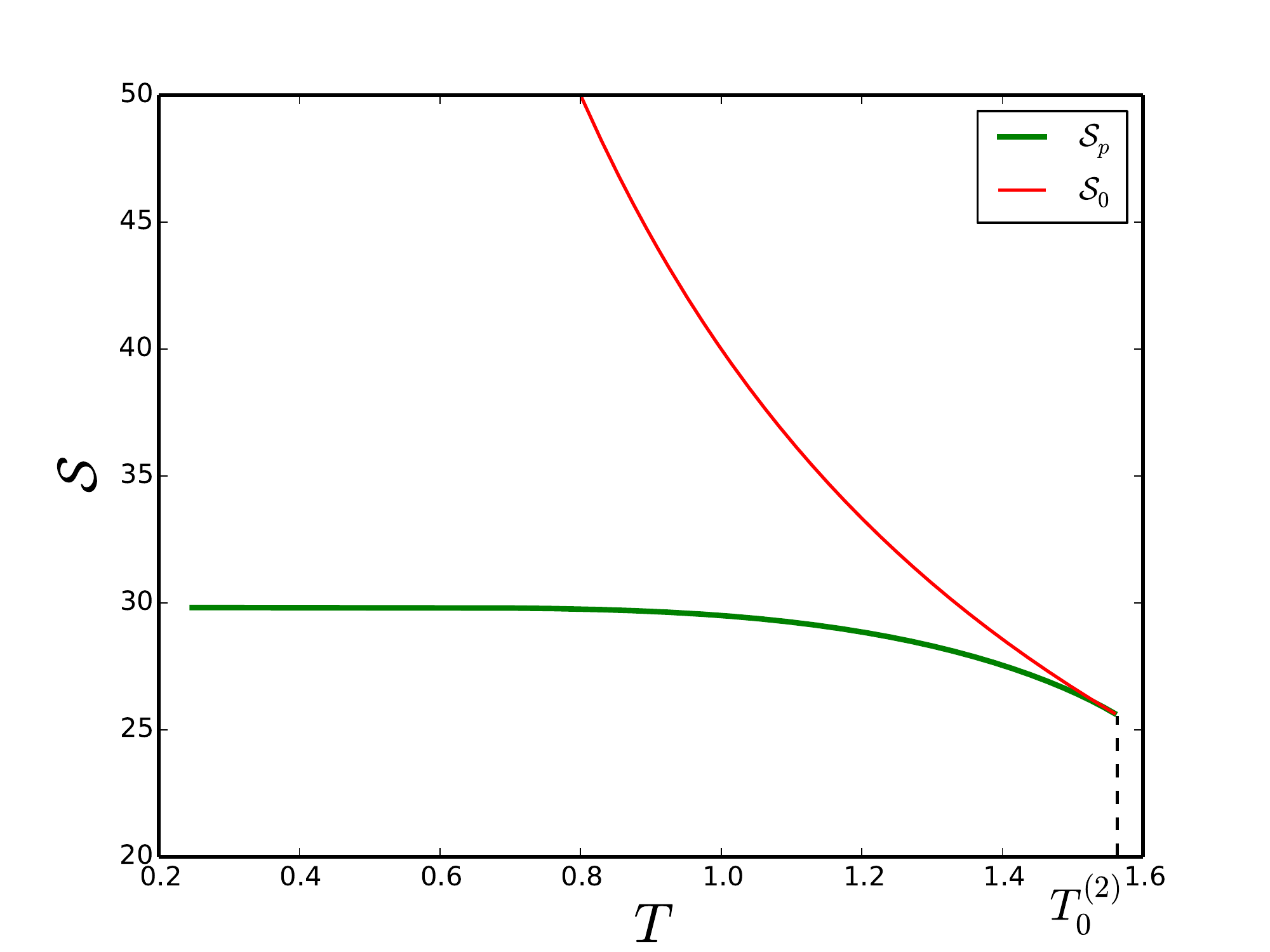}
\caption{ Color online: The plot of the thermon action Eqn.\eqref{liangth} and thermodynamic action Eqn.\eqref{thermonda} against temperature for the biaxial model. Left:  $s=10$, $K_1=1$, $\lambda=0.9$ (first-order transition), where $B$ is the vacuum instanton action. Right: $s=10$, $K_1=1$, $\lambda=0.4$ (second-order transition). Reproduced with permission from \cite{mull00}.}
\label{biathermon}
\end{figure}

Integrating once we obtain:
\bea
\frac{1}{2}m({\phi}_p)\dot{{\phi}}_p^2-U({\phi}_p)=-\mathcal{E},
\eea
where $\mathcal{E}$ is the integration constant.
The corresponding periodic instanton solution of this equation yields 
\begin{equation}
\phi_p = \arcsin\bigg[\frac{1-k^2\sn^2(\omega_p\tau,k)}{1-\lambda^2k^2\sn^2(\omega_p\tau,k)}
\bigg]^{\frac{1}{2}},
\end{equation}
where
\begin{align}
k^2&=\frac{n^2-1}{n^2-\lambda},\quad n^2=\frac{ K_2s^2}{\mathcal{E}}; \quad \omega_p^2=\omega_0^2(1-\frac{\lambda}{n^2});\quad \omega_0^2 = 4K_1K_2s^2 .
\end{align}
 The classical action for this trajectory is found to be
\begin{align}
\mathcal{S}_p&=\frac{4\omega}{\lambda K_1}\mathcal{I}(k^2\lambda,k) +\beta(\mathcal E-U_\text{min}),\label{liangth}\\  
\mathcal{I}(k^2\lambda,k)&=[\mathcal{K}(k)-(1-k^2\lambda)\Pi(k^2\lambda,k)],
\label{thermonbia}
\end{align}
where $\mathcal{K}$ and $\Pi$ are the complete elliptic integrals of the first and the third kinds respectively, and the period of oscillation is given by
\bea
\beta(\mathcal{E})=\frac{2}{\sqrt{K_1}\sqrt{K_2s^2-\mathcal{E}\lambda}}\mathcal{K}(k).
\eea
 Near the top of the barrier $\mathcal{E}\rightarrow U_\text{max}= K_2 s^2$ and $\beta(\mathcal{E}) \rightarrow \beta_0^{(2)}= 2\pi/\omega_b$, with $\omega_b$ defined by Eqn.\eqref{omega} and near the bottom of the barrier $\mathcal{E}\rightarrow U_\text{min}=0$ and $\beta(\mathcal{E}) \rightarrow \infty$, then $\mathcal{S}_p$ reduces to the usual vacuum $(T=0)$ instanton solution, Eqn.\eqref{act4}. Fig.\eqref{biathermon} shows the thermon and the thermodynamic actions with the crossover temperatures indicated, which is of similar trend to that of Fig.\eqref{unithermon} .
 \subsubsection{Free energy and crossover temperatures}
The ground state crossover temperature is determined from $T_0^{(1)}= \Delta U/\mathcal{S}_p(\mathcal E\to U_\text{min}) = K_2s^2/\mathcal{S}_p(\mathcal E\to U_\text{min})$. For the second order phase transition we have $T_0^{(2)}= \omega_b/2\pi$. In the limit $\lambda\rightarrow 0$, one finds that $T_0^{(1)}/T_0^{(2)}=\pi/4\approx 0.785$. The exact  free energy follows from Eqn.\eqref{liangth}:
\begin{align}
\frac{F}{\Delta U}=1-P+\frac{4\theta\sqrt{(1-\lambda)[1-\lambda(1-P)]}}{\pi\lambda}\mathcal{I}(k^2\lambda,k),
\label{liangfre}
\end{align}
where $\theta=T/T_0^{(2)}$, and $T_0^{(2)}=sK_1\sqrt{\lambda(1-\lambda)}/\pi$. Near the top of the barrier $P\ll1$, the free energy reduces to \cite{zha3}
\begin{align}
&F/\Delta U \cong
 1 + \left(\theta-1\right)P +\frac{\theta}{4\left(1-\lambda\right)}\left(\frac{1}{2}-\lambda\right)P^2 
+ \frac{\theta}{8\left(1-\lambda\right)^2}\left(\lambda^2-\lambda+\frac{3}{8} \right)P^3 + O(P^4).
\label{biafre}
\end{align}
\begin{figure}[ht]
\centering
\includegraphics[width=3.5in]{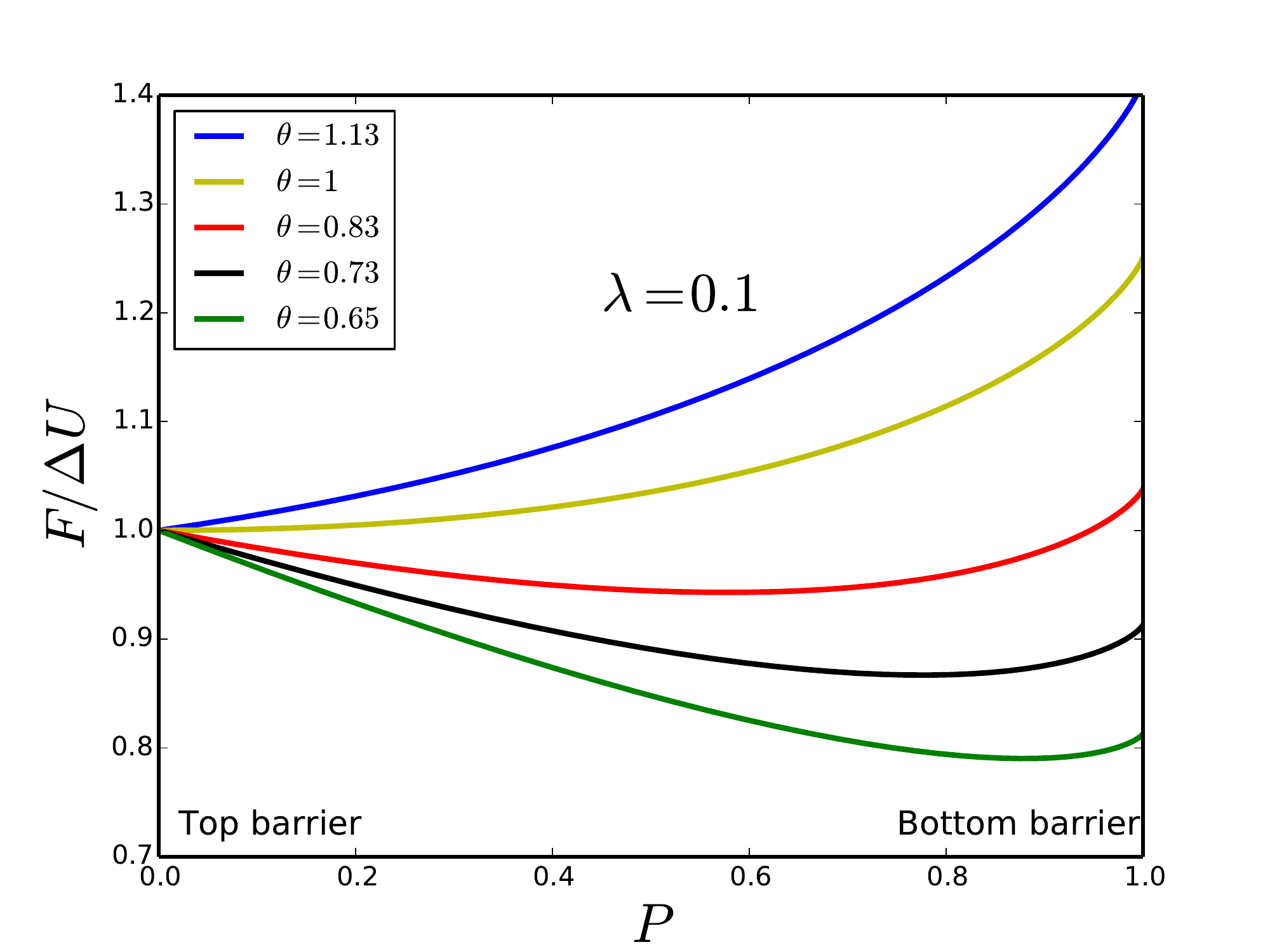}
\includegraphics[width=3.5in]{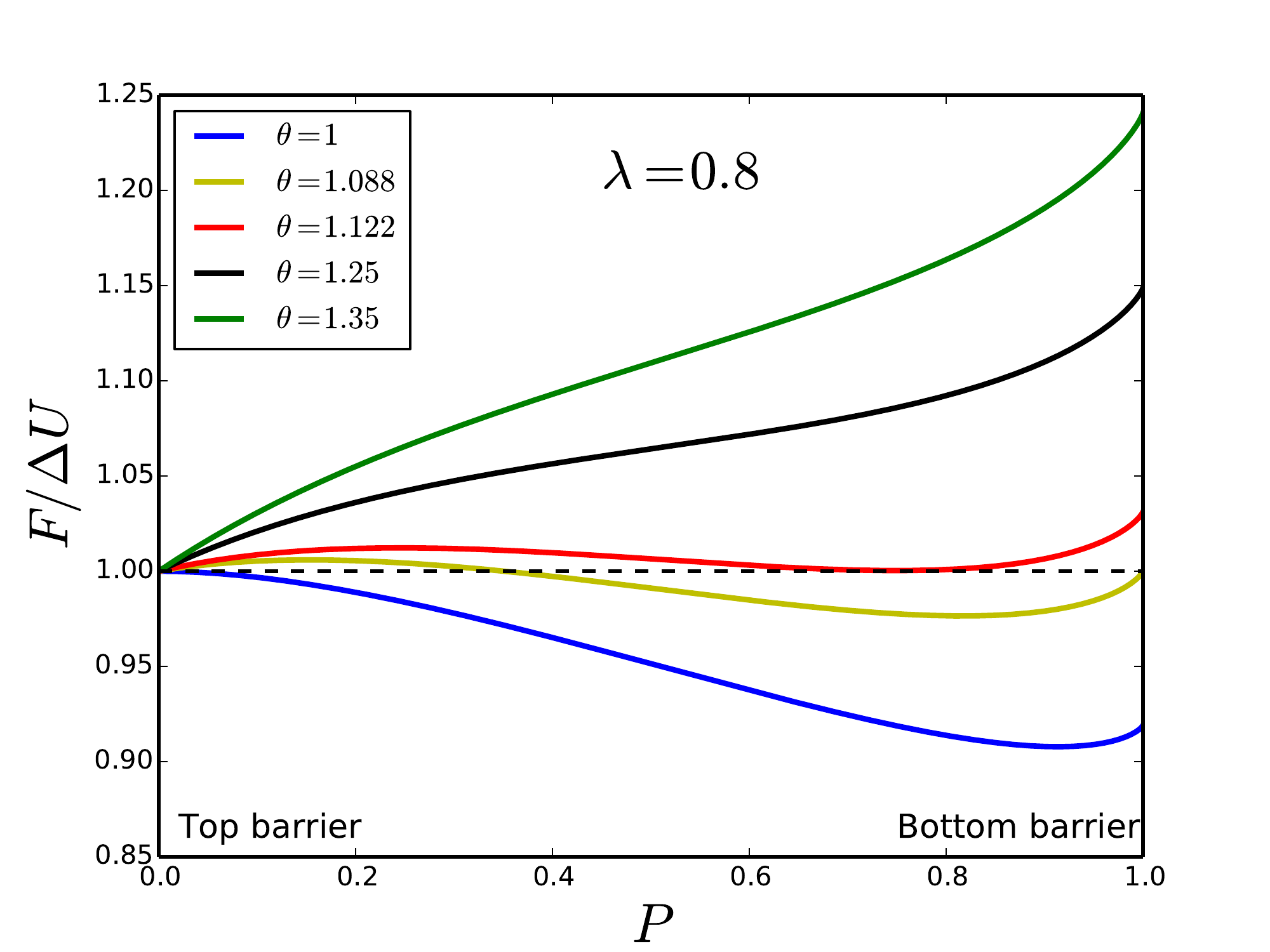}
\caption{ Color online. The effective free energy of the escape rate vs P. Left: for $\lambda=0.1$, second-order transition. Right: for $\lambda=0.8$, first-order transition. }
\label{pot11}
\end{figure}

The coefficient of the $P^2$ changes sign when $\lambda>\frac{1}{2}$, which corresponds to the regime of first-order phase transition. In this analysis, the mass is coordinate dependent, therefore the coefficient of $\phi^4$ in the series expansion near $\phi_s=\frac{\pi}{2}$ cannot determine the condition for any type of quantum-classical phase transitions, thus Eqn.\eqref{mull} becomes indispensable. Using Eqn.\eqref{mull} with $x_s=\phi_{\text{max}}=\frac{\pi}{2}$ one obtains \cite{mull4}
\bea
\mathscr C= K_2s^2\frac{1-2\lambda}{1-\lambda}, 
\eea
where $\mathscr C$ is equivalent to the coefficient of the $P^2$ in Eq.\eqref{biafre}.  It is evident that $\mathscr C <0$ for $\lambda>\frac{1}{2}$, corresponding to the regime of first-order phase transition. At the phase boundary $\mathscr C=0$, which yields the critical value $\lambda_c=\frac{1}{2}$.
The plot of Eqn.\eqref{liangfre} in the whole range of energy is shown in Fig.\eqref{pot11}. For $\lambda=0.8$, that is first-order transition, the actual crossover from thermal to quantum regime is estimated as $T_0^{(1)}= 1.122T_0^{(2)}$ corresponding to the point where two minima have the same free energy. At $\lambda=0.1$, there is only one minimum of $T$ for all $T>T_0^{(2)}$, i.e $\theta>1$ at the top of the barrier. For $\theta<1$, the minimum continuously shifts to the bottom of the barrier with lowering temperatures. This corresponds to a second-order phase transition. The ratio of the anisotropy constants $\lambda$ in this model plays a similar role as $h_x$ in the uniaxial model. In general, the splitting term in the Hamiltonian is responsible for the dynamics of the system, and leads to the phase transition in large spin systems. However, the sharp first-order phase transition in this model is not as result of the flatness of the barrier top as in the case of the uniaxial model.  It should be noted that the mass of the particle at the top of the barrier $m\lb\frac{\pi}{2}\rb=[2K_1(1-\lambda)]^{-1}$ is heavier than that of the bottom of the barrier $m(0)=[2K_1]^{-1}$. The former is responsible for the sharp first-order crossover.  
\begin{figure}
\centering
\includegraphics[width=3.5in]{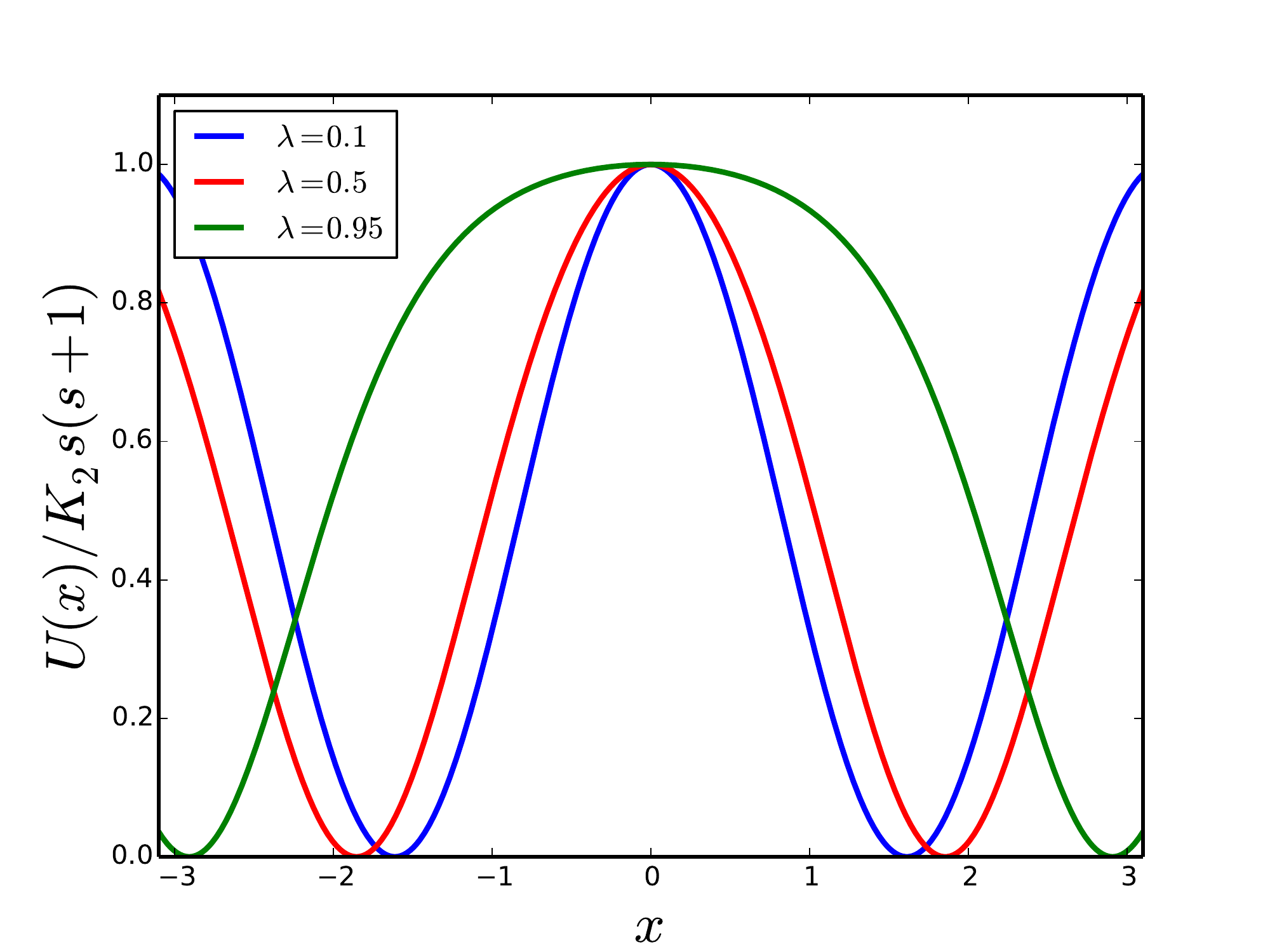}
\caption{Color online: The plot of the potential in Eq.\eqref{effmull} for several values of $\lambda$.}
\label{mull_pont}
\end{figure}
A constant mass in this model can be achieved through the effective potential method. The spin Hamiltonian  corresponds to the effective potential and the mass \cite{mull3}:
\begin{align}
U(x)= K_2s(s+1)\frac{\cn^2(x,\lambda)}{\dn^2(x,\lambda)};\quad m=\frac{1}{2K_1};
\label{effmull}
\end{align}
where $x_s=x_{\text{max}}=0$. It is now trivial  to check that Eq.\eqref{effmull} yields the same result in the large spin limit $s(s+1)\sim s^2$. Since the mass is now a constant,  the flatness of the barrier as $\lambda\to 1$ leads to a sharp first-order phase transition (see Fig.\eqref{mull_pont}).

\subsection{Phase transition in easy $z$-axis biaxial spin model with a magnetic field}
\subsubsection{Introduction}

In the presence of a magnetic field, other interesting features arise. In this case one can study how the crossover temperatures vary with the magnetic field at the phase boundary. 
\begin{figure}[ht]
\centering
\includegraphics[width=3.5in]{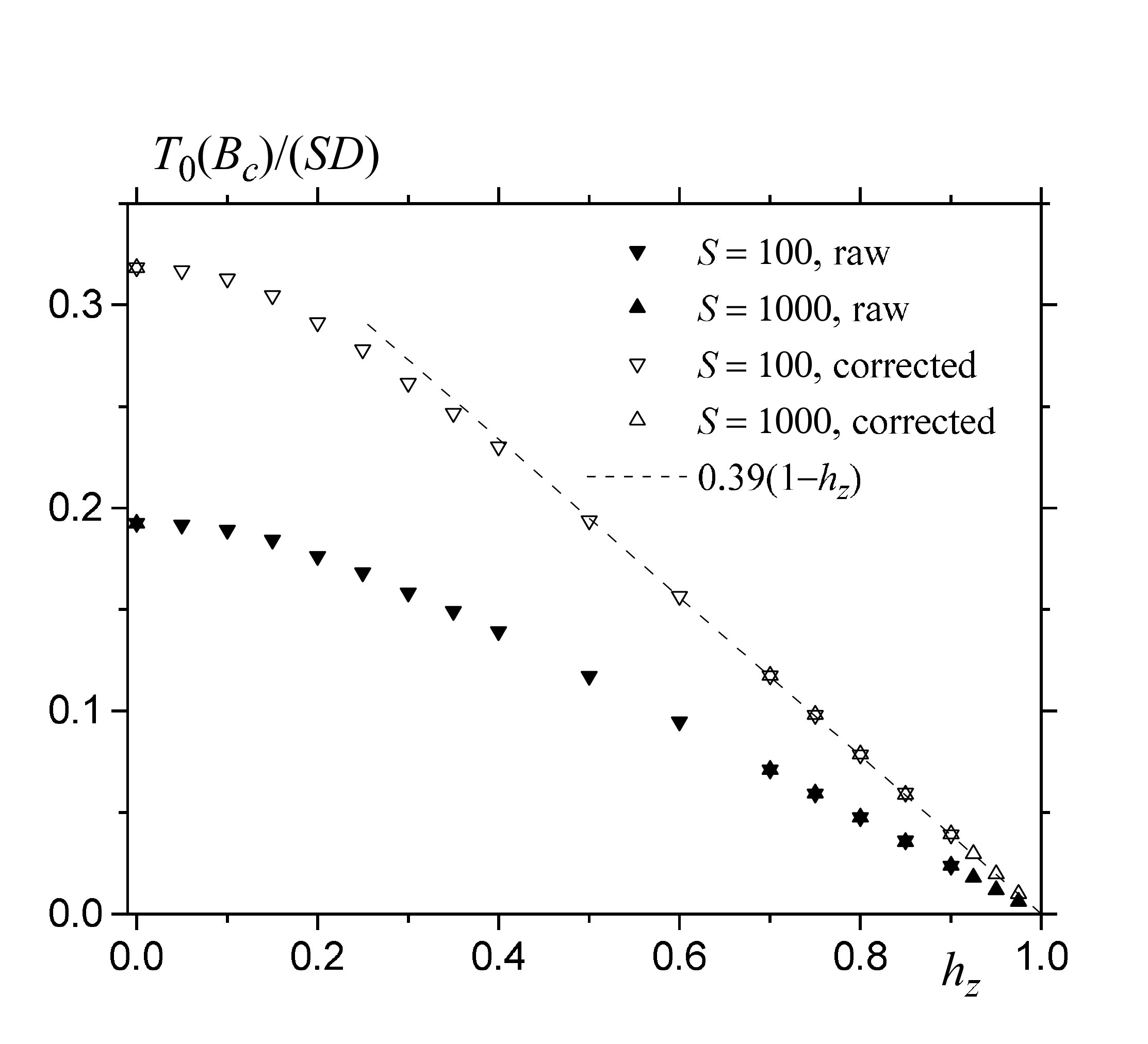}
\includegraphics[width=3.5in]{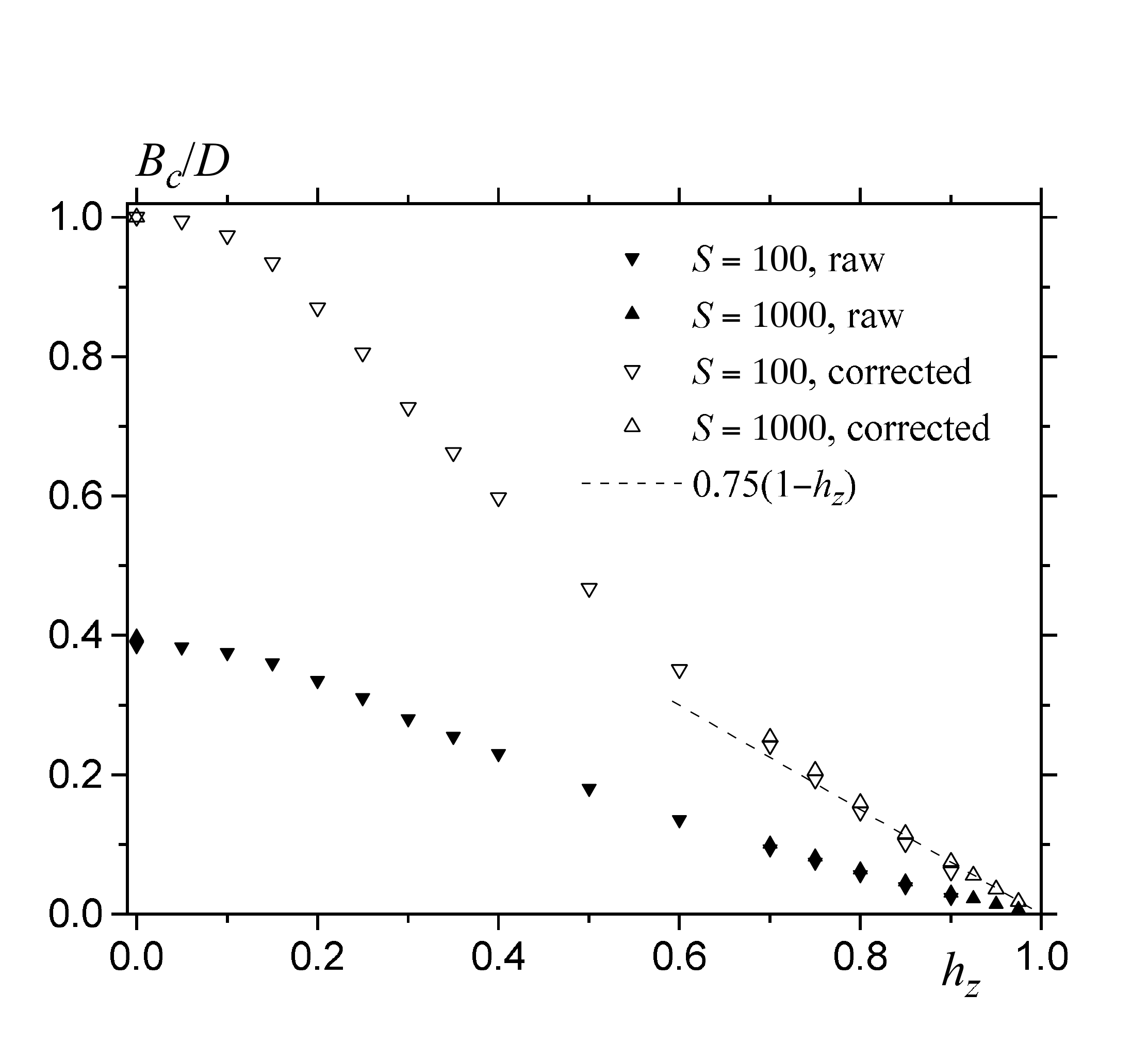}
\caption{ Left: The crossover temperature $T_0^{(c)}$ at the phase boundary between first- and second-order transitions for the model  $\hat H =-D\hat{S}_z^2+B\hat{S}_x^2-H_z\hat{S}_z$ . Right: The phase boundary between the first- and the second-order phase transitions for the same model using perturbation theory, where $h_z= H_z/2Ds$. Adapted with permission from \onlinecite{chud4} }
\label{pot14}
\end{figure}
Many biaxial spin models in the presence of  an external magnetic with different easy axes  directions  have been studied by different approaches, although these systems are related by their anisotropy constants. The early work on these models began with \textcite{mull2}. They studied a Hamiltonian of the form:
\bea
\hat H = K(\hat{S}_z^2+\lambda\hat{S}_y^2)-2\mu_B h_y \hat{S}_y, \quad \lambda <1,
\eea
by spin coherent state path integral formalism. This Hamiltonian possesses an easy $x$-axis, $y$-medium axis and $z$ hard axis. They explicitly demonstrated numerically the influence of the magnetic field on the crossover temperatures, and the period of oscillation. This analysis is however valid only in the regime $\lambda<1$. In the case of a field parallel  to the $z$-axis, their approach will  break down, since such magnetic field pushes the spin away from the $x$-$y$ plane. 
\textcite{chud4, chud41} have studied two biaxial spin  models of the form
\bea
 \hat H =-D\hat{S}_z^2+B\hat{S}_x^2-H_x\hat{S}_x,
 \label{ch1}
 \eea by direct numerical method and \bea 
 \hat H =-D\hat{S}_z^2+B\hat{S}_x^2-H_z\hat{S}_z,
 \label{ch2}
 \eea 
by perturbation theory with respect to $B$. However, the perturbative approach is less justified in the large $B$ limit. These two models have an easy $z$ axis, $x$ hard axis and $y$ medium axis with the magnetic field applied along the hard and easy axes respectively. 
The first model, i.e Eqn.\eqref{ch1} is related to the model in Eqn.\eqref{biam} sec\eqref{solo} by rescaling the anisotropy constants\footnotemark[4]. It is realized in Fe$_8$ molecular cluster with $s=10$, $D=0.229 K$ and $B=0.092 K$. The second model, Eqn.\eqref{ch2} has a magnetic field along the easy axis which creates a bias potential minima. For this model the effect of the external magnetic field on the crossover temperatures was explicitly demonstrated by perturbation theory. In Fig.\eqref{pot14} we show the phase boundary and its crossover temperature $T_0^{(c)}$ obtained via perturbation theory for the model in Eqn.\eqref{ch2}. 
\subsubsection{Effective potential method}
Based on the results obtained from perturbation theory and numerical methods, \textcite{kim1} considered the effective potential method of the model:
\begin{align}
\hat H &=-K_{\parallel}\hat{S}_z^2+K_{\perp}\hat{S}_y^2-H_x\hat{S}_x -H_z\hat{S}_z.
\label{kimh1}
\end{align}
 For  $H_x=0$, this model is exactly the model in Eqn.\eqref{ch2}, and for $H_z=0$ the magnetic field $H_x$ is along the medium axis. It is related to Eqn.\eqref{ch1} only by the rotation of axis $\hat{S}_y\rightarrow \hat{S}_x$.  If one introduces the spin wave function in Eqn.\eqref{u3}, then using the generating function
\bea
\Phi(x)=\sum_{m=-s}^{s}\frac{c_me^{mx}}{\sqrt{(s-m)!(s+m)!}},
\eea
and the particle wave function function \begin{align}
\Psi(x)=e^{-y(x)}\Phi(x),\end{align}
where $y(x)$ is determined in the usual way\footnotemark[10]. The corresponding effective potential and the coordinate dependent mass  are  given by

\begin{align}
u(x)&=\frac{1}{1+k\cosh 2x}[(1-k)(h_x\sinh x-h_z)^2-2h_x(1+k)\cosh x-2h_zk\sinh 2x+k(1-\cosh 2x) ]\label{eq1},\\ m(x)&=\frac{1}{K_{\parallel}(2+k_t)(1+k\cosh 2x)},
\label{eeqq1}
\end{align}
where $U(x)=\tilde{s}^2K_{\parallel}u(x)$, $k=k_t/(2+k_t)$ and $k_t= K_{\perp}/K_{\parallel}$, $h_{x,z}= H_{x,z}/2K_{\parallel}\tilde{s} $. The large $s$ limit, {\it i.e.,} $s\gg1$ has been used, thus terms independent of $s$ drop out in the potential. 
\begin{figure}[ht]
\centering
\includegraphics[width=3.5in]{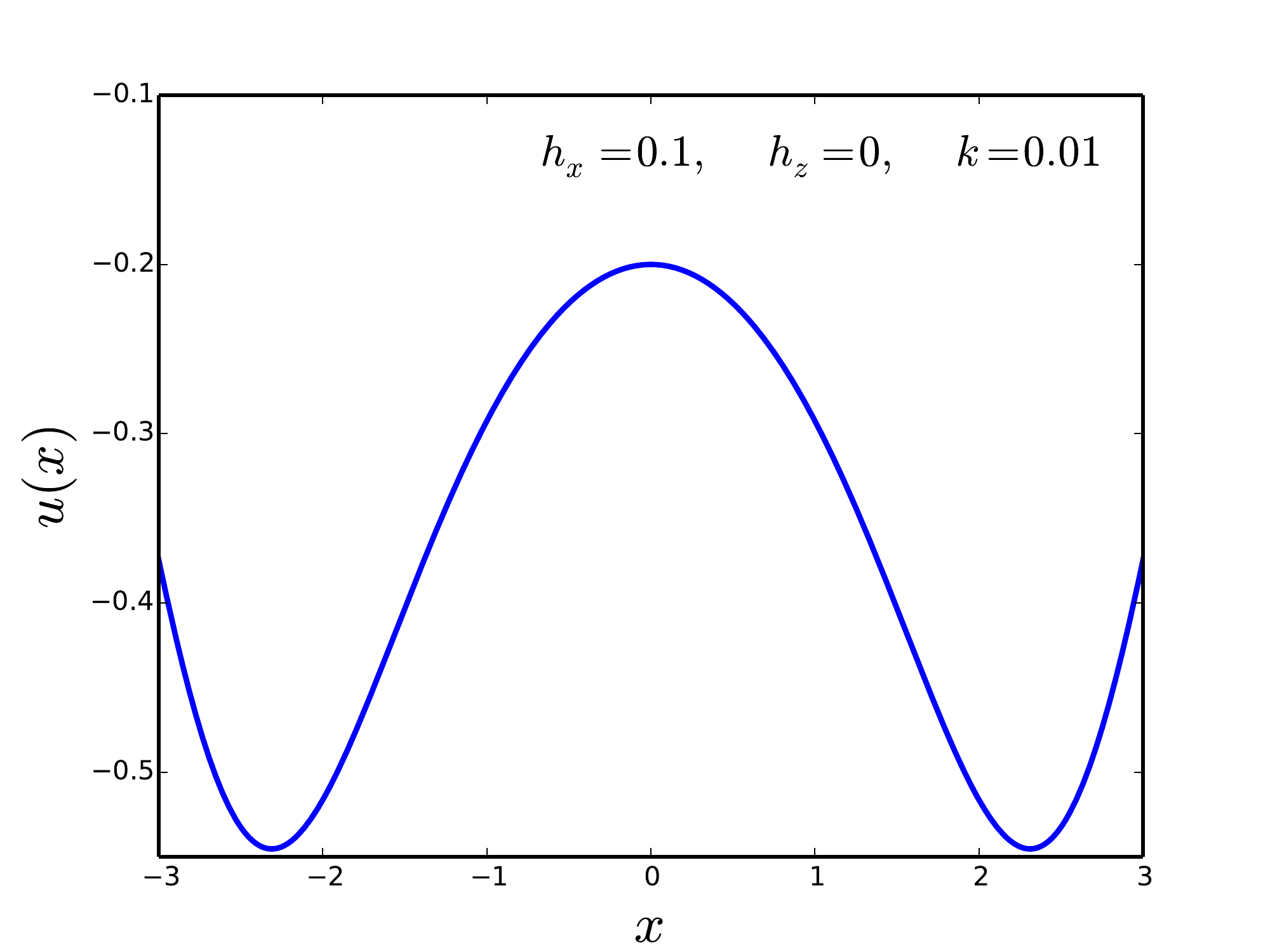}
\includegraphics[width=3.5in]{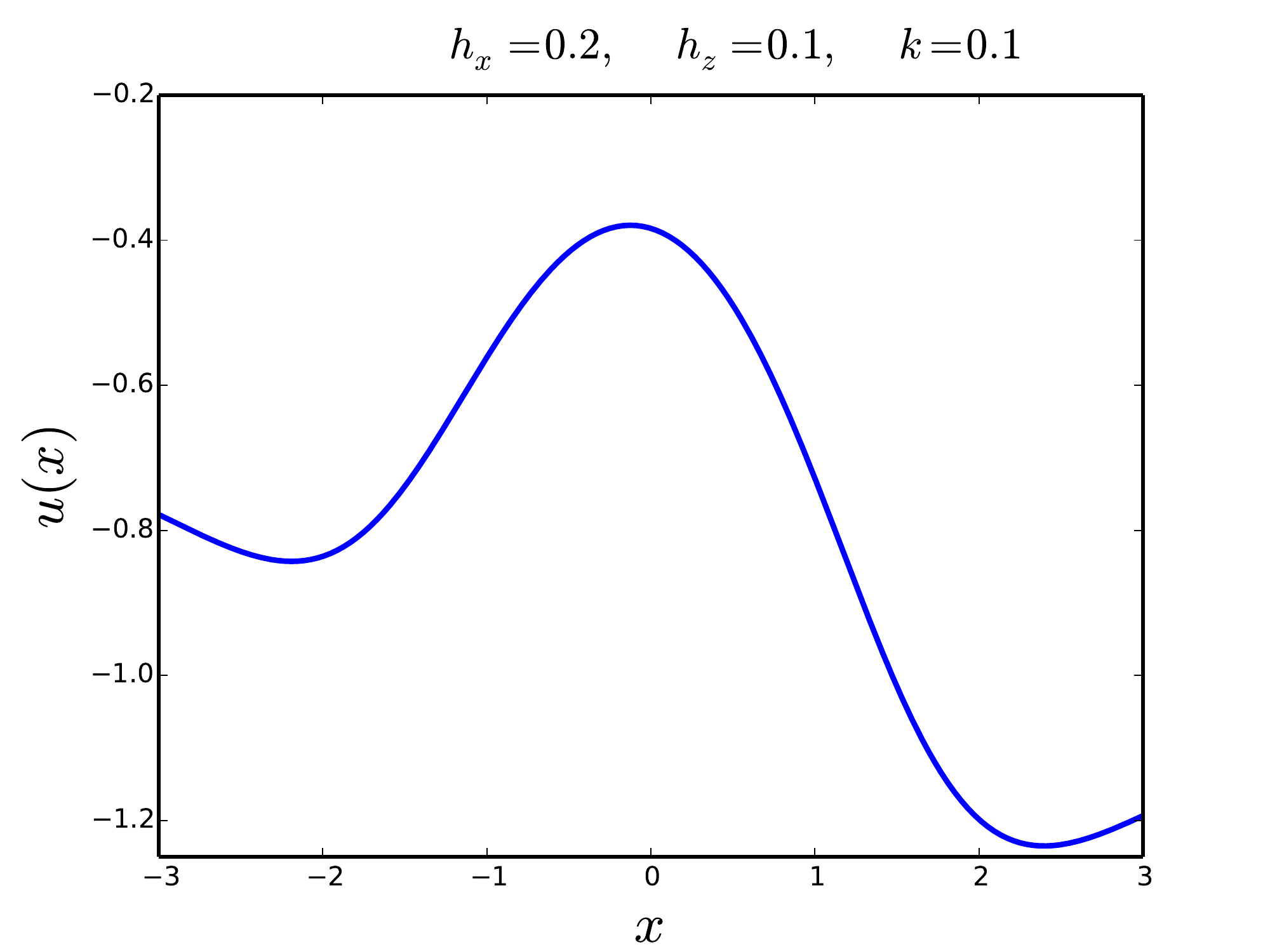}
\caption{ Color online: The plot of the effective potential in Eqn.\eqref{eq1} with different parameters.}
\label{pot15}
\end{figure}
\subsubsection{Phase boundary and crossover temperatures}
\label{kimphase}
We consider two cases $h_z=0$, $h_x\neq0$, and $h_x=0$, $h_z\neq0$. Let us consider the first case $h_z=0$ and $h_x\neq0$, in this case the potential reduces to
\begin{align}
u(x)&=\frac{(1+k)(h_x\cosh x-1)^2}{1+k\cosh 2x},
\label{kimbia}
\end{align}

where a constant of the form $(1+h_x^2)$ has been added to normalize the potential to zero at the minimum  $x_{\text{min}}=\pm\arccosh\lb\frac{1}{h_x}\rb$ . This  potential is now an even function with  a maximum  at $x_s=x_{\text{max}}=0$ . The barrier height is $\Delta U = K_{\parallel}\tilde{s}^2(1-h_x)^2$ (see Fig.\eqref{pot15}).  In the limit $k_t\rightarrow 0$, Eqn.\eqref{eeqq1} and \eqref{kimbia} reduce to that of the uniaxial model studied in Sec.\eqref{phaseuni}. \begin{figure}[ht]
\centering
\includegraphics[width=3.5in]{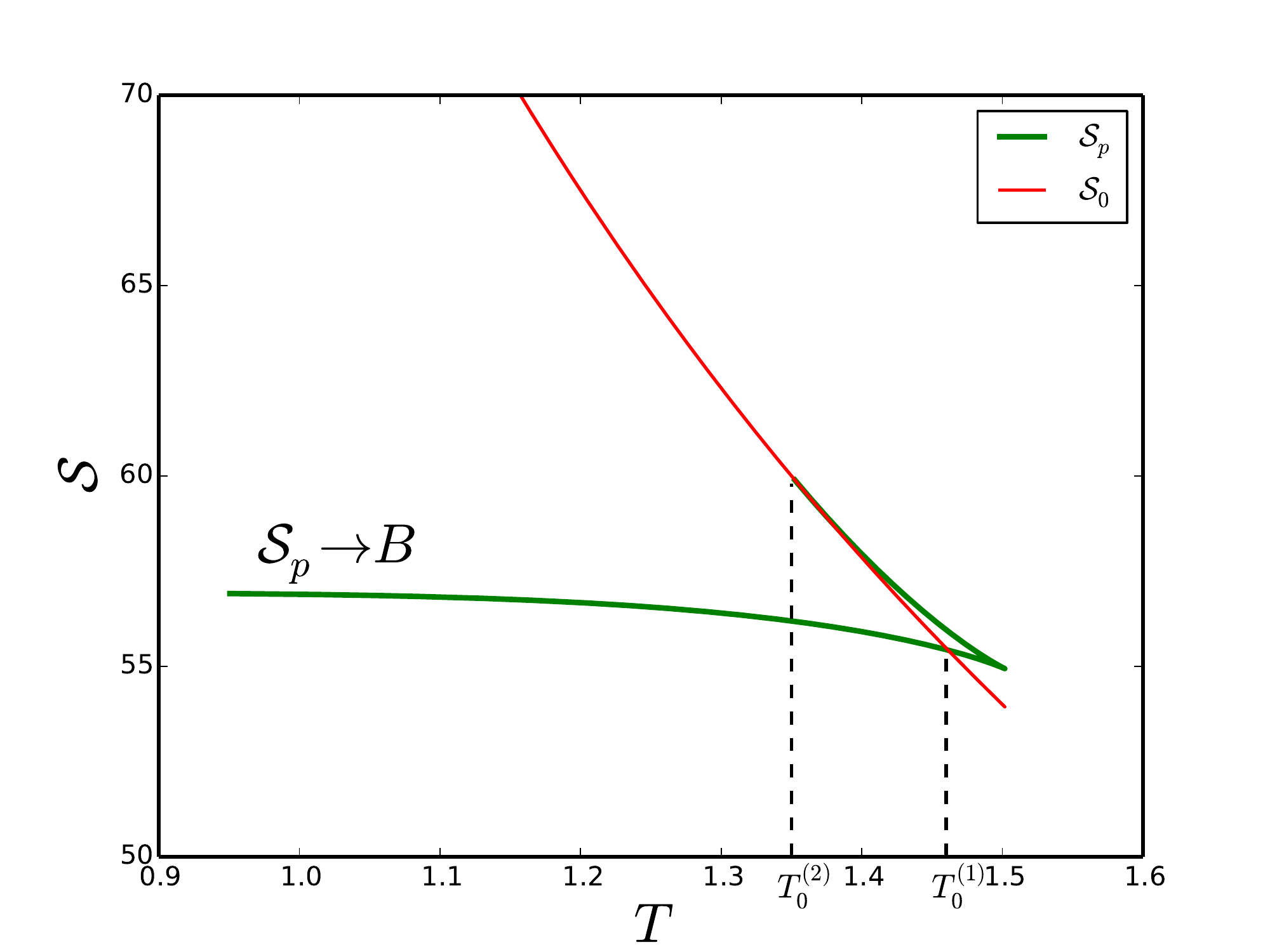}
\includegraphics[width=3.5in]{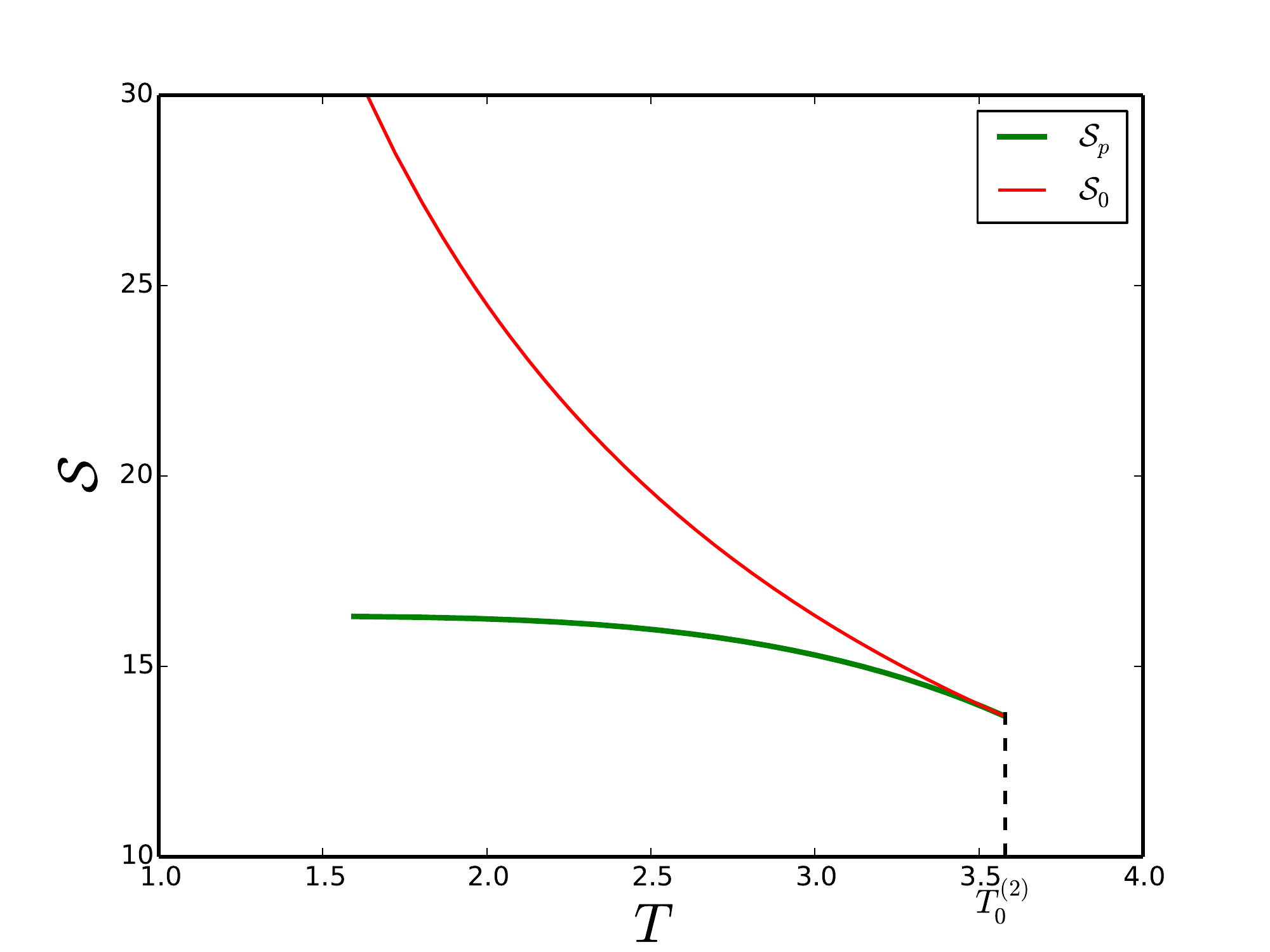}
\caption{ Color online: The plot of the thermon action Eqn.\eqref{kim_bia1} and thermodynamic action Eqn.\eqref{thermonda} against temperature for the biaxial model with magnetic field $h_x$. Left:  $s=10$, $K_{||}=1$, $k_t=0.1$, $h_x=0.1$ (first-order transition), where $B$ is the vacuum instanton action. Right: $s=10$, $K_{||}=1$, $k_t=1.5$, $h_x=0.3$ (second-order transition). }
\label{kim_bia}
\end{figure}
The thermon action is given by
\begin{align}
\mathcal{S}_p&= 2\tilde{s}\sqrt{\frac{2}{2+k_t}}\int_{-x_1}^{x_1}dx\frac{\sqrt{c_1\cosh^2x -c_2\cosh x +c_3}}{1+k-2k\cosh^2x}+\beta\Delta U(1-P),
\label{kim_bia1}
\end{align}
where
\begin{align}
c_1&= (1+k)h_x^2-2k(1-h_x)(1-P);\quad c_2=2h_x(1+k);\quad
c_3=1+k-(1-k)(1-h_x)^2(1-P).
\end{align}
The turning points $\pm x_{1}$ are found by setting the term in the square root to zero. Integrating the classical equation of motion one finds that the denominator in Eqn.\eqref{kimbia} cancels the mass in Eqn.\eqref{eeqq1}. Thus, this action Eqn.\eqref{kim_bia1} corresponds to the action of the periodic instanton trajectory in Eqn.\eqref{uniperiod} but with $\tilde{\mathcal E}=\mathcal E/K_{||}\tilde{s}^2(1+k)$ and $\omega^2=(K_{||}\tilde{s})^2(1+k)(1+k_t)(1-(h_x-\sqrt{\tilde{\mathcal E}}))^2/2 $.  Mathematically, the integral in Eqn.\eqref{kim_bia1} can be evaluated exactly in terms of complete elliptic integrals\cite{abra,by}. At the bottom of the barrier $P\to 1$, the exact vacuum instanton action is given by \cite{kim1}
\begin{align}
S(U_{\text{min}})&= 2 B= 2\tilde{s}\bigg[ \ln\lb\frac{\sqrt{1+k_t}+\sqrt{(1-h_x^2)}}{\sqrt{1+k_t}-\sqrt{(1-h_x^2)}}\rb-\frac{2h_x}{\sqrt{k_t}}\arctan\lb \frac{\sqrt{k_t(1-h_x^2)}}{(1+k_t)h_x^2}\rb\bigg].
\label{kimm_act}
\end{align}
This expression is consistent with the small barrier action in Eqn.\eqref{eq4}, if one uses the relation of the anisotropy constants. In Fig.\eqref{kim_bia} we have shown the plot of this action Eqn.\eqref{kim_bia1} and the thermodynamic action Eqn.\eqref{thermonda} against temperature. We notice that the plot of these two actions has a similar trend to every other model, which indicates the presence of the quantum-classical phase transitions in each of the models. The free energy can also be obtained from Eqn.\eqref{kim_bia1}.  \textcite{kim1} expanded this integral in Eqn.\eqref{kim_bia1} around $x_s=x_{\text{max}}$ in terms of $P$ for a general potential and coordinate dependent mass:
\begin{align}
\mathcal{S}_p= \pi\sqrt{\frac{2m(x_s)}{U^{\prime\prime}(x_s)}}\Delta U[P + b P^2+ O(P^3)]+\beta\Delta U(1-P),
\end{align}
where
\begin{align}
b&= \frac{\Delta U}{16U^{\prime\prime}}\bigg[\frac{12U^{\prime\prime\prime\prime} U^{\prime\prime}  + 15(U^{\prime\prime\prime })^2 }{2(U^{\prime\prime })^2}+3\lb \frac{m^{\prime}}{m}\rb \lb \frac{U^{\prime\prime\prime}}{U^{\prime\prime}}\rb  +\lb \frac{m^{\prime\prime}}{m}\rb-\frac{1}{2}\lb \frac{m^{\prime}}{m}\rb^2\bigg]_{x=x_s},
\label{eq2}
\end{align}
$U^{\prime\prime}(x_s)=- K_{\parallel}\tilde{s}^2u^{\prime\prime}(x_s)/2!$, $U^{\prime\prime\prime}(x_s)=  K_{\parallel}\tilde{s}^2u^{\prime\prime\prime}(x_s)/3!$, and $U^{\prime\prime\prime\prime}(x_s)= K_{\parallel}\tilde{s}^2u^{\prime\prime\prime\prime}(x_s)/4!$.

In analogy with Landau's theory of phase transition,  $b$ corresponds to the coefficient of $\psi^4$ in the Landau free energy expression Eqn.\eqref{land}, and $b<0$ determines the condition for first-order transition while $b=0$ determines the boundary between the first- and the second-order transition. One can check that  this condition, Eqn.\eqref{eq2} yields exactly the same result as Eqn.\eqref{mull}. Applying the criterion, Eqn.\eqref{eq2} one obtains the phase boundary between first- and second-order:
\begin{align}
h_x^{\pm}= \frac{1-14k_t+k_t^2\pm (1+k_t)\sqrt{1+34k_t+k_t^2}}{8(1-k_t)}.
\label{phab}
\end{align}
 At $k_t=0$, corresponding to the uniaxial limit, the phase boundary reduces to $h_x^c=h_x^+=1/4$, which is exactly the result obtained before. At $h_x=0$, one obtains $k_t=1$ which also corresponds to the result of the biaxial model without magnetic field, since the anisotropy constants are related by $K_{\parallel}=D_2$ and $K_{\perp}=D_1-D_2$. In the small anisotropy limit $k_t\ll1$, the phase boundary behaves linearly as $h_x^+\approx (1+3k_t)/4$.  \begin{figure}[ht]
\centering
\includegraphics[width=3.5in]{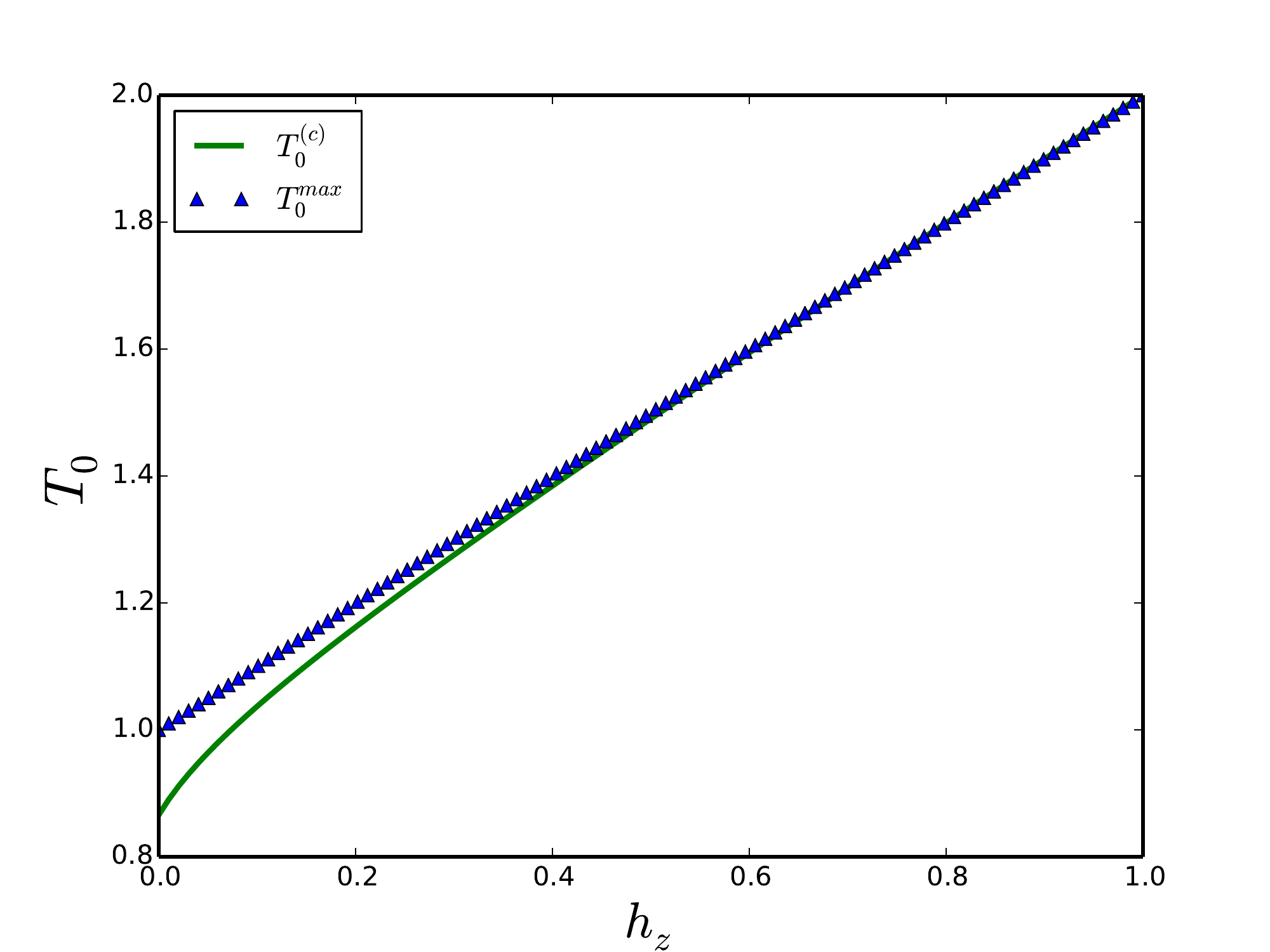}
\caption{ Color online: The second-order crossover temperature at the phase boundary and its maximum at $h_z=0$. $T_0 = 2\pi T_0/\tilde{s}K_{\parallel}$. Reproduced with permission from \onlinecite{kim1}}
\label{pot17}
\end{figure}
The approximate form for the first-order crossover temperature is estimated as $T_0^{(1)}=\Delta U/S(U_{\text{min}})$. Notice that this model does not have a large barrier so the concept of large and small barriers does not apply here. In Sec.\eqref{larb} we will present a complete phase diagram for small and large barriers for the model in Eqn.\eqref{eze10}. For the case of second-order transition the crossover temperature and its maximum are  given by
\begin{align}
T_0^{(2)}&=\frac{\tilde{s}K_{\parallel}}{\pi}\sqrt{(k_t+h_x)(1-h_x)};\quad
T_0^{\text{max}}=\frac{\tilde{s}K_{\parallel}}{2\pi}(1+k_t).
\end{align}
\begin{figure}[ht]
\centering
\includegraphics[width=3.5in]{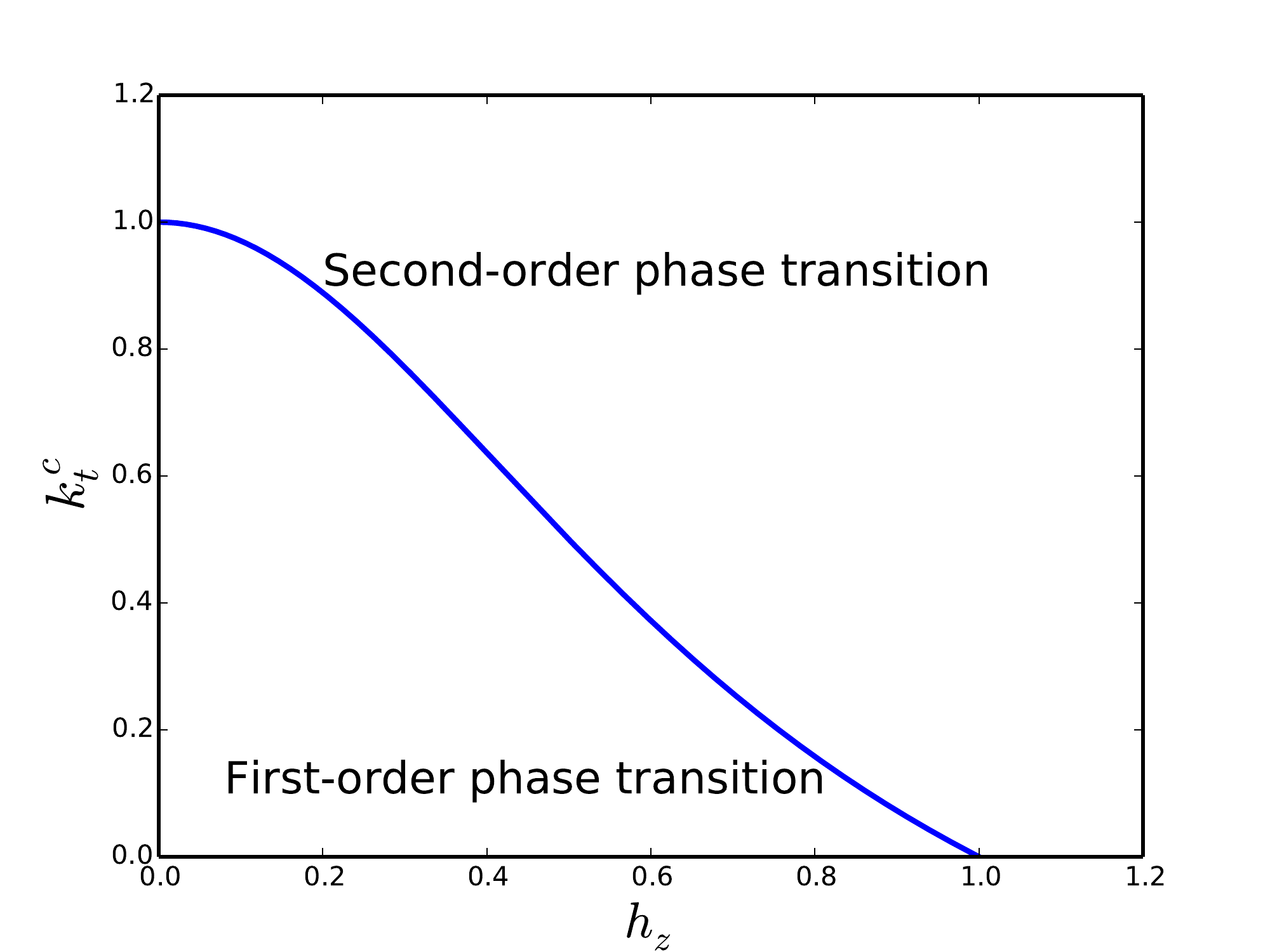}
\includegraphics[width=3.5in]{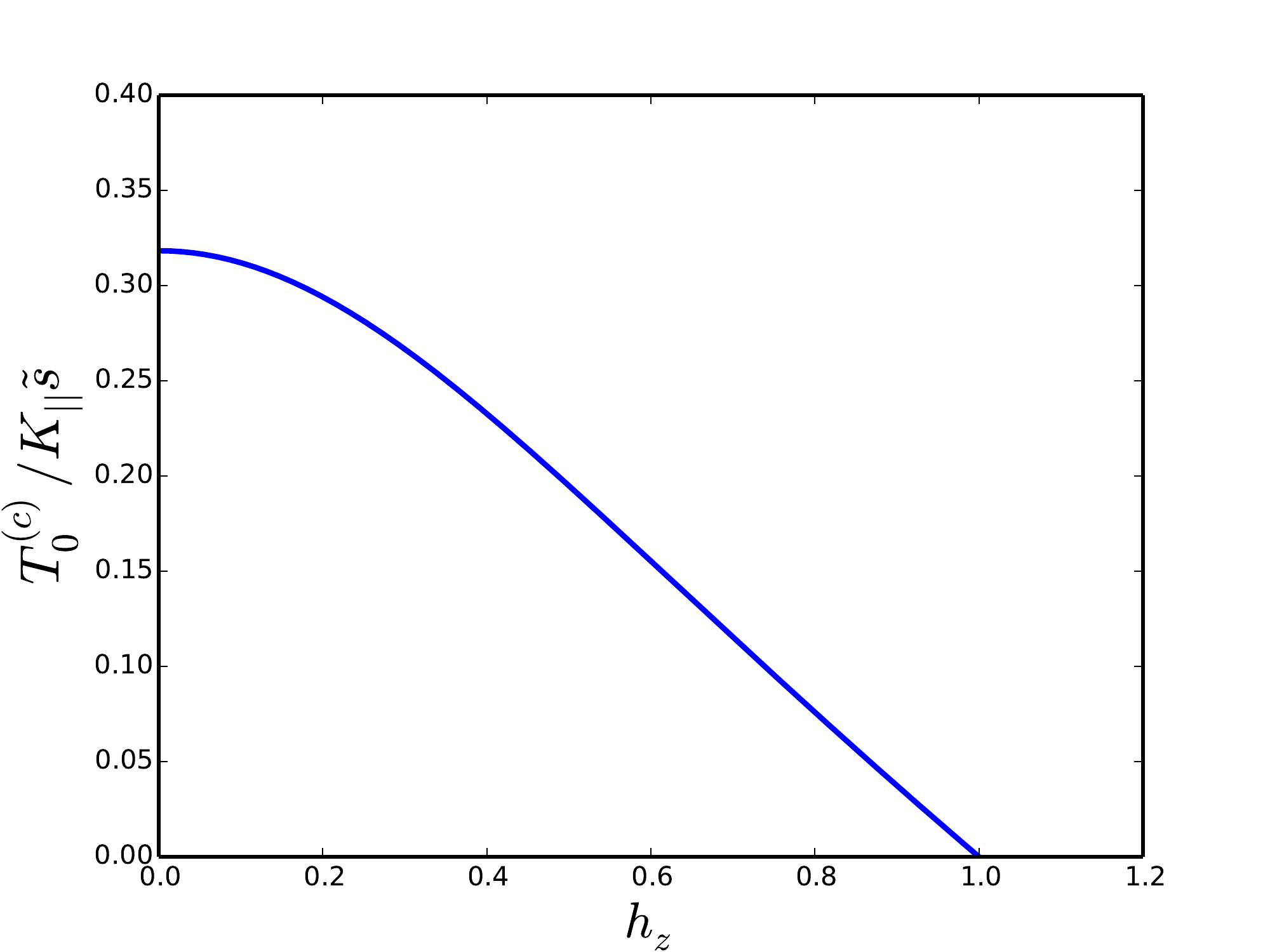}
\caption{ Color online: Left: The phase boundary between the first- and second-order transitions at $h_x=0$. Right: The crossover temperature $T_0^{(c)}$at the boundary between first- and second-order transitions. Reproduced with permission from \onlinecite{kim1}}
\label{pot19}
\end{figure}

Using the value of $h_x=h_x^+$ at the phase boundary, Eqn.\eqref{phab}, one obtains the  crossover temperature at the phase boundary $T_0^{(c)}$.  As shown in Fig.\eqref{pot17} the difference between these temperatures vanishes at $k_t=1$ which is the critical value at the phase boundary for $h_x=0$. 

\begin{figure}[ht]
\centering
\includegraphics[width=3.5in]{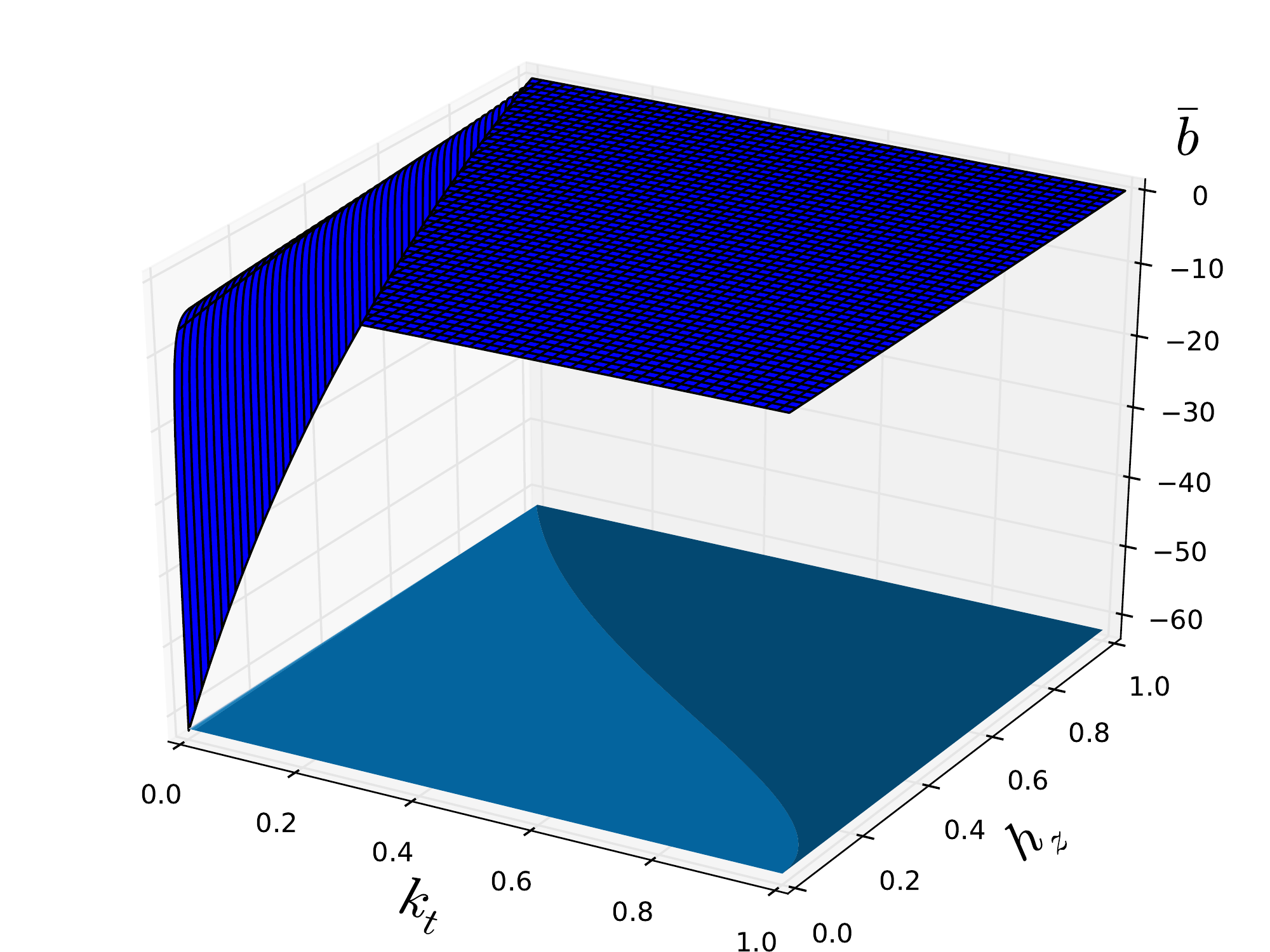}
\caption{ Color online: Three dimensional plot of the Landau coefficient $\bar{b}=b/K_{||}\tilde{s}^2$. In this figure $\bar{b}<0$ is for first-order transition, $\bar{b}>0$ is for second-order transition, and the phase boundary $\bar{b}=0$  is placed on the lower  two dimensional plane for proper view of the upper plane.}
\label{bia_lan}
\end{figure}

For the second case $h_x=0$ and $h_z\neq0$, the potential Eqn.\eqref{eq1} is an odd function with a bias minima, and the barrier height is $\Delta U = K_{\parallel}\tilde{s}^2(1-h_z)^2$. The maxima is located at $x_s=x_{\text{max}}=\frac{1}{2}\ln\lb\frac{1+h_z}{1-h_z}\rb$. Direct application of Eqn.\eqref{eq2} yields 
\begin{align}
b =K_{||}\tilde{s}^2\frac{(2+k_t)}{16k_t(1+h_z)^2}\lb h_z^2(1+2k_t)-(1-k_t)\rb,
\label{270}
\end{align}
At the phase boundary $b=0$ which yields
\begin{align}
k_t^c= \frac{1-h_{cz}^2}{1+2h_{cz}^2}.
\label{eq6}
\end{align}
The plot of Eq.\eqref{270} is shown in Fig.\eqref{bia_lan}, indicating the regions of the phase transitions based on the sign of $b$. 
For small field parameter $h_z\ll1$, the critical value decreases as $k_t^c\approx 1-3h_{cz}^2$. One obtains that the second-order crossover at the phase boundary is
\begin{equation}
T_0^{(c)}= \frac{ K_{\parallel}\tilde{s}}{\pi}\frac{1-h_{cz}^2}{\sqrt{1+2h_{cz}^2}},
\label{eq7}
\end{equation}
which has the form $T_0^{(c)}/K_{\parallel}\tilde{s}\approx (1-2h_z^2)/\pi$ for $h_z\ll1$. The plot of Eqn.\eqref{eq6} and \eqref{eq7} is shown in Fig.\eqref{pot19}. It clearly shows the consistency  of the result with that of perturbation theory in Fig.\eqref{pot14}. At $k_t=0$, there is no tunneling due to the following:  Quantum mechanically, in this limit the Hamiltonian commutes with $\hat{S}_z$, thus there is no splitting term since $\hat{S}_z$ is conserved quantity. In the effective potential method, this implies that the barrier becomes infinitely thick and the spin cannot tunnel.
\subsubsection{An alternative model}
\label{chocop}
 It is sometimes difficult to deal with a particle Hamiltonian with a position dependent mass. It is possible to get a particle with a constant mass from the model in Eqn.\eqref{kimh1} using another approach. 
Let us consider the model\footnote{An alternative choice is $\hat H =K_1\hat{S}_z^2+K_2\hat{S}_y^2-H_x\hat{S}_x$. Setting $H_x=0$ in Eqn.\eqref{csp1}, these models are related by $K_1=A$ and $K_2=A-B$ \cite{mull3}.} 
\begin{align}
\hat H &=-A\hat{S}_z^2-C\hat{S}_x^2-H_x\hat{S}_x -H_z\hat{S}_z.
\label{csp1}
\end{align}
This model is exactly the same as Eqn.\eqref{kimh1} if we set $A=K_{\parallel}+K_{\perp}$ and $C=K_{\perp}$. Introducing an unconventional generating function of the form
\bea
\mathcal {G}(x)=\sum_{m=-s}^{s}\frac{c_m}{\sqrt{(s-m)!(s+m)!}}\lb\frac{\sn x+1}{\cn x}\rb^s,
\eea
and the particle wave function function \begin{align}
\Psi(x)=e^{-y(x)}\mathcal {G}(x).\end{align}
As in the previous analysis, $y(x)$ is determined by the usual procedure\footnotemark[10]. One finds that the corresponding effective potential and the mass  are  given by \cite{csp1,csp}, $U(x)=\tilde{s}^2Au(x)$
\begin{align}
u(x)&=\frac{1}{4\dn^2x}[(\alpha_x\sn x-\alpha_z\cn x)^2-4b-4(b\alpha_z\sn x+\alpha_x \cn x)], \quad m= \frac{1}{2A},
\label{chang}
\end{align}
where the large $s$ limit $s\sim s+1 \sim (s+\frac{1}{2})=\tilde{s}$ has been used. $b=C/A$ and $\alpha_{x,z}=H_{x,z}/{sA}$, the modulus of the elliptic functions is $k^2=1-b$. The maximum of the potential is at $x_s=x_{\text{max}}=0$ for $\alpha_z=0$ . 
\subsubsection{Phase boundary and crossover temperatures}

Since the mass is now a constant and the potential is even for $\alpha_z=0$, the criterion for the first-order transition, Eqn.\eqref{mull} is determined only by the fourth derivative of the potential at $x_s$ or by considering where the coefficient of the fourth order expansion changes sign near $x_s$. For $\alpha_z=0$ we find 
\begin{align}
U(x)&\approx U(0) + A\tilde{s}^2[-\frac{1}{4}(2-2b-\alpha_x)(2b+\alpha_x)x^2+ \frac{1}{24}(\alpha_x-\alpha_x^+)(\alpha_x-\alpha_x^-) x^4 + O(x^6)].
\label{expa1}
\end{align}

The vanishing of the coefficient of $x^4$ determines the phase boundary
\begin{equation}
\alpha_{cx}^{\pm}= \frac{1-16b_c(1-b_c)\pm\sqrt{1+32b_c(1-b_c)}}{4(1-b_c)},
\label{cspcon}
\end{equation}
which is exactly the result obtained by \textcite{csp}. Eqn.\eqref{cspcon} is consistent with Eqn.\eqref{phab} by noticing that $b=k_t/(1+k_t)$ and $\alpha_x=2(1-b)h_x$. The second-order crossover temperature is given by
\bea
T_0^{(2)} = \frac{A\tilde{s}}{2\pi}\sqrt{(2b+\alpha_x)(2(1-b)-\alpha_x)}.
\label{cspcon1}
\eea
\begin{figure}[ht]
\centering
\includegraphics[width=3.5in]{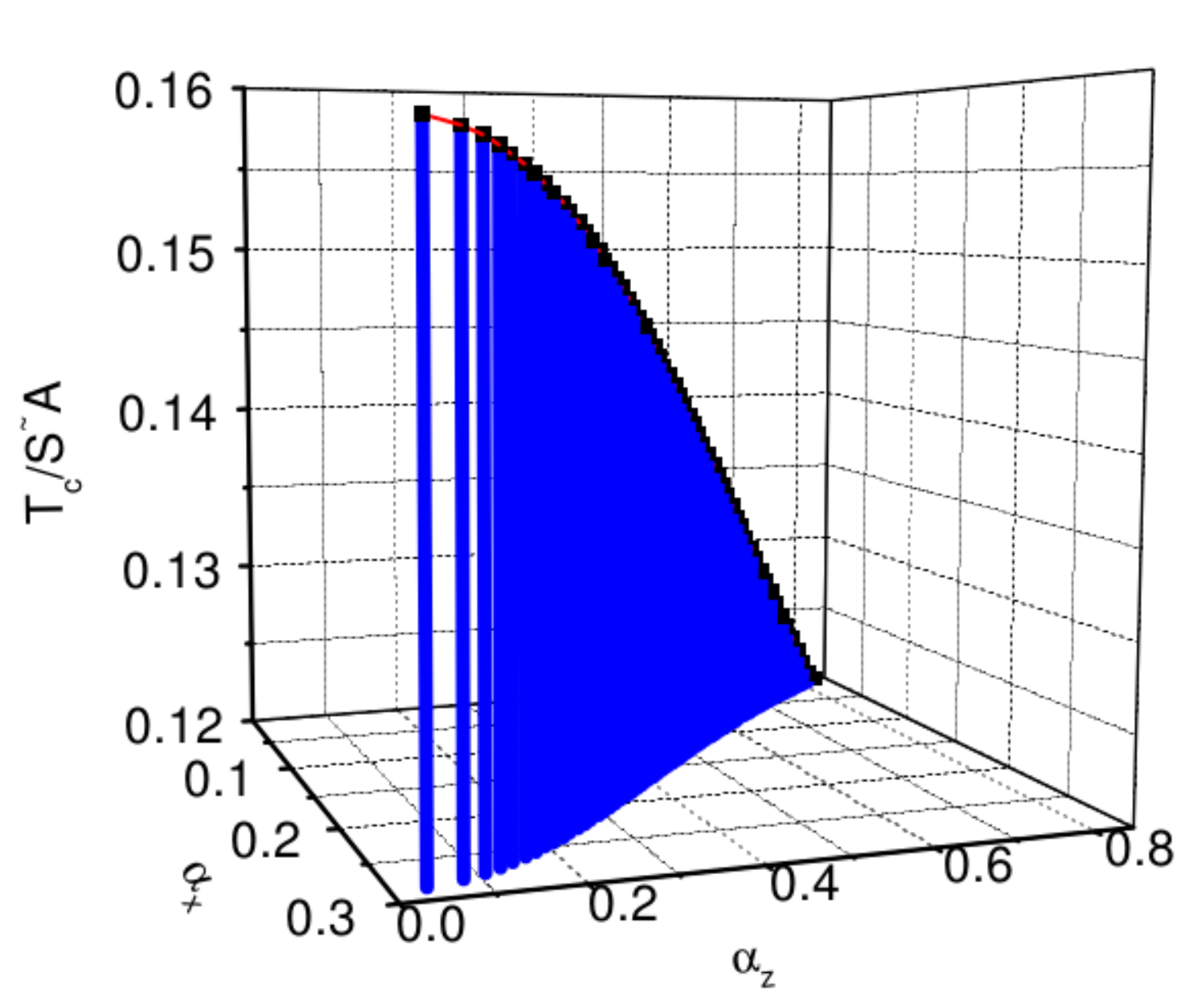}
\caption{ Color online: 3D numerical plot of the crossover temperature $T_c(\alpha_{cz},\alpha_{cx})/\tilde{s}A$ at the boundary as a function of field parameters $\alpha_{cx}$ and $\alpha_{cz}$ with $b_c=0.29$. From \textcite{csp}}
\label{choo}
\end{figure}
Plugging Eqn.\eqref{cspcon} into Eqn.\eqref{cspcon1} one obtains the crossover temperature at the phase boundary. For $\alpha_x=0$ in Eqn.\eqref{chang}, the maximum occur at $x_s=\sn^{-1}[-\alpha_z/2(1-b)]$, the potential is no longer an even function, therefore the coefficient of the fourth order expansion near $x_s$ cannot determine the regime of first-order transition. With the help of Eqn.\eqref{conmull2} one obtains the phase boundary and the crossover temperature at the phase boundary \cite{csp1}
\begin{align}
\alpha_{cz}&= 2(1-b_c)\sqrt{\frac{1-2b_c}{1+b_c}};\quad T_0^{c}=\frac{2\tilde{s}A\sqrt{3}b_c}{\pi}\sqrt{\frac{1-b_c}{1+b_c}}.
\label{chang1}
\end{align}
These expressions are consistent with Eqn.\eqref{eq6} and \eqref{eq7}. At $\alpha_{cz}=0=\alpha_{cx}^{+}$, one finds that $b_c=1/2$ which corresponds to the result of Sec.\eqref{kimphase} with the anisotropy constants  related by $A=K_1$ and $C=K_1-K_2$, implying that $\lambda=1-b$. In the limit of small anisotropy $b_c\ll1$, one finds $\alpha_{cz}\approx 2(1-5b_c/2)$ and $T_0^{c}\approx {2\tilde{s}A\sqrt{3}b_c}/{\pi}$.  The phase diagram of these expressions are related to Fig.\eqref{pot17} and Fig.\eqref{pot19}. For iron cluster Fe$_8$,  $s=10$, $A =0.316 K$, and $C=0.092 K$\cite{san,bar} one finds that $T_0^{c}=0.79 K$. In Fig.\eqref{choo} we have shown a $3$-dimensional plot of $T^c(\alpha_{cx},\alpha_{cz})$. It is evident that $T^c$ decreases as $\alpha_{z}$ increases, while it increases with increasing $\alpha_{x}$.
\subsection{Phase transition in easy $x$-axis biaxial spin model with a medium axis magnetic field}
\label{larb}
\subsubsection{Effective potential of medium axis magnetic field model}
For the model we considered in Sec.\eqref{med}, that is
\begin{equation}
\hat H = D_1\hat{S}_z^2 + D_2\hat{S}_x^2-H_x\hat{S}_x.
\label{rmp2}
\end{equation}
The effective potential and the mass were obtained as
\begin{align}
U(x) &= \frac{ {D_2}\tilde{s}^2[\cn(x)-\alpha_x]^2}{\dn^2(x)},\quad m=\frac{1}{2D_1}.
\label{rmp3}
\end{align}
\begin{figure}[ht]
\centering
\includegraphics[width=3.5in]{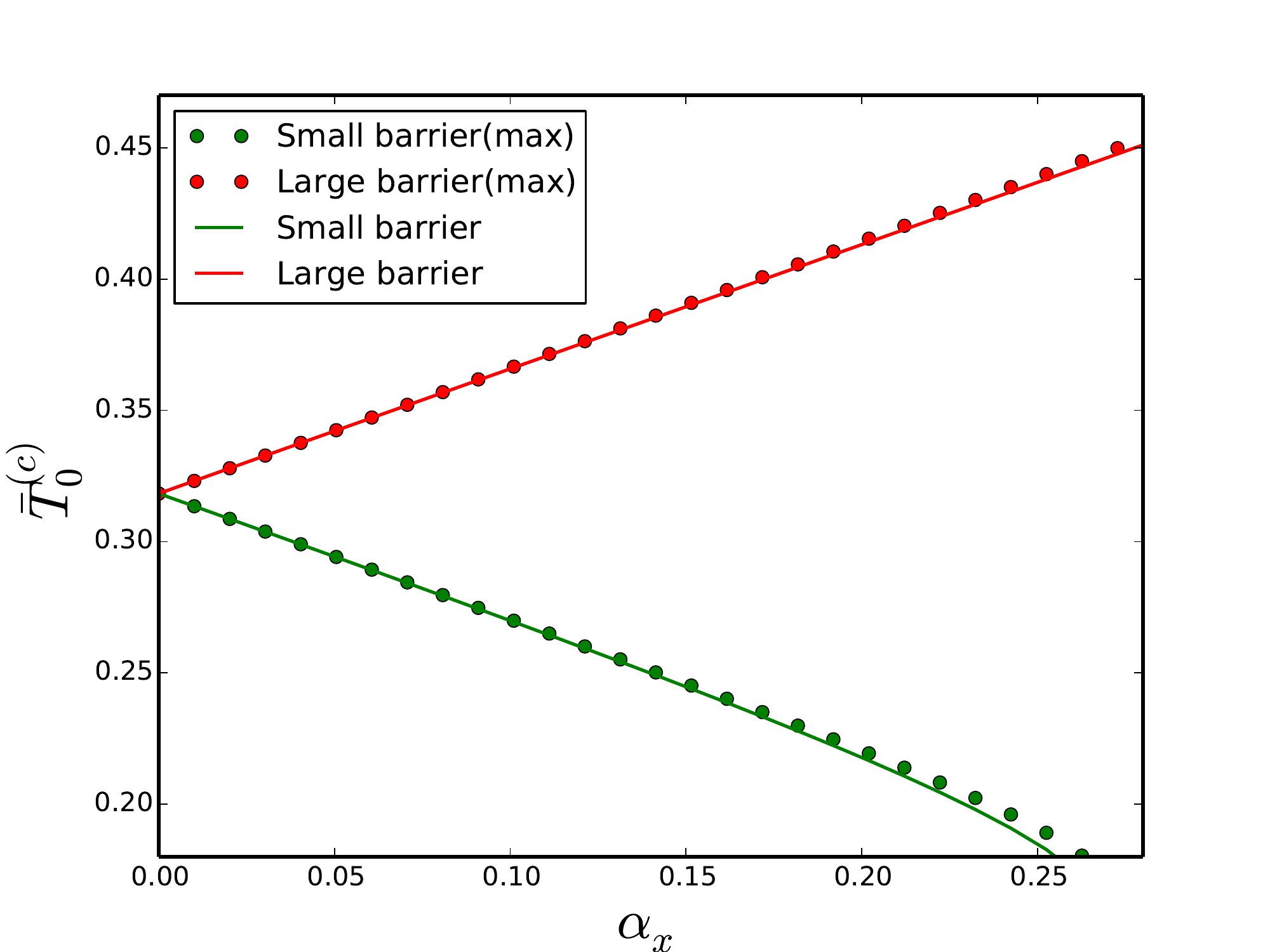}
\label{phase0}
\includegraphics[width=3.5in]{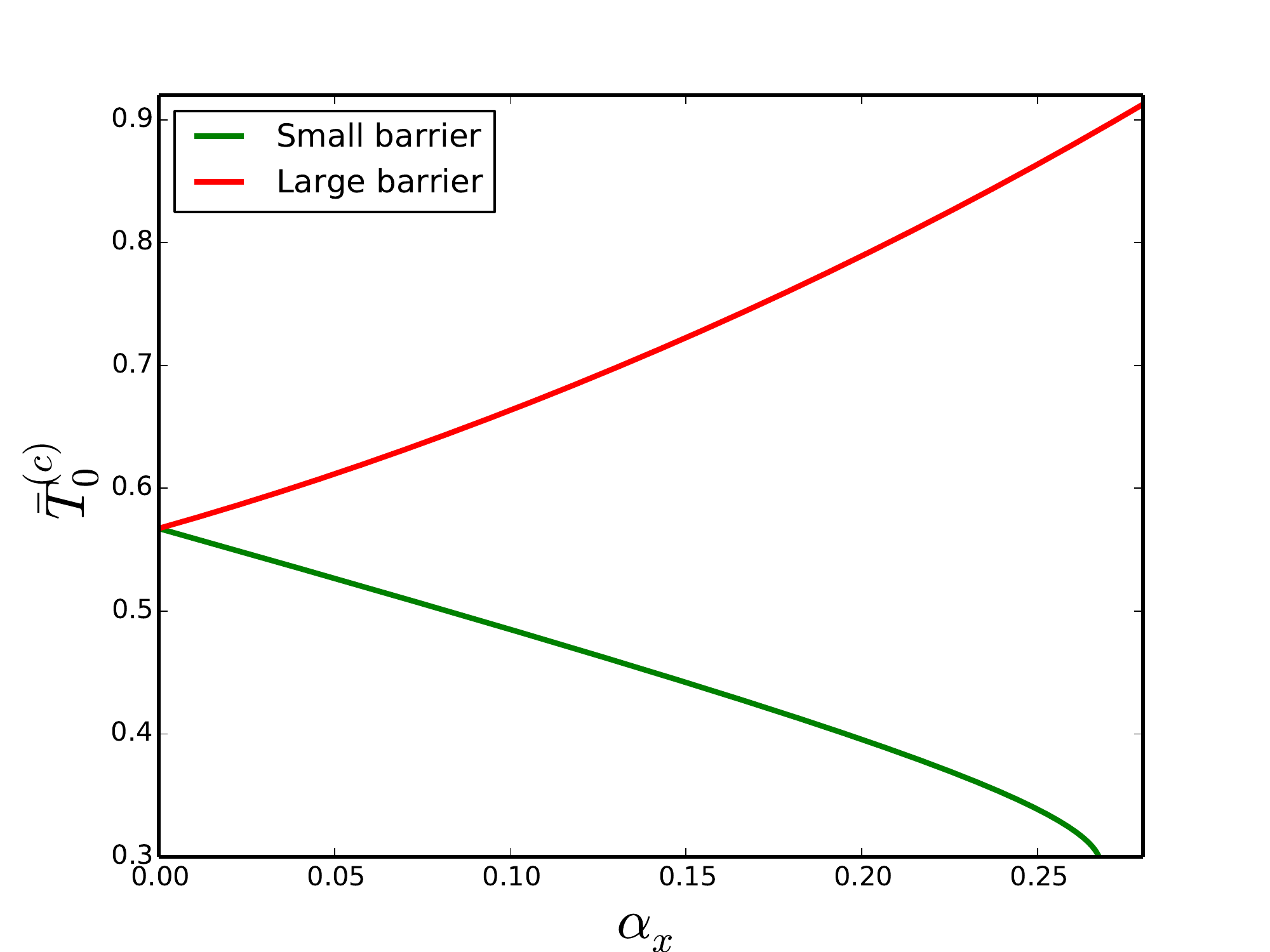}
\label{phase1}
\caption{ Colour online: Dependence of the crossover temperatures on the magnetic field at the phase boundary.  Left:  Second-order  (solid line) and its maximum (dashed line) for the small and large barrier. Right: First-order for the small and the large barrier. These graphs are plotted with $T_0^{(c)}= T_0^{(c)}/  {D_2}\tilde{s}$. Adapted with permission from \onlinecite{sm3}}
\label{pot20}
\end{figure}
This Hamiltonian is related to Eqn.\eqref{kimh1} for $H_z=0$ if one sets $D_1=K_{\parallel}+K_{\perp}$ and $D_2=K_{\parallel}$, it is also related to  Eqn.\eqref{csp1} for $H_z=0$ if one sets $D_1=A$ and $D_2=A-B$, but unlike these models, we saw that the potential,  Eqn.\eqref{rmp3} has large and small barriers (see Fig.\eqref{pot6}) located at  $x_{lb}=\pm2(2n+1)\mathcal{K(\kappa)}$ and $x_{sb}=\pm4n\mathcal{K(\kappa)}$ respectively, with the barrier heights given by Eqn.\eqref{ba}. The phase transition of the escape rate was studied by \textcite{mull2} using spin coherent state path integral. In this review we will consider it in the effective potential method. 
\subsubsection{Phase boundary and crossover temperatures}
Using Eqn.\eqref{mull} and the maximum points $x_{lb},x_{sb}$, the boundary between the first and second-order  transition  for small and large barriers are found to be \cite{sm3}
\begin{align}
\lambda_{sb}^{\pm}(\alpha_x)=\frac{3 - 4 \alpha_x +  \alpha_x^2 \pm (1-\alpha_x)\sqrt{1 - 4\alpha_x +  \alpha_x^2 }}{4(1- 2 \alpha_x +  \alpha_x^2)},\label{rmp4}
\\
\lambda_{lb}^{\pm}(\alpha_x)=\frac{3 + 4 \alpha_x +  \alpha_x^2 \pm (1+\alpha_x)\sqrt{1 + 4\alpha_x +  \alpha_x^2 }}{4(1+2 \alpha_x +  \alpha_x^2)}.\label{rmp5}
\end{align}
For small barrier one can check that Eqn.\eqref{rmp4} is consistent with Eqn.\eqref{cspcon} and Eqn.\eqref{phab}. The crossover temperature for the first-order transition is estimated as
$T_0^{(1)}= \Delta U/2B$, which can be obtained from Eqn.\eqref{ba} and \eqref{eq4}. At the phase boundary we find that $T_0^{(c)}\approx {D_2}\tilde{s}/(\ln[(3+2\sqrt{2})e^{\pm \frac{3\alpha_x}{\sqrt{2}}}])$ for $\alpha_x\ll1$, where the upper and lower signs correspond to   small and large barrier respectively. Both temperatures coincide at $ \alpha_x=0,\lambda=\frac{1}{2}$ with $T_0^{(c)}={D_2}\tilde{s}/\ln(3+2\sqrt{2})$. For the case of second-order transition, the crossover temperature and its maximum are found to be
\begin{align}
  T_{0}^{(2)}&=  \frac{ D_2\tilde{s}\sqrt{ (1\pm \alpha_x)}}{ \pi}\lb\frac{1-\lb 1\pm\alpha_x\rb\lambda }{\lambda }\rb^{1/2}, \label{rpm6}
\\ 
  T_{0}^{(\text{max})}&=  \frac{D_2\tilde{s}}{ 2\pi\lambda}.
\label{sec}
\end{align}
where the upper and lower signs correspond to the large and small barriers respectively. At the phase boundary one finds that Eqn.\eqref{rpm6} behaves as $T_{0}^{(c)}\approx D_2\tilde{s}(1\pm\frac{3}{2}\alpha_x)/{\pi}$ for $\alpha_x\ll1$, which coincides at $\alpha_x=0,\lambda=\frac{1}{2}$ as shown in Fig.\eqref{pot20}. \begin{figure}[ht]
\centering
\includegraphics[width=4in]{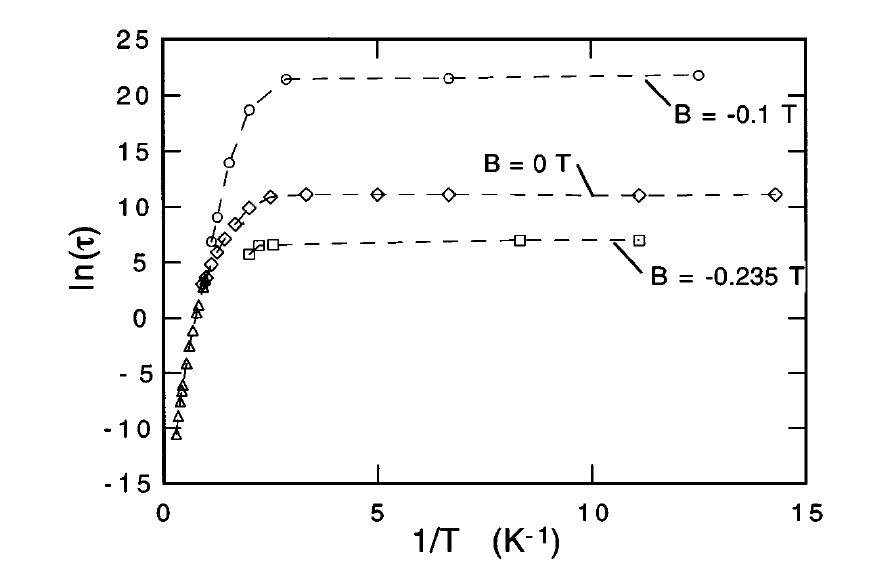}
\caption{ Temperature dependence on the relaxation time $\tau$. Points below $400$ mK show temperature independent quantum tunneling. Adapted with permission from \onlinecite{san}}
\label{pot34}
\end{figure}
 The evidence of this crossover temperatures has predicted in Fe$_8$ molecular cluster with $s=10$. There are $21\times 21$ matrices with $2s+1$ states which can be found by the so-called exact numerical  diagonalization. The energy barrier of this system is much smaller than that of  Mn$_{12}$Ac.  In the low-temperature limit, specifically for $T<0.4 K$, only the two lowest energy level with $M=\pm s$ are occupied and tunneling is possible between these two states. For this system  experimentally measured relaxation rate showed a temperature independent rate below $400$mK which suggests the evidence of spin tunneling across its anisotropy energy barrier \cite{san} (see Fig.\eqref{pot34}).
 
 \subsubsection{Free energy}
 
In the presence of a magnetic field, the Euclidean action cannot be obtained exactly or analytically \cite{mull2}, thus it is studied numerically. The periodic instanton action or the thermon action is given by
\bea
\mathcal{S}_p(\mathcal E)=2\sqrt{2m}\int_{x_1} ^{x_2} dx\sqrt{U(x)-\mathcal E} + \beta(\mathcal E -U_\text{min}).
\eea
\begin{figure}[ht]
\centering
\includegraphics[width=3.5in]{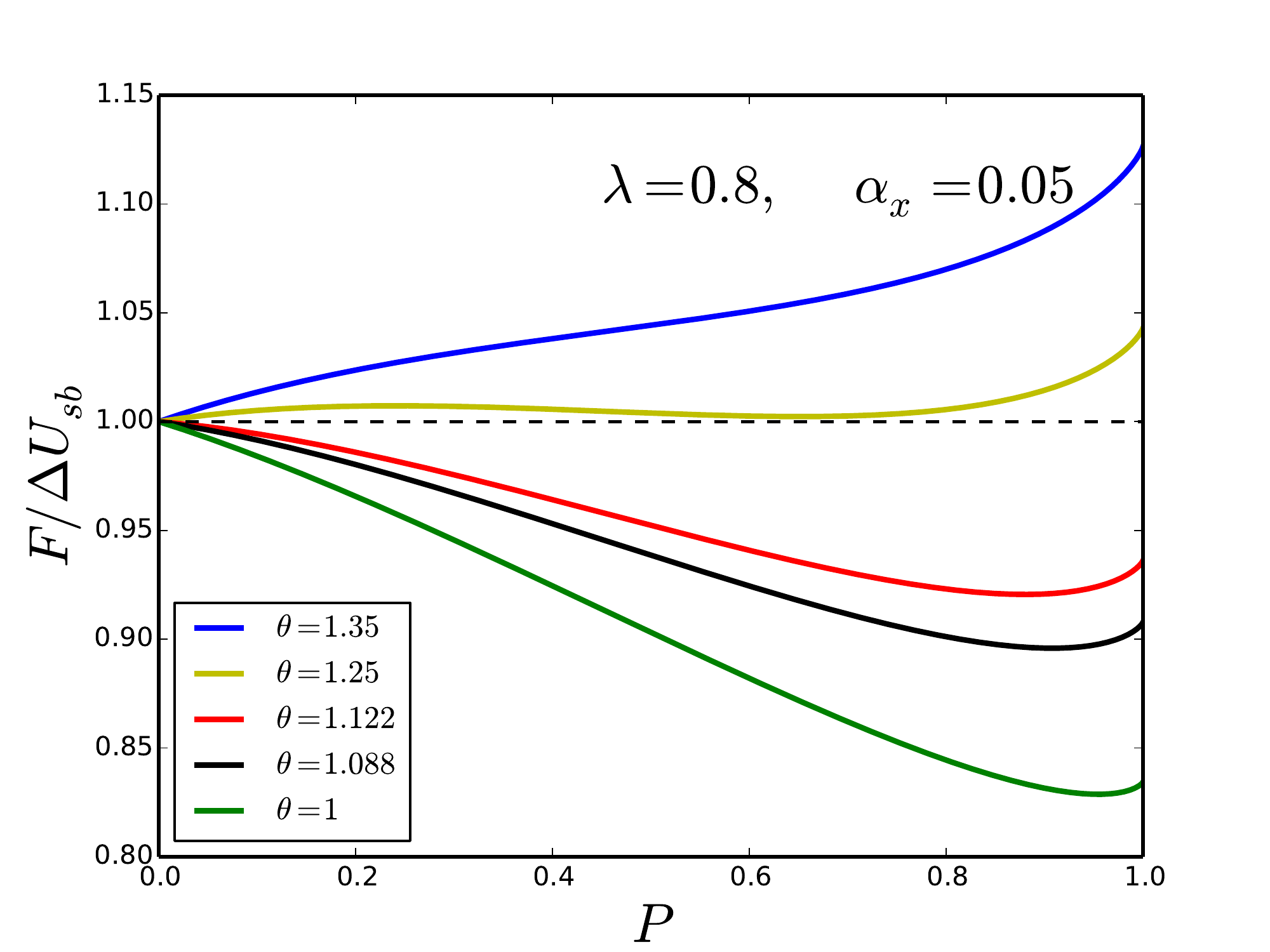}
\caption{ Color online: The numerical plot of the free energy for $\kappa=0.8$ and $\alpha=0.05$. }
\label{free_bia}
\end{figure}
 Setting $y=\sn(x,\lambda)$ and using Eqn.\eqref{eze8} we have
 \begin{align}
\mathcal{S}_p(P)&=2\tilde{s}\sqrt{\kappa}\tilde{S}(P)+\beta\Delta U(1-P),
\end{align}
\begin{align}
\tilde{S}(P)=\int_{y_1}^{y_2}dy\bigg[\frac{\lb\sqrt{1-y^2}-\alpha_x\rb^2-Q(1-\kappa y^2)}{(1-\kappa y^2)^2(1-y^2)}\bigg]^{\frac{1}{2}},
\end{align}
where $Q=(1-\alpha_x)^2(1-P)$,
and the turning points $y_1$ and $y_2$ are determined by setting the numerator in the square bracket to zero.
The free energy  can then be written as
\begin{align}
\frac{F(P)}{\Delta U_{sb}}=1-P+\frac{2\theta\sqrt{1-\kappa(1-\alpha)}}{\pi\lb 1-\alpha\rb^{3/2}}\tilde{S}(P),
\end{align}
where $\theta=T/T_{0}^{(2)}$, $T_0^{(2)}$ is given in Eqn.\eqref{rpm6}, and $\Delta{U}_{sb}$ is given in Eqn.\eqref{ba}.

In Fig.\eqref{free_bia} we have shown the numerical plot of the free energy with some of the  temperature parameters in \cite{zha3}, and the same dimensionless anisotropy constant $\lambda=0.8$, but in the presence of a small magnetic field $\alpha_x=0.05$. We notice that  the phase transition from classical to quantum  regime (where two minima of a curve have the same free energy) has been shifted to  $\theta=1.25$ or $T_0^{(1)}=1.25 T_0^{(2)}$, which is larger than the zero magnetic field value  $T_0^{(1)}=1.122 T_0^{(2)}$ in Fig.\eqref{pot11}(b) \cite{zha3}. Thus, the magnetic field increases the crossover temperature for this model as we found in the previous model in Sec.\eqref{chocop}. However, for large barrier we expect the crossover temperature to decrease. Thus, the large barrier plays a similar role as the longitudinal field $H_z$ in Sec.\eqref{chocop}.
 
\subsection{Phase transition in exchange-coupled dimer model}
\subsubsection{Model Hamiltonian}
In Sec.\eqref{sec:anti}, we reviewed the problem of an antiferromagnetic exchange-coupled dimer model via spin coherent state path integral formalism. In this section we will study the effective potential method of the model. In the presence of a staggered magnetic field applied along  easy $z$-axis, the Hamiltonian is given by  
\begin{align}
\hat{H}&=J\hat{\bold{S}}_A \cdot \hat{\bold{S}}_B  - D(\hat{S}_{A,z}^{2} +\hat{S}_{B,z}^{2}) +g\mu_B h (\hat{S}_{A,z}-\hat{S}_{B,z}),\label{1}\end{align}
 where $J>0$  is antiferromagnetic exchange coupling respectively , $D> J>0$ is the easy $z$-axis anisotropy,  and $h$ is the external magnetic field, $\mu_B$ is the Bohr magneton and $g=2$ is the spin $g$-factor. 
 \subsubsection{Effective potential}
 The spin wave function in this case can be written in a more general form as    
 
\beq
\psi =\psi_A\otimes\psi_B =\sum_{\substack {\sigma_A=-s_A \\ \sigma_B=-s_B}}^{s_A,s_B} \mathcal{C}_{\sigma_A,-\sigma_B} \mathcal{F}_{\sigma_A,-\sigma_B},
\label{2.2}
\eeq
where 
\beq
 \mathcal{F}_{\sigma_A,-\sigma_B} = \binom{2s_A}{s_A+\sigma_A}^{-1/2}\binom{2s_B}{s_B-\sigma_B}^{-1/2}\ket{ \sigma_A,-\sigma_B}.
\eeq
Following the same procedures outlined above, one finds that the effective potential $U(r)$  and the coordinate dependent reduced mass $\mu(r)$ are given by \cite{sm4}
\begin{align}
& U(r) =  \frac{2D\tilde{s}^2[2\alpha^2+\kappa (1-\cosh r) + 2\alpha\kappa \sinh r]  }{\lb 2 +\kappa +\kappa\cosh r\rb}, \label{2.21}\\
&\mu(r)= \frac{1}{2D\lb 2 +\kappa +\kappa\cosh r\rb}.\label{2.22}
\end{align}
In order to arrive at these equations we have used the fact that the two giant spins are equal $s_A=s_B=s$
and the approximation  $ s(s+1) \sim \tilde{s}^2 = (s+ \frac{1}{2})^2$ , where $\kappa = J/D$  and $\alpha = g\mu_Bh/2D\tilde{s}$. The variable $r$  denotes the relative coordinate of the particles, the center of mass coordinate does not contain  any information about the system.

\subsubsection{Periodic Instanton at zero magnetic field }
\begin{figure}[ht]
\centering
\includegraphics[width=3.5in]{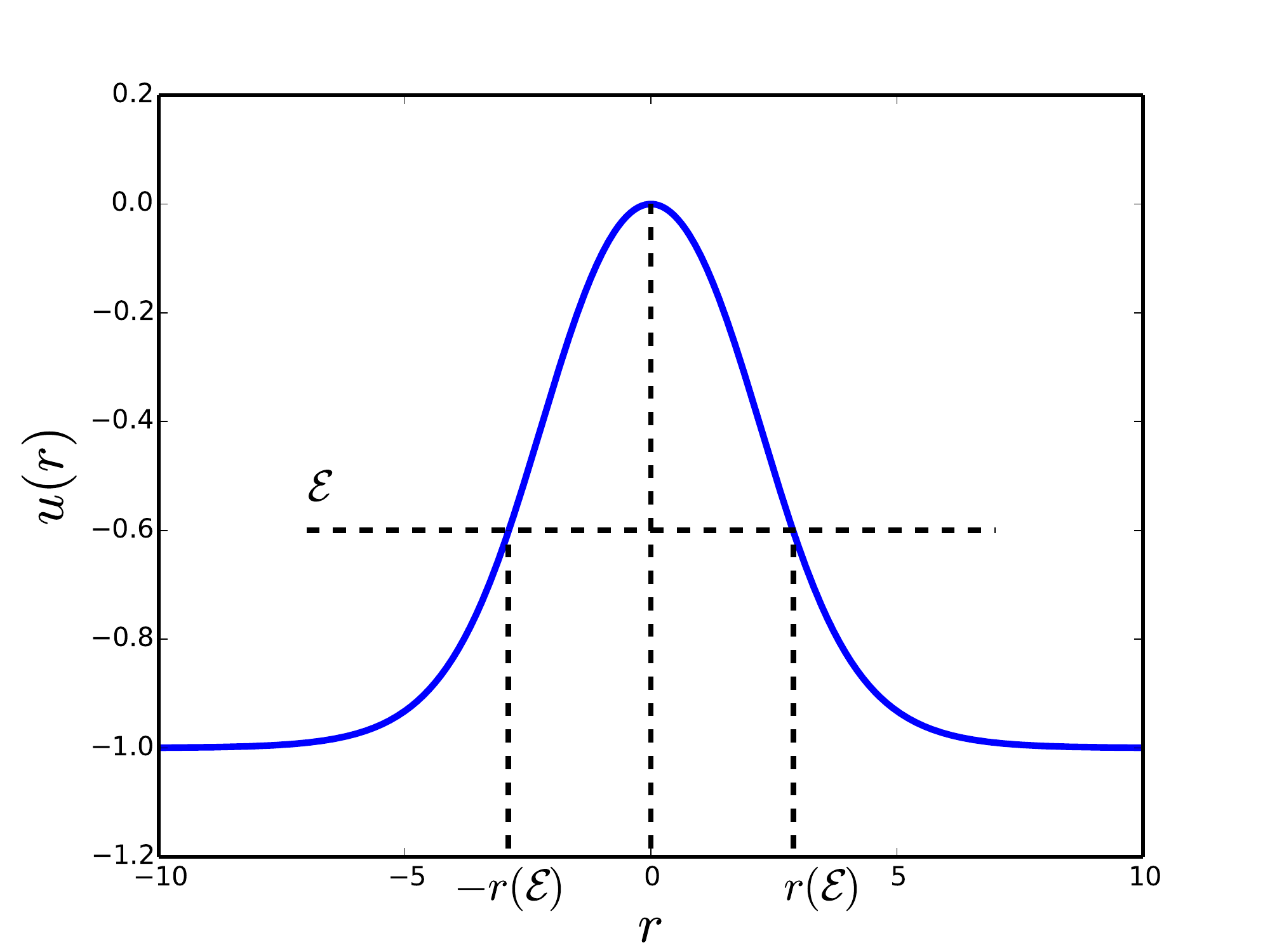}
\caption{ Color online: The plot of the potential for $\kappa=0.6$. The minimum energy is $u_\text{min}=-1$, and the maximum is $u_\text{max}=0$. Thus, $\Delta U= U_\text{max}-U_\text{min}=2Ds^2 $. }
\label{per1}
\end{figure}

\begin{figure}[ht]
\centering
\includegraphics[width=3.5in]{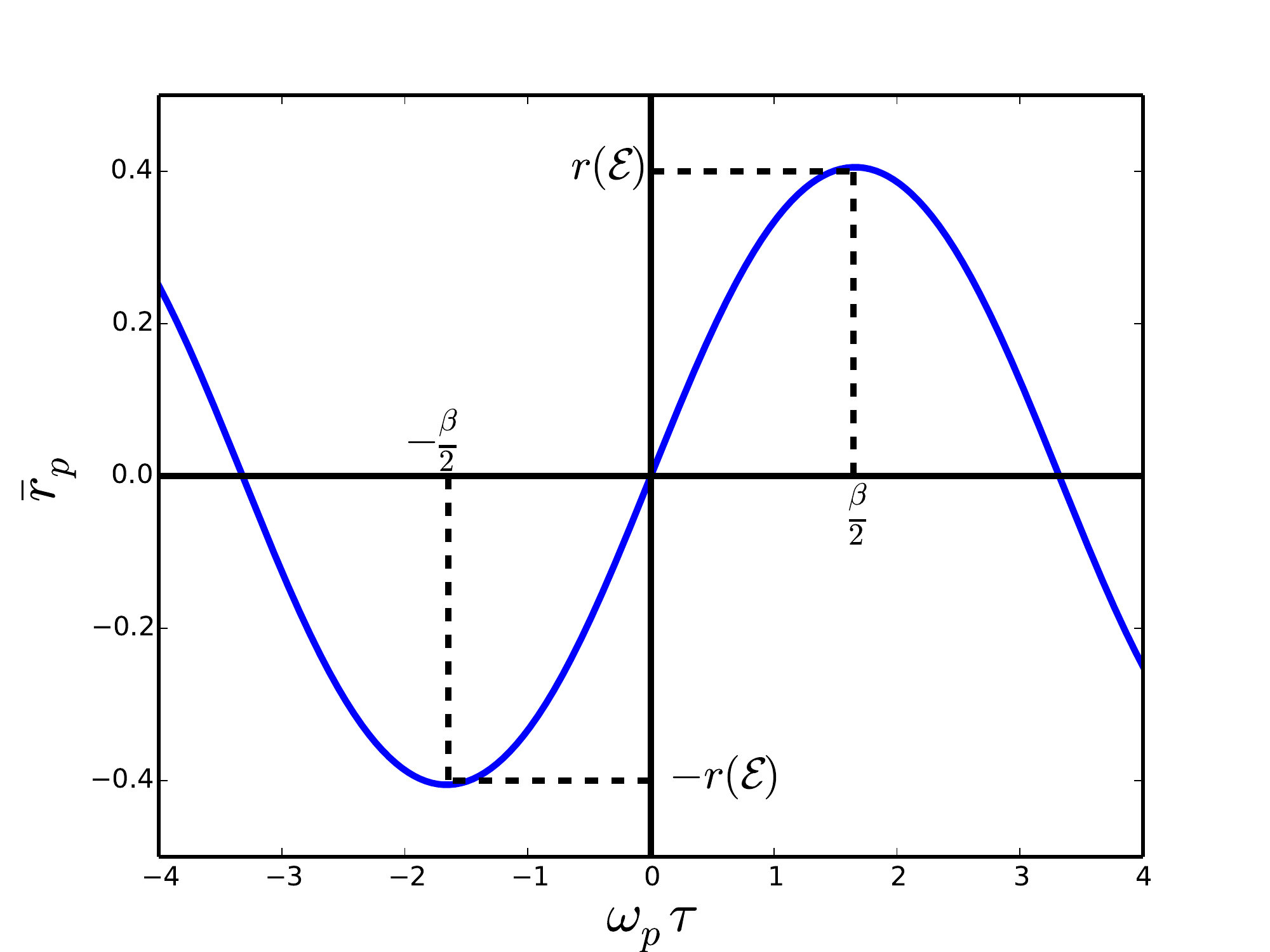}
\caption{ Color online: The periodic instanton trajectory with $\lambda=0.2$. The turning points $\pm r(\mathcal E)$ are shown in Fig.\eqref{per1}.}
\label{per}
\end{figure}

In the absence of a magnetic field, the effective potential is now of the form 
\begin{align}
U(r)&= 2Ds(s+1)u(r), \quad u(r) = \frac{\kappa(1-\cosh r)}{\lb 2 +\kappa +\kappa\cosh r\rb} .
\label{dimer19}
\end{align}
Since we are considering large spin systems, the  coefficient $s(s+1)$ will be approximated as $s^2$. The potential is now symmetric with degenerate minima, and hence the turning points are $\pm r(\mathcal{E})$ with the maximum of the barrier height located at $r_b=0$ as shown in Fig.\eqref{per1}. 
The action associated with the thermon action is given by
\bea
S(\mathcal{E})=2\int_{-r(\mathcal{E})}^{r(\mathcal{E})} dr \sqrt{2\mu(r)\lb U(r)-\mathcal{E}\rb}.
\label{act}
\eea
This action can be integrated exactly for all possible values  of the energy without computing the periodic instanton trajectory explicitly \cite{sm4}. In this paper, we will obtain this action by first calculating the periodic instanton trajectory corresponding to the action. The Euclidean Lagrangian  is given by
\bea
L_E= \frac{1}{2}\mu(r)\dot{r}^2+U(r).
\eea 
The Euler-Lagrange equation of motion gives
\bea
\mu(\bar{r}_p)\ddot{\bar{r}}_p+\frac{1}{2}\frac{d\mu(\bar{r}_p)}{d\bar{r}_p}\dot{\bar{r}}_p^2-\frac{dU}{d\bar{r}_p}=0.
\eea
Integrating once we obtain
\bea
\frac{1}{2}\mu(\bar{r}_p)\dot{\bar{r}}_p^2-U(\bar{r}_p)=-\mathcal{E},
\eea
where $\mathcal{E}$ is the integration constant. Thus, the periodic instanton trajectory can be found from the solution of this equation:
\begin{align}
\tau &= \int_{0}^{\bar{r}_p}dr\sqrt{\frac{\mu(r)}{2(U(r)-\mathcal{E})}}
=\frac{1}{\sqrt{2}\omega_b}\int_{0}^{\bar{r}_p} dr\frac{1}{\sqrt{a+b-2b\cosh^2\lb\frac{r}{2}\rb}},
\label{peom}
\end{align}
where $\omega_b=2Ds\sqrt{\kappa}$ is the frequency of oscillation at the well of the inverted potential of Fig.\eqref{per1}, $a=1-(2+\kappa)\mathcal{E}^{\prime}$, $b=1+\kappa \mathcal{E}^{\prime}$, and $\mathcal{E}^{\prime}= \mathcal{E}/2Ds^2\kappa$.
In terms of a new variable $y= \cosh\lb\frac{r}{2}\rb$, we have
\begin{align}
\omega_b\tau &=\frac{1}{\sqrt{b}}\int_{1}^{\bar{y}_p} dy\frac{1}{\sqrt{(y^2-1)(\frac{a+b}{2b}-y^2)}},
\label{firkin}
\end{align}
where  $\bar{y}_p=\cosh\lb\frac{\bar{r}_p}{2}\rb$. 
Introducing another change of variable:
\bea
x^2=\frac{y^2-1}{\lambda^2y^2},  \quad \lambda^{2}=\frac{a-b}{a+b}.
\eea
The integral in Eqn.\eqref{firkin} becomes
\begin{align}
\omega_b\tau &=\sqrt{\frac{2}{a+b}}\int_{0}^{\bar{x}_p} dx\frac{1}{\sqrt{(1-x^2)(1-\lambda^2x^2)}}=\sqrt{\frac{2}{a+b}} F(\bar{\theta}_p,\lambda)=\sqrt{\frac{2}{a+b}}\sn^{-1}(\sin\bar{\theta}_p, \lambda),
\label{firkin2}
\end{align}
where 
\bea
\bar{x}_p=\sin\bar{\theta}_p=\sqrt{\frac{\bar{y}_p^2-1}{\lambda^{ 2}\bar{y}_p^2}}=\frac{1}{\lambda}\tanh\lb\frac{\bar{r}_p}{2}\rb,
\label{firkin1}
\eea
and $F(\bar{\theta}_p,\lambda)$ is an incomplete elliptic integral of first kind  with  modulus $\lambda$ and $\bar{\theta}_p$ Substituting Eqn.\eqref{firkin1} into Eqn.\eqref{firkin2}, and solving for $\bar{r}_b $ we obtain the periodic instanton:
\begin{align}
\bar{r}_p(\tau)=2\arctanh[\lambda\sn(\omega_p\tau,\lambda)], \quad \omega_p=\omega_b\sqrt{\frac{a+b}{2}}.
\label{gaco}
\end{align}
It is required that as $\tau \to \pm \frac{\beta}{2}$, the periodic instanton trajectory must tend to the classical turning points defined in Eqn.\eqref{act}. In other words,  $\bar{r}_p \to \pm r(\mathcal E)=\pm \arccosh\lb\frac{a}{b}\rb$ as $\tau \to \pm \frac{\beta}{2}$ as depicted in Fig.\eqref{per}. This demands that $\sn(\omega_p\tau,\lambda) \to \pm 1$ as $\tau \to \pm \frac{\beta}{2}$.
Using the fact that $\mu(\bar{r}_p)$ and  $\dot{\bar{r}}_p^2$ are given by
\begin{align}
\mu(\bar{r}_p)& = \frac{\dn^2(\omega_p\tau,\lambda)}{4D[1+\kappa-\lambda^{2}\sn^2(\omega_p\tau,\lambda)]};\quad \dot{\bar{r}}_p^2=(2\lambda\omega_p)^2 \frac{\cn^2(\omega_p\tau,\lambda)}{\dn^2(\omega_p\tau,\lambda)},
\label{thermm}
\end{align} 
and making the transformation $x=\sn(\omega_p\tau,\lambda)$, the action for the periodic instanton path can be computed as
\begin{align}
S_p&=\int_{-\frac{\beta}{2}}^{\frac{\beta}{2}} d\tau \mu(\bar{r}_p)\dot{\bar{r}}_p^2+\beta(\mathcal E-U_\text{min}) 
=2s\sqrt{2(a+b)\kappa}[\mathcal{K}(\lambda)-(1-\gamma^2) \Pi(\gamma^2,\lambda)]+\beta(\mathcal E-U_\text{min}),
\label{acttic2}
\end{align}
where $ \gamma^2=\lambda^{2}(1+\kappa)^{-1}$.
The functions $\mathcal{K}(\lambda)$ and $\Pi(\gamma^2,\lambda)$ are known as the complete elliptic integral of first and third kinds respectively.
\subsubsection{ Vacuum instanton at zero magnetic field}
\begin{figure}[ht]
\centering
\includegraphics[width=3.5in]{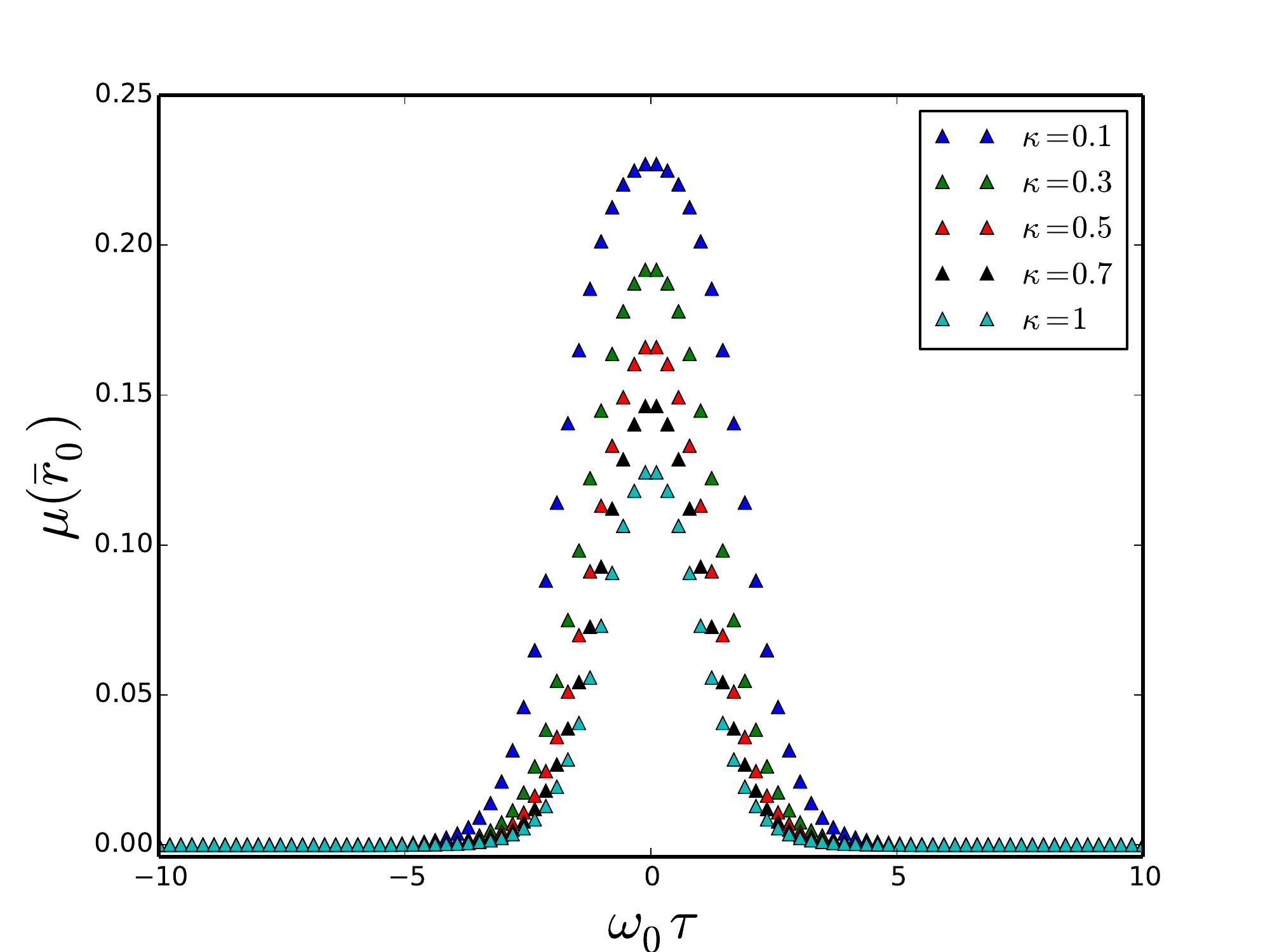}
\caption{ Color online: Dependence of the dimensionless anisotropy constant on the vacuum instanton mass, with $D=1$. }
\label{intmass}
\end{figure}

\begin{figure}[ht]
\centering
\includegraphics[width=3.5in]{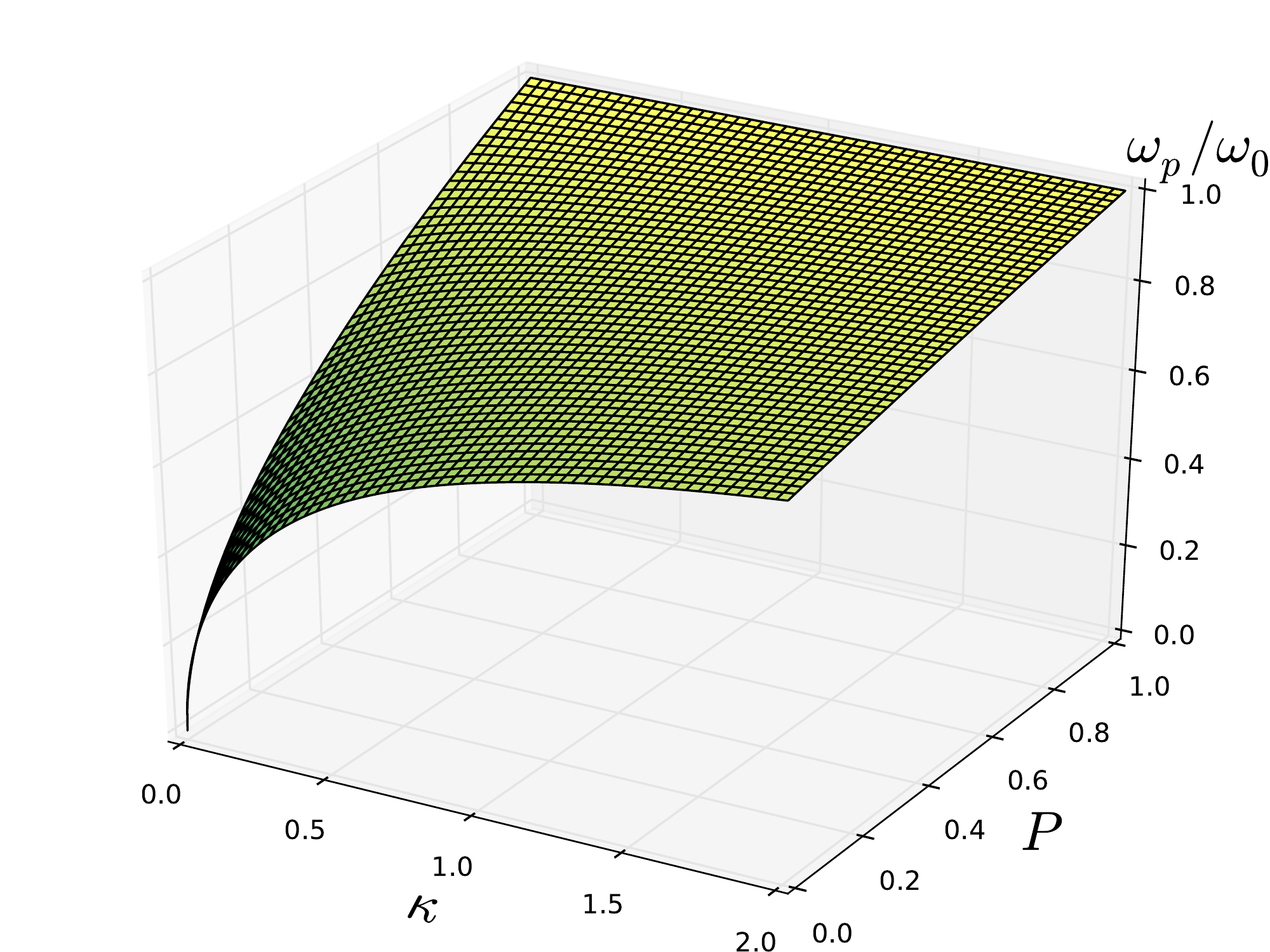}
\label{freq}
\includegraphics[width=3.5in]{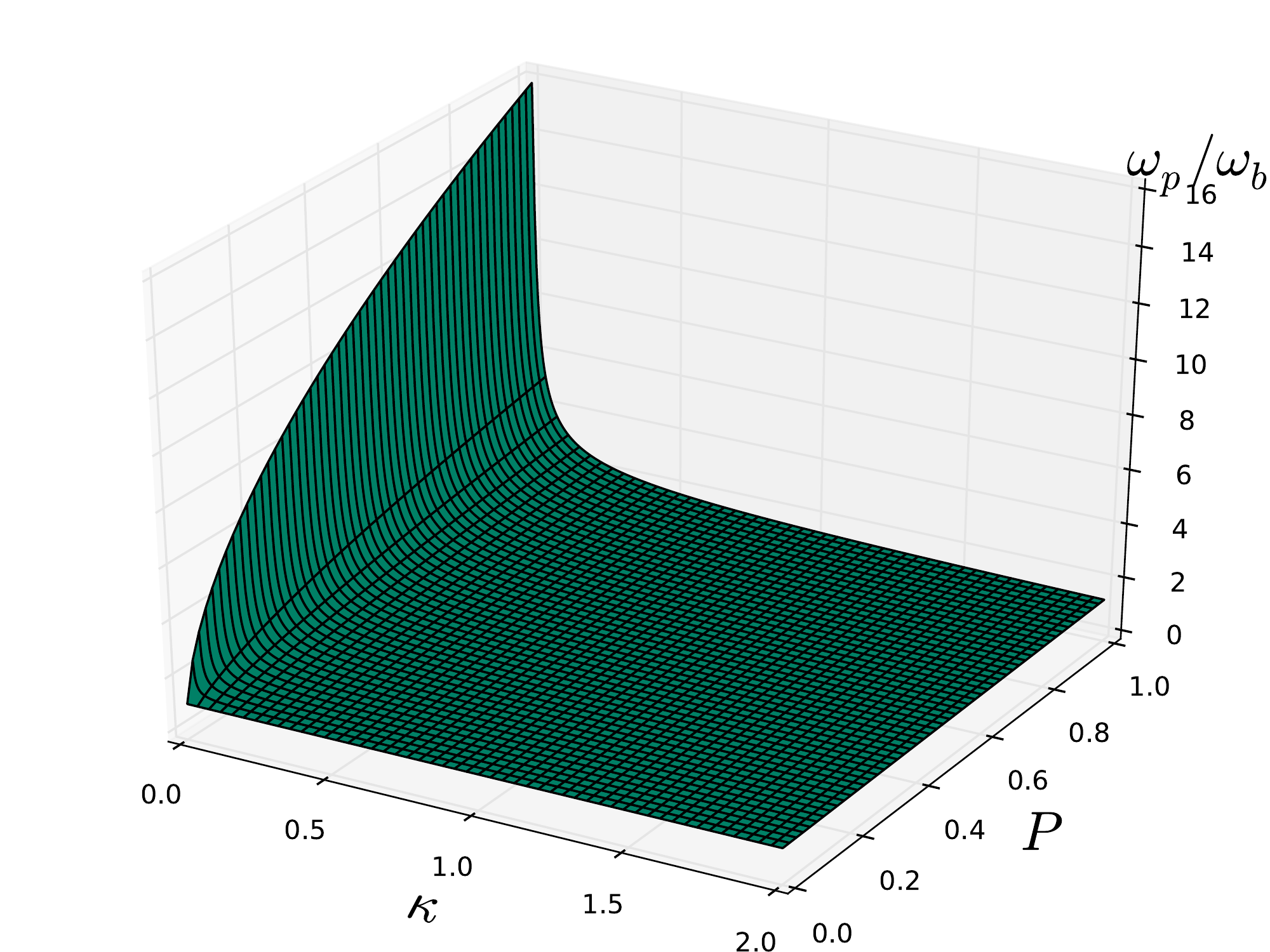}
\label{freq1}
\caption{ Color online: The 3-d plot of the ratio of the periodic instanton frequency in Eqn.\eqref{gaco} to that of vacuum instanton in \eqref{vaint}, and to that of sphaleron in Eqn.\eqref{gaco1}. At the bottom of the barrier $\omega_p\to\omega_0$, and at the top of the barrier $\omega_p\to\omega_b$.}
\label{freqq}
\end{figure}

 Since vacuum  instanton occurs at zero temperature  $T \to 0$, which implies that $\beta\rightarrow \infty$, the energy of the particle must be close to the minima of potential yielding tunneling between degenerate ground states. Near the bottom of the barrier $\mathcal{E}\to U_\text{min}= -2Ds^2$, $a\to 2(1+\kappa)/\kappa$ and $b\to 0$, thus $\lambda\to 1$, we get $\sn(v,1)\to\tanh v$. The periodic instanton trajectory, Eqn.\eqref{gaco} reduces to  a vacuum instanton:

\bea
 \bar{r}_p(\tau)\to\bar{r}_0=2\omega_0\tau, \quad  \omega_p\to\omega_0=2Ds\sqrt{1+\kappa}.
\label{vaint}
\eea
As $\tau \to \pm \infty $,  $\bar{r}_0\to \pm \infty$, which corresponds to  the minima of the zero magnetic field potential.  A particle sitting at the minimum of this potential is massless, $\mu(\bar{r}_0\to\infty)=0$, but the vacuum instanton mass is not zero. It is given by
\begin{align}
\mu(\bar{r}_0)& = [2D(2+\kappa+\kappa\cosh(2{\omega}_0\tau))]^{-1}.
\label{vaintmass}
\end{align} 
In Fig.\eqref{intmass}, we have shown the dependence of the dimensionless anisotropy constant on the vacuum instanton mass. 
Near the top of the barrier $\mathcal{E}\to U_\text{max}= 0$, $a\to 1$ and $b\to 1$, thus $\lambda \to 0$, the periodic instanton reduces to a sphaleron ( static, unstable, finite-energy solutions of the
classical equations of motion) at the top of the barrier:
\bea
 \bar{r}_p(\tau)\to r_b=0, \quad \omega_p\to \omega_b=2Ds\sqrt{\kappa}.
 \label{gaco1}
\eea
The mass of the sphaleron is given by
\bea
\mu(r_b)=[{4D(1+\kappa)}]^{-1}.
\eea
 In Fig.\eqref{freqq} we have shown the plot of the ratio of the frequencies $\omega_p/\omega_0$ and $\omega_p/\omega_b$ against energy for several values of $\kappa$. The action associated with the vacuum instanton trajectory can be obtained by expanding the elliptic integrals in Eqn.\eqref{acttic2} near the bottom of the potential $\lambda \to 1$, or simply by computing the action associated with the vacuum instanton path in Eqn.\eqref{vaint}. Using Eqn.\eqref{vaintmass}  and the fact that $\dot{\bar{r}}_0^2$ is given by
\begin{align}
\dot{\bar{r}}_0^2=(2\omega_0)^2.
\end{align} 
One can easily confirm by direct integration that  the vacuum  instanton action is given by
\begin{align}
B&=\int_{-\infty}^{\infty} d\tau \mu(\bar{r}_0)\dot{\bar{r}}_0^2
=4s\arctanh\lb\frac{1}{\sqrt{1+\kappa}}\rb=2s\ln\lb\frac{\sqrt{1+\kappa}+1}{\sqrt{1+\kappa}-1}\rb.
\label{vacint}
\end{align}
This is the exact vacuum instanton action. In the  perturbative limit $J\ll D$,  which implies that $\kappa\ll1$, Eqn.\eqref{vacint} reduces to
\bea
B\approx 2s\ln\lb\frac{4}{\kappa}\rb = 2s\ln\lb\frac{4D}{J}\rb.
\eea
This is the same action that was obtained  by spin coherent state path integral in Eqn.\eqref{aka}, but the imaginary term in the spin coherent state path integral result, which is responsible for different ground state behaviour of integer and half-odd integer spins has disappeared. This is one of the disadvantages of mapping a spin system to a particle system. 
\subsubsection{Free energy and phase transition at zero magnetic field}
We will now investigate the phase transition of the escape rate using the free energy method . Having obtained the  periodic instanton action for all possible values of the energy, that is Eqn.\eqref{acttic2},
the free energy associated with the escape rate  can then be written as 
\begin{align}
\frac{F}{\Delta U}= 1-P+\frac{4}{\pi}\theta \sqrt{\kappa(\kappa+P)} [\mathcal{K}(\lambda)-(1-\gamma^2) \Pi(\gamma^2,\lambda)],
\label{freennn}
\end{align}

where $\theta =T/T_{0}^{(2)}$ is a dimensionless temperature quantity, and $T_{0}^{(2)}=Ds\sqrt{\kappa}/\pi$ .
The  modulus of the complete elliptic integrals $\lambda$ and the  elliptic characteristic $\gamma$ are related to $P$ by
\begin{align}
\lambda^{ 2} = \frac{(1+\kappa)P}{\kappa+P},\quad \gamma^2=\frac{P}{\kappa+P}.
\label{eneq}
\end{align}
 \begin{figure}[ht]
\centering
\includegraphics[width=3.5in]{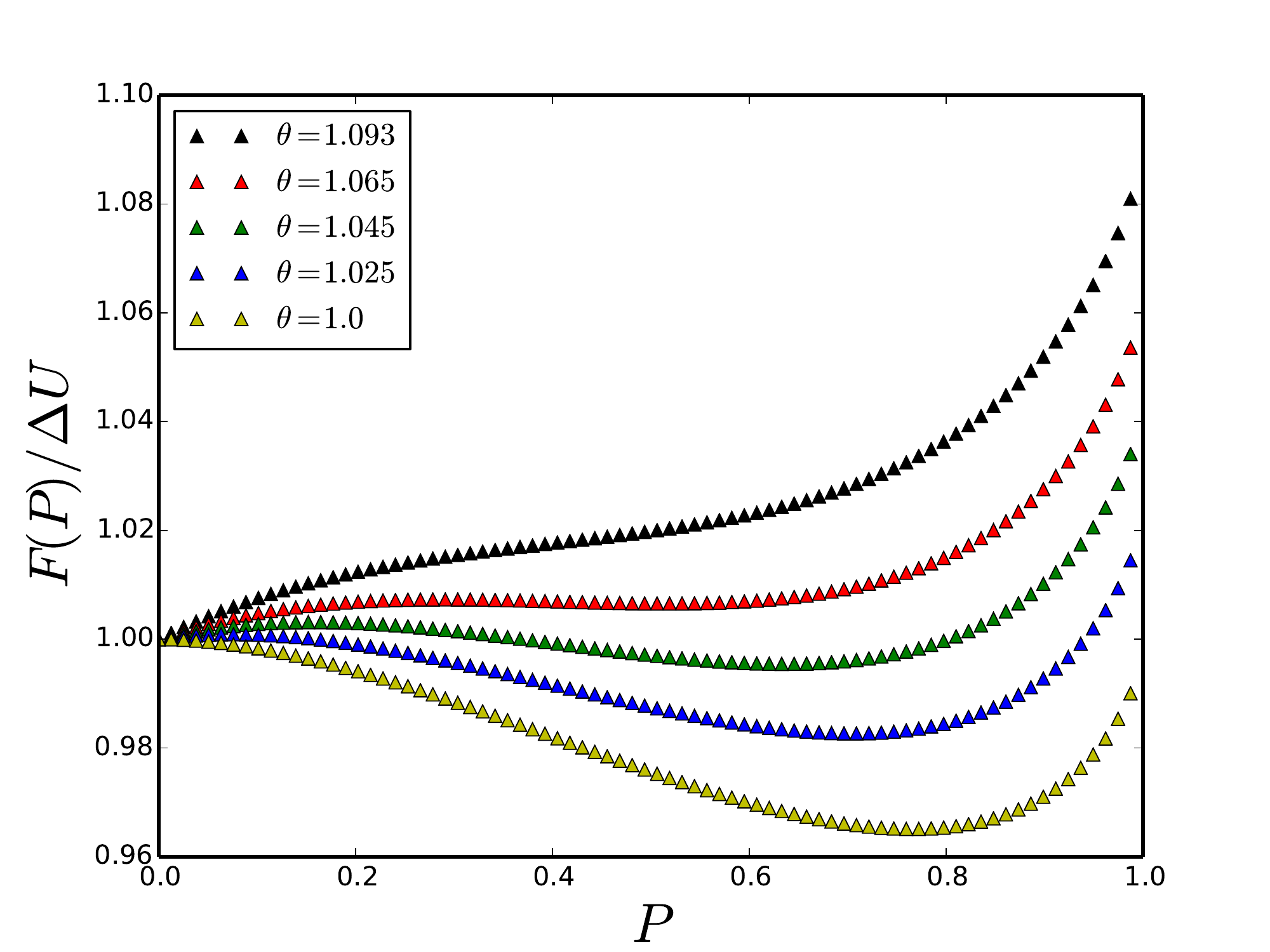}
\caption{ Color online: The effective free energy of the escape rate vs. P for $\kappa=0.4$ and several values of $\theta = T/T^{(2)}_0$, first-order transition. }
\label{f_en}
\end{figure}

In Fig.\eqref{f_en} we have shown the plot of the free energy against $P$ for $\kappa=0.4$ (first-order transition). In the top three curves, the minimum of the free energy is at the top of the barrier, $P=0$. For $\theta= 1.054$ or $T_0^{(1)}=1.054T_0^{(2)}$, two minima have the same free energy. This corresponds to the crossover temperature from classical to quantum regimes.  As the temperature decreases from this crossover temperature, a new minimum of the free energy is formed, this new minimum becomes lower than the one at $P=0$.
We have pointed out that phase transition occurs near the top of the potential barrier, so it is required that we expand this free energy close to the barrier top. Thus, near the top of the barrier $P\rightarrow 0$, the complete elliptic integrals can be expanded up to order $P^3$.  The full simplification of Eqn.\eqref{freennn} yields
 \begin{align}
 \frac{F}{\Delta U}&= 1+(\theta-1)P+\frac{\theta(\kappa-1)}{8\kappa}P^2 +\frac{\theta(3\kappa^2-2\kappa+3)}{64\kappa^2}P^3. 
 \label{fredd}
 \end{align}
Similar to the case of uniaxial spin model in a transverse magnetic field \cite{chud2,chud3}, this free energy looks more like the Landau's free energy, which suggests that we should compare the two free energies. The Landau's free energy  has the form:
\bea 
F = F_0+a\psi^2+b\psi^4 +c\psi^6.
\label{landd}
\eea

The coefficient of $P$ in Eqn.\eqref{fredd} is equivalent to the coefficient $a$ in Landau's free energy. It changes sign at the phase temperature $T=T_0^{(2)}$. The phase boundary between the first- and the second-order phase transitions depends on the coefficient of $P^2$, it is equivalent  to the coefficient $b$ in Eqn.\eqref{landd}. It changes sign at $\kappa=1$. Thus $\kappa<1$ indicates the regime of first-order phase transition. 
 The period of oscillation $\beta(\mathcal{E})$ is found to be
\begin{align}
\beta(\mathcal{E})&
= \frac{2\sqrt{2}}{Ds\sqrt{(a+b)\kappa}}\mathcal{K}(\lambda).
\end{align}
\begin{figure}[ht]
\includegraphics[width=3.5in]{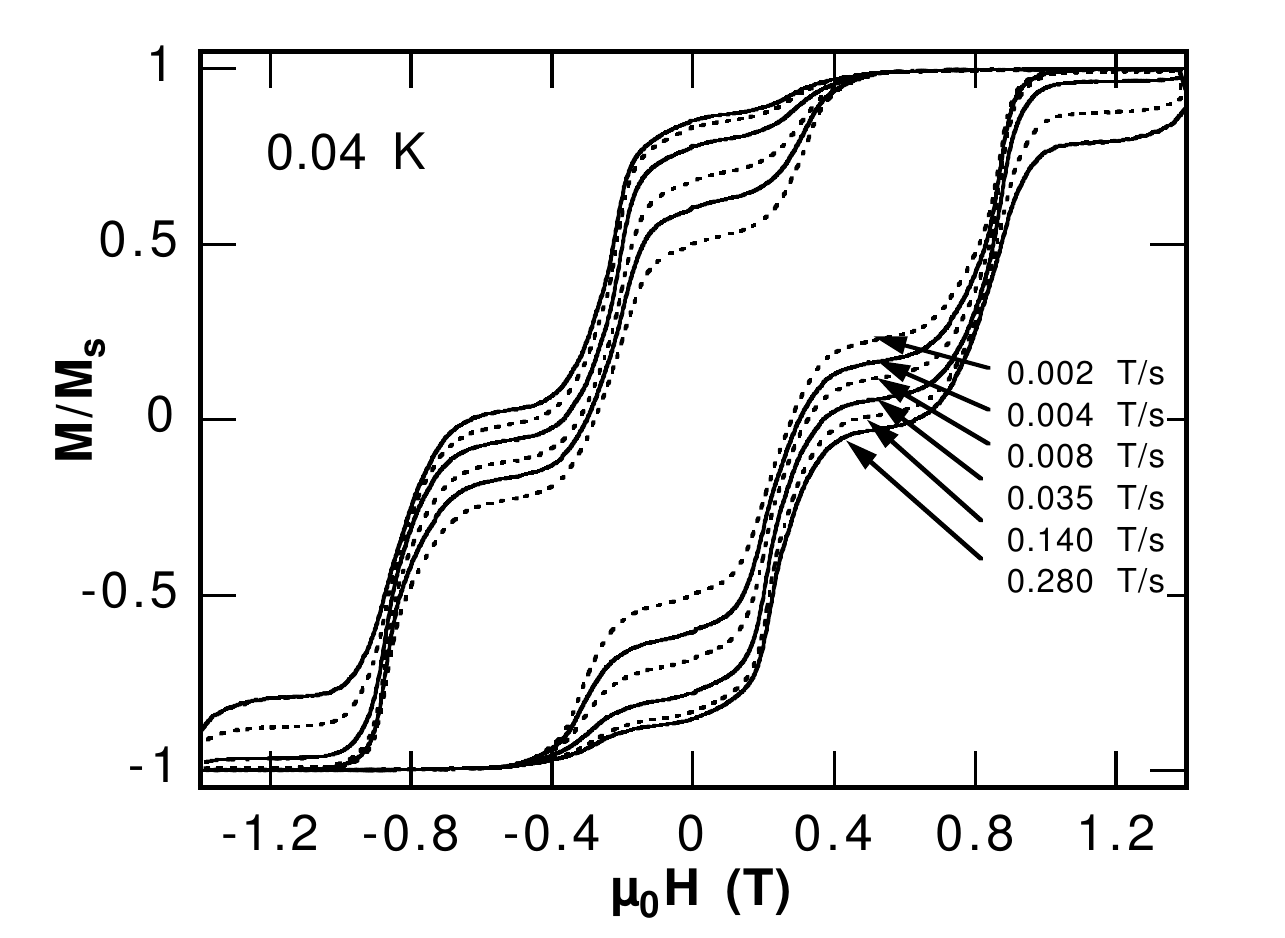}
\caption{ Hysteresis loops for the [Mn$_4$]$_2$ dimer at several field sweep rates and 40 mK. The tunnel transitions are labeled from 1 to 5 corresponding to the plateaus. Adapted with permission from \onlinecite{tiron}} 
\label{pot24}
\end{figure}
\subsubsection{Free energy with magnetic field}
In the previous section we considered the phase transition of the interacting dimer model at zero magnetic field. In this section  we will study the influence of the staggered magnetic field on the phase boundary, the crossover temperatures and the free energy.  
\begin{figure}[ht]
\centering
\includegraphics[width=3.5in]{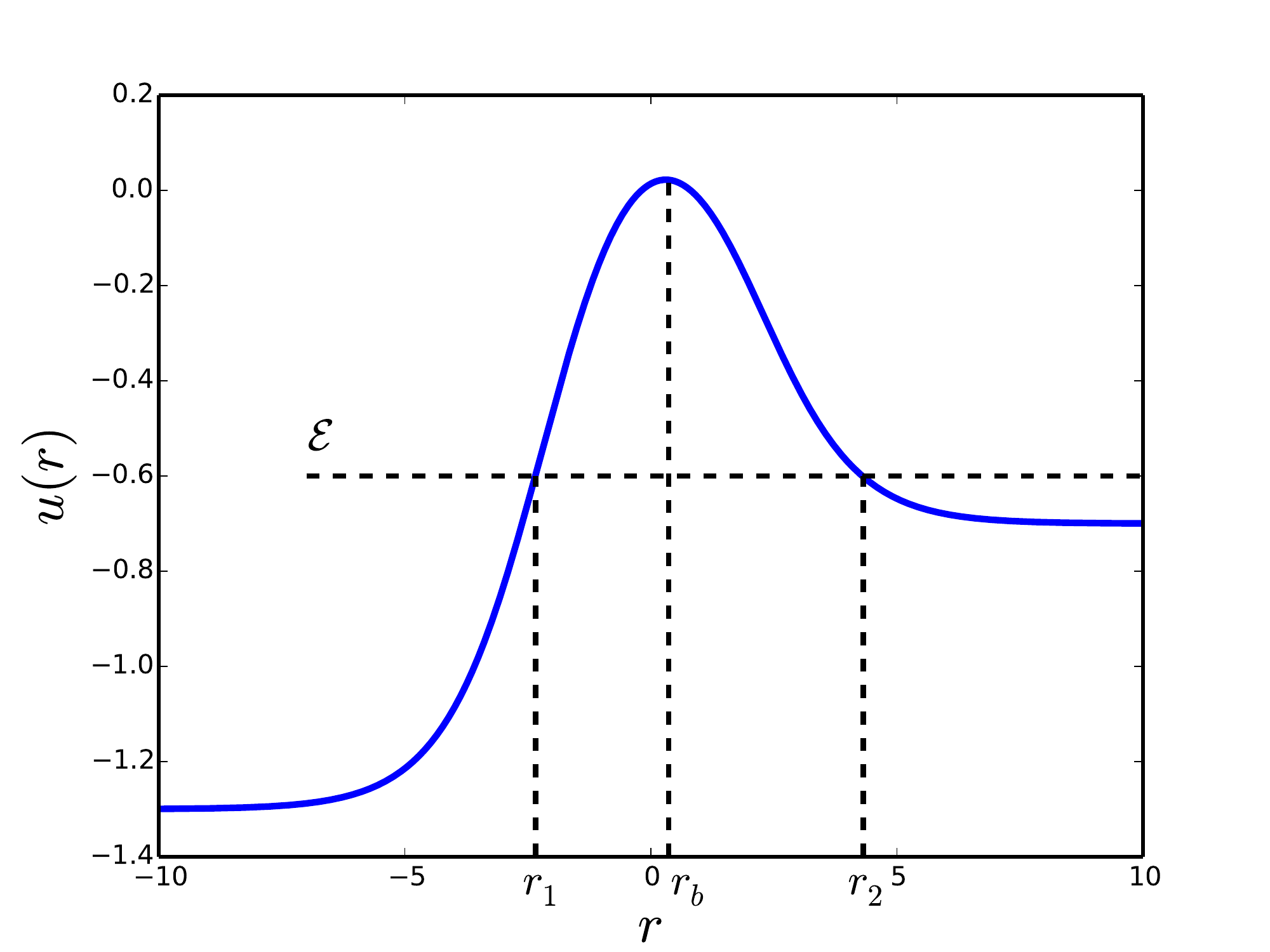}
\caption{ Color online: The plot of the  effective potential and its inverse as a function of $r$ for $\kappa=0.6$ and $\alpha=0.15$.}
\label{interact}
\end{figure}
\begin{figure}
\centering
\includegraphics[width=3.5in]{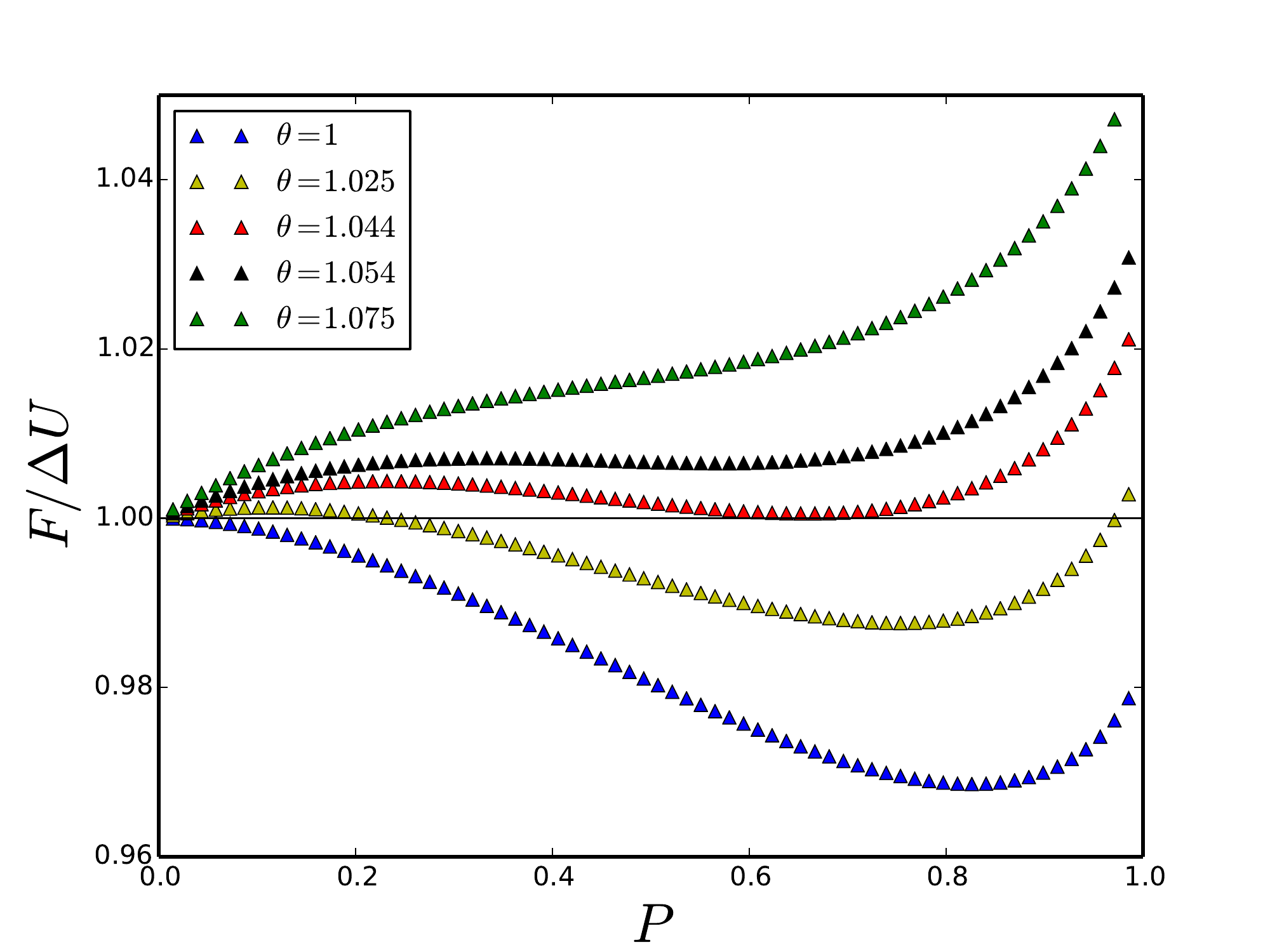}
\caption{Color online: The numerical plot of the free energy with $\kappa=0.4$ and $\alpha=0.15$. The phase  transition from thermal to quantum regimes occurs at $\theta= 1.044$, which is smaller than that of zero magnetic field, $\theta= 1.054$.}
\label{magfree}
\end{figure}
These analyses will be based on the potential and the position dependent mass in Eqn.\eqref{2.21} and Eqn.\eqref{2.22}. In  Fig.\eqref{interact} we have shown the plot of this potential for some values of the parameters. The potential has  a maximum at 
\bea r_b = \ln\lb\frac{1+\alpha}{1-\alpha}\rb,
\label{dimer17}
\eea 
and the height of the potential barrier is given by
\bea
\Delta U = U_{\text{max}}-U_{\text{min}}=2D\tilde{s}^2 \lb 1-\alpha
\rb^2.
\label{dimer18}
\eea

In the presence of a magnetic field, the periodic instanton action or thermon action is given by
\begin{align}
\mathcal{S}_p&= 2\tilde{s}\sqrt{2}\tilde{S}(P)+\beta(\mathcal E-U_\text{min});\quad
\tilde{S}(P)=\int_{r_1}^{r_{2}} dr \frac{\sqrt{a_1-a_2\cosh r+a_3\sinh r}}{2+\kappa(1+\cosh r)};
\label{dimm}
\end{align}
where 
\begin{align}
a_1&=2 \alpha^2 + \kappa - (2 + \kappa) (\alpha^2 - P (1 - \alpha)^2);\quad
a_2=\kappa (1 + \alpha^2 - P (1 -\alpha)^2); \quad
a_3=2\kappa\alpha.
\end{align}
     
 The turning points  are determined from the solution of the equation:
\bea
a_1-a_2\cosh r+a_3\sinh r=0.
\eea
At zero magnetic field $a_3=0$, the potential becomes symmetric hence $r_1=-r_2$. This action, however cannot be integrated exactly either by periodic instanton method or otherwise. Thus, we have to resort to numerical analysis.
The exact free energy can then be written as
\begin{align}
\frac{F}{\Delta U}=1-P +\frac{\theta}{\pi(1-\alpha)^2}\sqrt{2\kappa(1-\alpha^2)}\tilde{S}(P),\end{align}
where the barrier height $\Delta U$ is given in  Eqn.\eqref{dimer18}, $\theta= T/T_0^{(2)}$, and $T_0^{(2)}=\frac{D\tilde{s}}{\pi}\sqrt{\kappa(1-\alpha^2)}$ .  
In Fig.\eqref{magfree} we have shown the numerical plot of this free energy with $\kappa=0.4$ and $\alpha=0.15$. In this case, the minimum of the free energy remains at $\Delta U$ for the top three curves, however, the quantum-classical phase transition (where two minima of a curve have the same  free energy)  has been shifted down to  $T_0^{(1)}=1.044T_0^{(2)}$ due the the presence of a small magnetic field. Thus, the presence of a longitudinal staggered magnetic field in this model decreases the crossover temperatures as in the case of biaxial ferromagnetic spin models.
\subsubsection{Phase boundary and crossover temperatures}
 The phase boundary with the help of Eqn.\eqref{mull} yields
\bea
\alpha_c = \pm \lb{\frac{ 1-\kappa_c}{ 1+2\kappa_c}}\rb^{\frac{1}{2}}.
\label{3.4}
\eea
One finds that the second-order transition crossover temperature $T_{0}^{(2)}$ at the phase boundary yields
 \bea
T_{0}^{(c)} =\frac{D \tilde{s}\kappa_c}{\pi}\lb\frac{ 3 }{1+2\kappa_c}\rb^{\frac{1}{2}}.
\eea
For [Mn$_{4}$]$_{2}$ dimer, the parameters are: $s=9/2$, $D=0.75 K$, and $J=0.12 K$ \cite{hill,tiron}, one finds that the value of the crossover temperature at the phase boundary is $T_{0}^{(c)} =0.29 K$, which is much smaller than that of Fe$_8$ molecular cluster.
\begin{figure}
\includegraphics[width=3.5in]{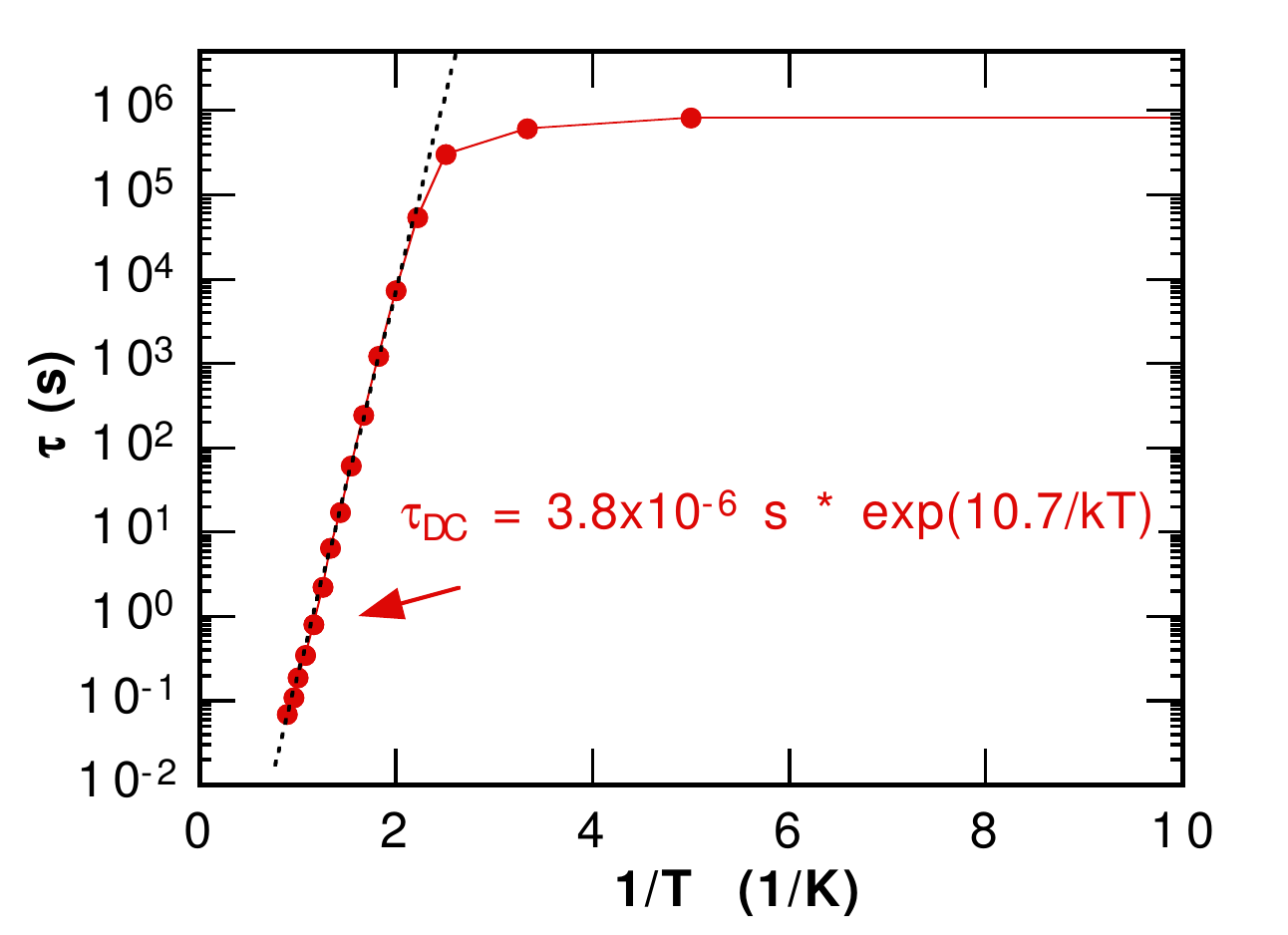}
\caption{ Color online: Arrhenius plot of the relaxations times $\tau$ vs. the inverse temperature for [Mn$_4$]$_2$ dimer with the model Hamiltonian $\hat{H}=J\hat{\bold{S}}_A \cdot \hat{\bold{S}}_B  - D(\hat{S}_{A,z}^{2} +\hat{S}_{B,z}^{2}) +g\mu_B\mu_0 h (\hat{S}_{A,z}+\hat{S}_{B,z}) $. Adapted with permission from \onlinecite{wern4}} 
\label{pot30}
\end{figure}
 In Fig.\eqref{pot30}, we show the experimental result of the Arrhenius plot of [Mn$_{4}$]$_{2}$ dimer. The plot shows that the relaxation rate is temperature-dependent above   ca. $0.3 K$ with $\tau_0=3.8\times 10^{-6}s$ and $\Delta U= 10.7 K$ and below ca. $0.3 K$, the relaxation rate is temperature-independent with a relaxation rate of $8\times 10^{5}s$ indicating the quantum tunneling of the spins between the ground states \cite{wern4}. The hysteresis loops in Fig.\eqref{pot24} show  the tunneling transitions through plateaus as obtained from experimental measurement. The step heights are temperature independent below $400$mK, which indicates quantum tunneling between the ground energy states.

\section{Conclusion and discussion}
\label{con}
In this review we discussed recent theoretical and experimental developments on macroscopic quantum tunneling and phase transitions in spin systems. 
We reviewed  different theoretical approaches to the problem of spin tunneling in single molecule magnets and exchange coupled dimer models. It is now understood that the suppression of tunneling at zero magnetic field for half-odd integer spin system is independent of the coordinate representation but only depends on the WZ or Berry phase term. This is related to Kramers degeneracy, and its experimental confirmation has been reported\cite{wern3}.  Theoretically, it is still an open problem to determine the necessary conditions in which classical degenerate ground state for half-odd integer spin implies  degenerate ground states in the pure quantum case. In the presence of a magnetic field along the spin hard anisotropy axis, tunneling is not suppressed for half-odd integer spins but rather oscillates with the field in accordance with the experimental observations.

Experimental and theoretical research on single-molecule magnets have focused on the search for other molecular magnets that exhibit tunneling and crossover temperatures. This research is expanding rapidly, and with the advance in technology, these molecular magnets have been used in the implementation of Grover's algorithm and magnetic qubits in quantum computing \cite{ll2001, tej1}. Other interesting areas include tunneling of Ne\'el vector in antiferromagnetic ring clusters with even number of spins \cite{me,ml2001, taft}. As far as we know the odd number of antiferromagnetic spin chain has not been reported. The present authors have suggested that this might give rise to solitons due to the spin frustration \cite{sm5}. Most experimental research has focused on organizing the SMMs into layers with the possibility of singling out the individual molecules\cite{leon, leon1}.
\section*{Acknowledgments} 
The authors would like thank NSERC of Canada for financial support. We thank Ian Affleck, Sung Sik Lee and Joachim Nsofini for useful discussions.

\bibliographystyle{apsrmp}

\end{document}